\documentclass{aa}  

\usepackage{graphicx}
\usepackage{txfonts}
\usepackage{natbib}
\usepackage[colorlinks=true,citecolor=blue,breaklinks=true,allcolors=blue]{hyperref}
\usepackage{xspace}
\usepackage{amsmath,amssymb}
\usepackage{enumitem}
\usepackage{upgreek}
\usepackage{tabularx}
\usepackage{float}
\usepackage{afterpage}   
\usepackage{arydshln}
\usepackage{nicefrac}
\usepackage{subcaption}

\usepackage{enumitem,kantlipsum}
\renewcommand{\theenumi}{\roman{enumi}}

\bibpunct{(}{)}{;}{a}{}{,}

\newcommand{\MJup}{\ensuremath{M_{\mathrm{Jup}}}\xspace}
\newcommand{\RJup}{\ensuremath{R_{\mathrm{Jup}}}\xspace}
\newcommand{\Teff}{\ensuremath{T_{\mathrm{e\!f\!f}}}\xspace}
\newcommand{\logg}{\ensuremath{\log\,(g)}\xspace}
\newcommand{\met}{\ensuremath{\mathrm{[M/H]}}\xspace}
\newcommand{\co}{\ensuremath{\mathrm{C/O}}\xspace}
\newcommand{\Av}{\ensuremath{A_{\mathrm{v}}}\xspace}
\newcommand{\fsed}{\ensuremath{f_{\mathrm{sed}}}\xspace}
\newcommand{\vsini}{\hbox{$v \sin (i)$}\xspace}
\newcommand{\mic}{\ensuremath{\upmu\mathrm{m}}\xspace}
\newcommand{\formosa}{\texttt{ForMoSA}\xspace}
\newcommand{\sono}{\texttt{Sonora Diamondback}\xspace}
\newcommand{\atmo}{\texttt{ATMO}\xspace}
\newcommand{\btex}{\texttt{BT-Settl}\xspace}
\newcommand{\btse}{\texttt{BT-SETTL$_{+ C/O}$}\xspace}
\newcommand{\exor}{\texttt{Exo-REM}\xspace}

\usepackage[dvipsnames]{xcolor}

\begin{document}

\title{The planetary-mass-limit VLT/SINFONI library\thanks{Based on observations collected at the European Organisation for Astronomical Research in the Southern Hemisphere under ESO programs 092.C-0535(A), 092.C-0803(A), 092.C-0809(A), 093.C-0502(A), and 093.C-0829(A\&B)}}
\subtitle{Spectral extraction and atmospheric characterization via forward modeling}

\author{P. Palma-Bifani \inst{\ref{inst:OCA},\ref{inst:LESIA}}, 
M. Bonnefoy \inst{\ref{inst:IPAG}}, 
G. Chauvin \inst{\ref{inst:MPIA}, \ref{inst:OCA}}, 
P. Rojo \inst{\ref{inst:DAS}},
P. Baudoz \inst{\ref{inst:LESIA}}, 
B. Charnay \inst{\ref{inst:LESIA}},
A. Denis \inst{\ref{inst:LAM}}, \\
K. Hoch \inst{\ref{inst:STSI}},
S. Petrus \inst{\ref{inst:nasa}},
M. Ravet \inst{\ref{inst:IPAG}, \ref{inst:OCA}},
A. Simonnin \inst{\ref{inst:OCA}}, and
A. Vigan \inst{\ref{inst:LAM}} 
} 

\institute{
Laboratoire Lagrange, Université Cote d’Azur, CNRS, Observatoire de la Cote d’Azur, 06304 Nice, France \label{inst:OCA}
\and
LIRA, Observatoire de Paris, Univ PSL, CNRS, Sorbonne Univ, Univ de Paris, 5 place Jules Janssen, 92195 Meudon, France \label{inst:LESIA}
\and
Université Grenoble Alpes, CNRS, IPAG, F-38000 Grenoble, France \label{inst:IPAG}
\and
Max-Planck-Institut fur Astronomie, Konigstuhl 17, 69117 Heidelberg, Germany\label{inst:MPIA}
\and
Departamento de Astronom\'ia, Universidad de Chile, Casilla 36-D, Santiago, Chile \label{inst:DAS} 
\and
Aix Marseille Univ, CNRS, CNES, LAM, Marseille, France \label{inst:LAM}
\and
Space Telescope Science Institute, 3700 San Martin Drive, Baltimore, MD 21218, USA \label{inst:STSI}
\and
NASA-Goddard Space Flight Center, Greenbelt, MD 20771, USA \label{inst:nasa}
}

\titlerunning{The VLT/SINFONI Library}
\authorrunning{P. Palma-Bifani et al.}
\date{Acceptance date: 25/06/2025}

\abstract
{Medium-resolution spectra (R$_{\lambda} \sim 1000 - 10 000$) at near-infrared wavelengths of young M-L objects enable the study of their atmospheric properties. Specifically, by unveiling a rich set of molecular features related to the atmospheric chemistry and physics.}
{
We aim to deepen our understanding of the M-L transition on planetary-mass companions and isolated brown dwarfs, and search for evidence of possible differences between these two populations of objects. To this end, we present a set of 21 VLT/SINFONI K-band ($1.95 - 2.45 \mathrm{\mu}$m) observations from five archival programs at R$_{\lambda} \sim 4000$. 
We aim to measure the atmospheric properties, like the \Teff, \logg, \met, and \co, to understand the similarities and differences between objects ranging from M5 to L5 in spectral type. 
}
{
We extracted the spectra of these targets with the \texttt{TExTRIS} code. Subsequently, we model them using \formosa, a Bayesian forward modeling tool for spectral analysis, exploring four families of self-consistent atmospheric models: \atmo, \btex, \exor, and \sono.
}
{
Here we present the spectra of our targets and the derived parameters from the atmospheric modeling. We confirm a drop in \Teff as a function of the spectral type of more than 500 K at the M/L transition. In addition, we report \co measurements for 3 companions (2M\,0103\,AB\,b, AB\,Pic\,b, and CD-35\,2722\,b), which add to the growing list of exoplanets with measured \co ratios. 
}
{
The VLT/SINFONI Library highlights two key points.
First, there is a critical need to further investigate the discrepancies among grids of spectra generated by self-consistent models, as these models yield varying results and do not uniformly explore the parameter space.
Second, we do not observe obvious discrepancies in the K-band spectra between companions and isolated brown dwarfs, which potentially suggests that these super-Jupiter objects formed through a similar process; however, this warrants further investigation.
}
   

\maketitle

\nolinenumbers
\section{Introduction}\label{sec:intro}

\begin{figure*}[ht!]
\centering
\includegraphics[width=0.7\hsize]{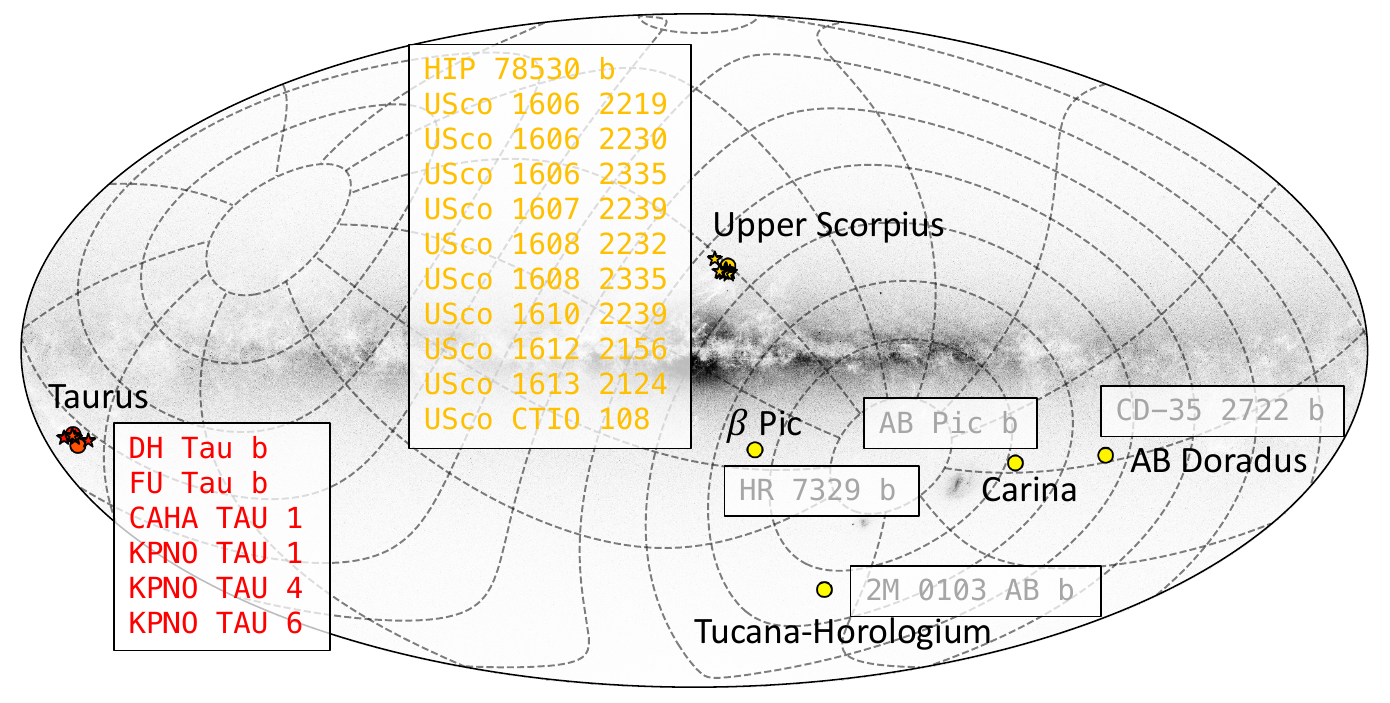}
\caption{Sky map showing right ascension and declination curves together with our 21 targets. The circle marks represent companions, while the star marks indicate isolated objects. The names of the different associations are indicated in black.
}
\label{fig:sample}
\end{figure*}

The discovery of the brown dwarf Teide~1 by \citet{Rebolo1995} marked a milestone in astronomy, as it was the first confirmed object of its kind. A decade later, the first direct image of a giant exoplanet 2M1207~b by \citet{Chauvin2004}, followed. Remarkably, these two objects share the same spectral type (M9), highlighting right from the beginning the challenge in distinguishing between a planet and a brown dwarf. 
Initially, the distinction between giant planets and brown dwarfs was defined by a mass threshold \citep{Kumar2002ds}, with dividing lines at $\sim$13.6\,\MJup and $\sim$75\,\MJup, corresponding to the limits for deuterium and hydrogen fusion, respectively \citep{Burrows1997, chabrier2005, Hayashi1963}. However, uncertainties remain regarding the Deuterium boundary \citep{Spiegel2011, Molliere2012}, and we also know of isolated objects with masses below 13.6\,\MJup. Therefore, a more detailed definition is needed to classify exoplanets and brown dwarfs.

In this context, an alternative approach is classifying these objects based on their formation pathways \citep{Chabrier2014}, a physically meaningful distinction since formation processes relate to the origin and nature of these bodies. Objects near the planetary-mass limit can form through three main mechanisms: they can form by the direct-gravitational collapse of a molecular cloud, like stars, but without sufficient mass for sustained fusion \citep{Padoan2005}, in which case they are classified as brown dwarfs. Alternatively, they may form within the gravitational influence and from the protoplanetary material of a newly formed star. These exoplanets could emerge through a rapid, top-down process, such as gravitational instabilities \citep{Boss1997}, or a slower, bottom-up mechanism, like core or pebble accretion \citep{Pollack1996}.

Numerous studies have explored and probed the overlapping regime between brown dwarfs and planets, as \citet{Leconte2009}, \citet{Faherty2013}, \citet{Bonnefoy2014s}, \citet{Biller2017}, and \citet{Nielsen2019}, to mention just a few.
From these and other studies, it has been suggested that brown dwarfs can be considered as exoplanet analogues; however, we also know that they were likely formed through different mechanisms. Different formation pathways should leave distinct imprints on their atmospheres, which we could potentially measure. An early approach in this direction was that the value of the carbon-to-oxygen ratio (\co) could serve as a proxy for the formation location of an exoplanet \citep{oberg2011,Madhusudhan2012}, extended later to other elements such as the nitrogen-to-oxygen (N/O) or nitrogen-to-carbon (N/C) ratios \citep{Cridland2017,Turrini2021}, and the refractory-to-volatile ratio \citep{Lothringer2021}.
More recently, studies such as \citet{molliere2019} and \citet{Gandhi2023} have suggested that isotopic ratios of elements like hydrogen, carbon, oxygen, and nitrogen could provide even tighter constraints to planet formation models.
It has also been proposed that the overall system's architecture can provide insights into the formation mechanisms as well \citep[e.g.][]{Desgrange2023}.

To better understand these connections, many studies have focused on characterizing directly imaged objects at the planetary-mass limit, spanning spectral types from T to M, using instruments that operate at low-to-medium spectral resolutions within the wavelength range of $\sim$0.5 to $\gtrsim$2.5 \mic. 
For example, the NIRSPEC instrument at Keck has been used for several brown dwarf studies, including the 17 targets observed by \citet{Rice2009s} and the 228 targets observed by \citet{Martin2017s}. 
Additionally, the X-Shooter instrument at the Very Large Telescope (VLT) has proven to be efficient for studying young, low-gravity objects, as evidenced by the observations of 9 targets by \citet{Petrus2020s} and 20 targets by \citet{Manjavacas2020s}. The instrument used in this study is SINFONI, located at the VLT. SINFONI has been employed in previously published libraries, such as the nine targets observed by \citet{Patience2012s} and the 15 targets observed by \citet{Bonnefoy2014s}.
In this context, we can expect to see more such libraries in the near future, for example, for the instruments ERIS, GRAVITY+, and CRIRES+ at the ESO/VLT and ESO/VLTI.

However, even though we are now in an era where we have (and are still gathering) extensive data, atmospheric \citep[e.g.,][]{Molliere2022} and protoplanetary disk studies \citep[e.g.,][]{vanderMarel2021} have demonstrated that the picture is far more complex than initially anticipated. In fact, most studies converge on a critical point: further homogeneous characterization of exoplanets is essential for enabling robust statistical analyses of their properties, which are crucial for understanding how these planets formed and evolved \citep[e.g.][]{Marleau2019,Petrus2021,wang2021,Hoch2023,PalmaBifani2023}. In light of this, it is worth asking whether the ongoing search for parameters that can trace formation pathways remains effective. A more meaningful approach may lie in consistently modeling spectra of diverse exoplanets and planetary analogues using uniform assumptions and observational strategies, thereby beginning to unravel the true complexity of their formation histories.

In this spirit, we identified a previously overlooked dataset of planetary-mass objects observed with the VLT/SINFONI instrument dating back to 2014. This rich dataset includes observations of young companions and isolated objects near the planetary mass limit.
We begin this document by presenting the sample and giving a detailed description of the observations, data reduction, and spectral extraction process in Section \ref{sec:data}. 
In Section \ref{sec:models}, we outline the forward modeling approach and describe the atmospheric models used in our study and our fitting strategy. 
The main results are reported in Section \ref{sec:results}, and the interpretation is discussed in Section \ref{sec:discussion}, leading to our conclusions in Section \ref{sec:conclusions}.

\section{The VLT/SINFONI observations}\label{sec:data}

We identified an archival library of medium-resolution K-band spectra. The corresponding observing programs were executed a decade ago, but no publication has yet presented or homogeneously characterized this dataset.
In this section, we first describe the SINFONI instrument in Subsection \ref{ssec:instrument}.
We characterize the selected sample of objects in Subsection \ref{ssec:sample} and outline the data reduction steps in Subsection \ref{ssec:reduction}.
Finally, we present the complete compilation of the SINFONI Library spectra in Subsection \ref{ssec:spectra}.

\subsection{The SINFONI instrument}\label{ssec:instrument}

SINFONI was an integral field spectrograph (IFS) operational at the European Organisation for Astronomical Research in the Southern Hemisphere (ESO) from 2004 to 2021. It was decommissioned in 2021 and replaced by the ERIS instrument, which, in its IFS configuration, shares similar properties.
It combined two primary subsystems: the adaptive optics (AO) module, MACAO, and the IFS known as SPIFFI. This configuration enabled the simultaneous spectroscopy of 32$\times$64 spatial pixels (spaxels) with medium spectral resolving power. 
SINFONI covered the near-infrared atmospheric windows (J: 1.1–1.4 \mic, H: 1.45–1.85 \mic, K: 1.95–2.45 \mic) with the additional capability to observe the combined H+K bands at a lower spectral resolution.

The SPIFFI spectrograph utilized an image slicer to divide the field of view (FoV) into 32 slices (slitlets) and rearranged them into a one-dimensional pseudo-slit, which is dispersed onto a two-dimensional 2048$\times$2048 pixels Hawaii Rockwell focal plane array \citep{Eisenhauer2003}. Four gratings, corresponding to the J, H, K, and H+K bands, were available, with spectral resolution influenced by the selected pre-optics. 
The FoV was recomposed spatially into rectangular spaxels, with plate scales of $0.25/0.1/0.025$ arcsec/pixel, offering resolutions of $R_{\lambda} = \frac{\lambda}{\Delta \lambda} = 4490$, 5090, or 5950, respectively, in the K-band.

Regarding performance, the line spread function (LSF) of SINFONI varied with plate scales and gratings due to differences in how the pre-optics illuminated the gratings. Nyquist sampling was not achieved in all modes, necessitating careful calibration to ensure accurate spectral resolution.
The Hawaii 2RG detector also exhibited persistence effects, even when illuminated at levels below saturation. This limitation required meticulous calibration to mitigate artifacts and maintain data quality.

\subsection{The sample}\label{ssec:sample}

The observations used in this work are part of an archival library of medium-resolution VLT/SINFONI K-band spectra, observed approximately a decade ago as part of five different programs: 092.C-0803(A) and 093.C-0829(A\&B) (P.I.: Kopytova),  093.C-0502(A) and 092.C-0535(A) (P.I.: Radigan), and  092.C-0809(A) (P.I. Patience). 
As part of the appendix, Table \ref{tab:Obslog} provides the observational log, detailing parameters such as the number of exposures per sequence, integration times (DIT), and the number of integrations per exposure (NDIT), as well as the airmass and seeing ranges for each night.

While SPIFFI is an essential component of SINFONI, the use of MACAO, the adaptive optics (AO) system, depends on the observational strategy (seeing-limited versus diffraction-limited). In Table \ref{tab:Obslog}, we show that MACAO was off when the observations of the isolated targets and FU Tau b were taken, as these objects are not contaminated by the presence of a bright star close by. 
For the remaining companions, MACAO was utilized, with the host star serving as a natural guide star (NGS) for wavefront sensing, generally providing a better signal-to-noise ratio.
Furthermore, different plate scales were used for the various targets, which impacted the choice of aperture radius and the true spectral resolution.
In addition, for each science target, a telluric standard star (STD) was observed immediately before or after each sequence to ensure proper telluric contamination subtraction.

The selected sample consists of 21 planetary-mass objects, as listed in Figure \ref{fig:sample}. 
A broad overview of previous studies and the characteristics of each target is available in Appendix \ref{sec_a:pertarget}.
The sky map displayed in Figure \ref{fig:sample} was generated using the \texttt{mw-plot} Python package\footnote{\texttt{mw-plot} documentation: \url{https://milkyway-plot.readthedocs.io}}.
Figure \ref{fig:sample} illustrates that the sample primarily comprises two architectures: directly imaged companions (targets ending by letter b) and isolated brown dwarfs.
All our targets are young (1–30 Myr) and predominantly located in two star-forming regions, Taurus and Upper Scorpius \citep{SiciliaAguilar2013, Torres2008}. The exceptions are the four companions marked in yellow, which are recognized members of the $\beta$\,Pictoris, Carina, AB Doradus, and Tucana-Horologium moving groups \citep{Zuckerman2004, Zuckerman2004b, deSilva2009}.
In terms of distance, all objects are within 150\,pc from the Sun and visible from the Chilean Atacama desert, where the Very Large Telescope (VLT) is located.

In general, depending on the sky location, we expect different amounts of interstellar dust to cause extinction, which we have estimated based on the Gaia 3-dimensional extinction map from \citet{Lallement2022}.
The colors in Figure \ref{fig:sample} correspond to the amount of interstellar extinction (\Av), revealing three distinct groups: The scattered companions have \Av values smaller than 0.001 magnitudes, the targets at the Taurus region have values between 0.05 and 0.25 magnitudes, and the targets at the Upper Scorpius regions have values between 0.25 and 0.35 magnitudes, except for USco 1606-2219 and USco 1610-2239, whose extinctions are lower (see Table \ref{tab:Targets1}).
Together with the visible interstellar medium extinction (\Av), in Table \ref{tab:Targets1} we report all the previously mentioned information, such as the right ascension (RA), declination (Dec), association, parallax, distance to us, and age for each target.

\subsection{Data reduction and calibrations}\label{ssec:reduction}

\begin{figure}[ht]
\centering
\includegraphics[width=\hsize]{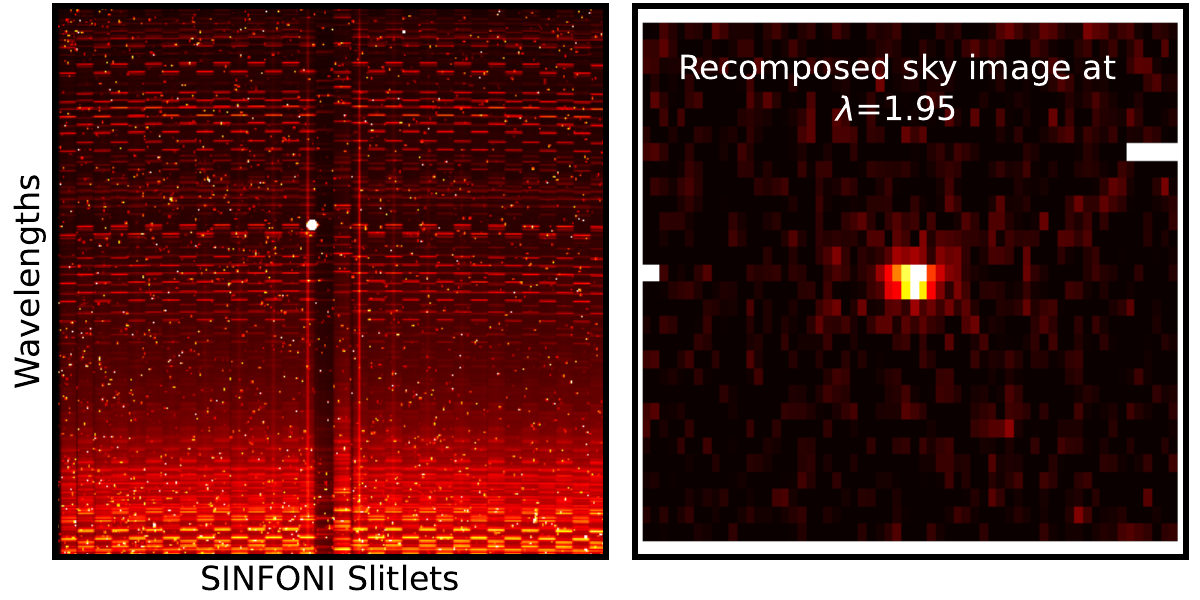}
\caption{Example of the first exposure of AB Pic b: The SINFONI raw observation (left), where each spectral slice (slitlet) is projected as vertical lines, compared to the reduced datacube delivered by EsoReflex for which we show the slice at $\lambda=1.95$ (right), which shows a clear detection of the source at the center.}
\label{fig:raw_reduced}
\end{figure}

We processed the raw data with the SINFONI handling pipeline v3.0.0\footnote{The SINFONI Pipeline: \url{https://www.eso.org/sci/software/pipelines/sinfoni}} through the \texttt{EsoReflex}\footnote{\texttt{EsoReflex} Software: \url{www.eso.org/sci/software/esoreflex}} environment. 
The SINFONI pipeline utilizes calibration frames (darks and flats) to perform basic adjustments on the raw science frames, correcting for bad pixels and pixel-to-pixel sensitivity variations as a function of wavelength. 
In practice, the pipeline identifies the slitlet positions on the frames via Gaussian fitting at each wavelength. From this, a data cube is built for each exposure, correcting for spatial and spectral distortion, as shown in Figure \ref{fig:raw_reduced}.

\begin{figure*}[ht]
\centering
\includegraphics[width=0.99\hsize]{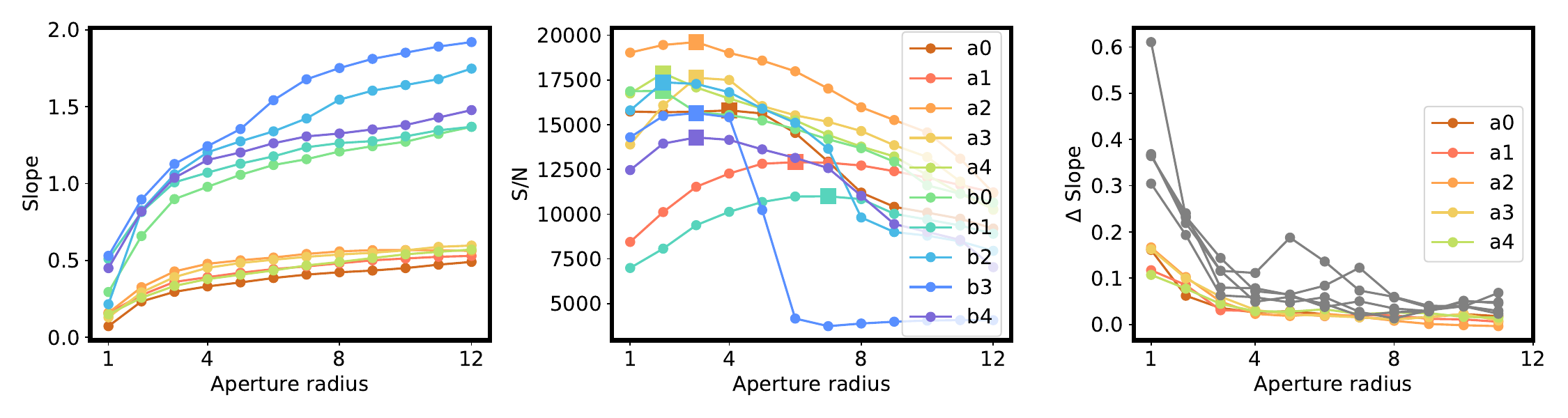}
\caption{Example of optimal aperture radius selection and spectra rejection criteria using the AB Pic b dataset. In each panel, each color represents a different exposure of AB Pic b.
\textbf{Left:} Here, we show the values for the fitted continuum slope between 2.05 and 2.2 \mic as a function of aperture radius. Two distinct domains are observed: for radii smaller than the optimal value, the slope increases steeply; for radii larger than the optimal value, the slope still varies but becomes more stable.
\textbf{Center:} Here, we show the signal-to-noise ratio (S/N) estimated within the same wavelength range as in the left panel, plotted as a function of aperture radius. For each observation, we highlighted the maximum S/N value with a square. The aperture radius corresponding to this maximum is our final selected radius each time.
\textbf{Right:} Here, we show the derivative of the slope (reported in the left panel) with respect to the aperture radius. Some targets exhibit a non-smooth, decreasing derivative; we reject those epochs. For AB Pic b, this resulted in the rejection of the entire dataset for the second night.
}
\label{fig:red_radselec}
\end{figure*}

Further corrections, including wavelength calibration, sky subtraction, and telluric removal, were necessary to extract a spectrum for each target.
In this study, we employed the \textit{Toolkit for Exoplanet deTection and chaRacterization with IfS} (TExTRIS), similarly to its implementation in previous works, such as \citet{Petrus2021, PalmaBifani2023, Kiefer2024}.
Hereafter, we detail the steps and specific post-processing techniques applied to extract spectra from each data cube.

We began by performing a wavelength calibration for each exposure using the OH emission lines.
Across the datasets, we identified constant spaxel-to-spaxel wavelength shifts of up to approximately $15\,\mathrm{km/s}$ relative to the telluric absorption lines.
We employed a cross-correlation method within a defined wavelength interval to correct these wavelength shifts and determine the offsets in a sequence of SINFONI data cubes. Then, we applied this offset to correct the corresponding cube wavelength solutions in the same way for the science and standard star cubes.

Before applying additional corrections, the exact position of each object's PSF center in each data cube must be determined.
We measured the motion of the point source affected by atmospheric refraction by fitting a 2D Moffat function, recovering the center coordinates in each cube as a function of spaxels and wavelength.
The wavelength dependency of the PSF center's motion is smooth and can be approximated by a polynomial of degree 2. This approximation allowed us to bin the data in wavelength before estimating the center position, thereby reducing computation time.
For the science objects, we binned every 10 points, while for the standard stars—which are brighter—we binned every 100 points.

We observed that sky emission lines heavily contaminated several of our data cubes. 
To correct this contamination, and because the sky emission lines are wavelength-dependent, every spaxel must be corrected equally. 
The degree of sky emission contamination was estimated from empty sky regions. 
For this, we applied a ring mask around the object, with a large inner radius, to ensure we were not capturing any wing of the object's PSF. We manually adjusted the inner and outer radii for each case and masked out all other elements. 
We applied an additional border mask when the object was near a border. 
We then measured the median value among the spaxels of the selected sky portion and subtracted this median from each spaxel, repeating the process for all wavelengths. 
When the sky emission is low, this correction subtracts zero from every spaxel, but when it is high, it suppresses the lines. 
We applied this correction only to the science data cubes.

Now, we were ready to select an aperture radius and extract the spectra from the science and standard star datacubes. We extracted a spectrum with error bars within a circular aperture in an IFU data cube. The error bars were computed from the estimated level of residuals around the circular aperture. For this, we used the \texttt{photutils}\footnote{\texttt{photutils} documentation: \url{https://photutils.readthedocs.io/en/stable/}} Python package, specifically the \texttt{CircularAperture} function.
We fixed the aperture radius at 5 pixels for the standard star cubes. For the science cubes, we estimated the optimal aperture radius, which depends on several factors such as the plate scale and MACAO usage, and maximized the signal-to-noise ratio of our data by performing spectral extractions with aperture radii ranging from 1 to 12 pixels (in integer increments). This approach allowed us to optimize the radius selection later on.

The preliminary spectrum of each data cube (both science and standard star) shows contamination by water bands from the Earth's atmosphere. To address this, we obtained a spectrum of the atmospheric transmissions for all nights using the standard star observations.
First, we corrected the standard star spectrum by removing the features of near-infrared hydrogen and other stellar lines. We identified and interpolated these lines across the affected spectral ranges. Next, we used the standard star's spectral type as input to calculate its effective temperature (\Teff) and then computed a blackbody spectrum at that specific \Teff. We derived a telluric spectrum by dividing the corrected standard star spectrum by the corresponding blackbody spectrum.
Finally, we divided the telluric spectrum for each data cube, effectively removing telluric contamination from the science observations. We applied the same step to each data cube and each spectral extraction, varying the aperture radius.
We then applied a Barycentric correction to each corrected spectrum computed using the \textit{hellcor} function of the Python \texttt{PyAstronomy} library \citep{Czesla2019}.

\begin{figure*}[ht!]
\centering
\includegraphics[width=\hsize]{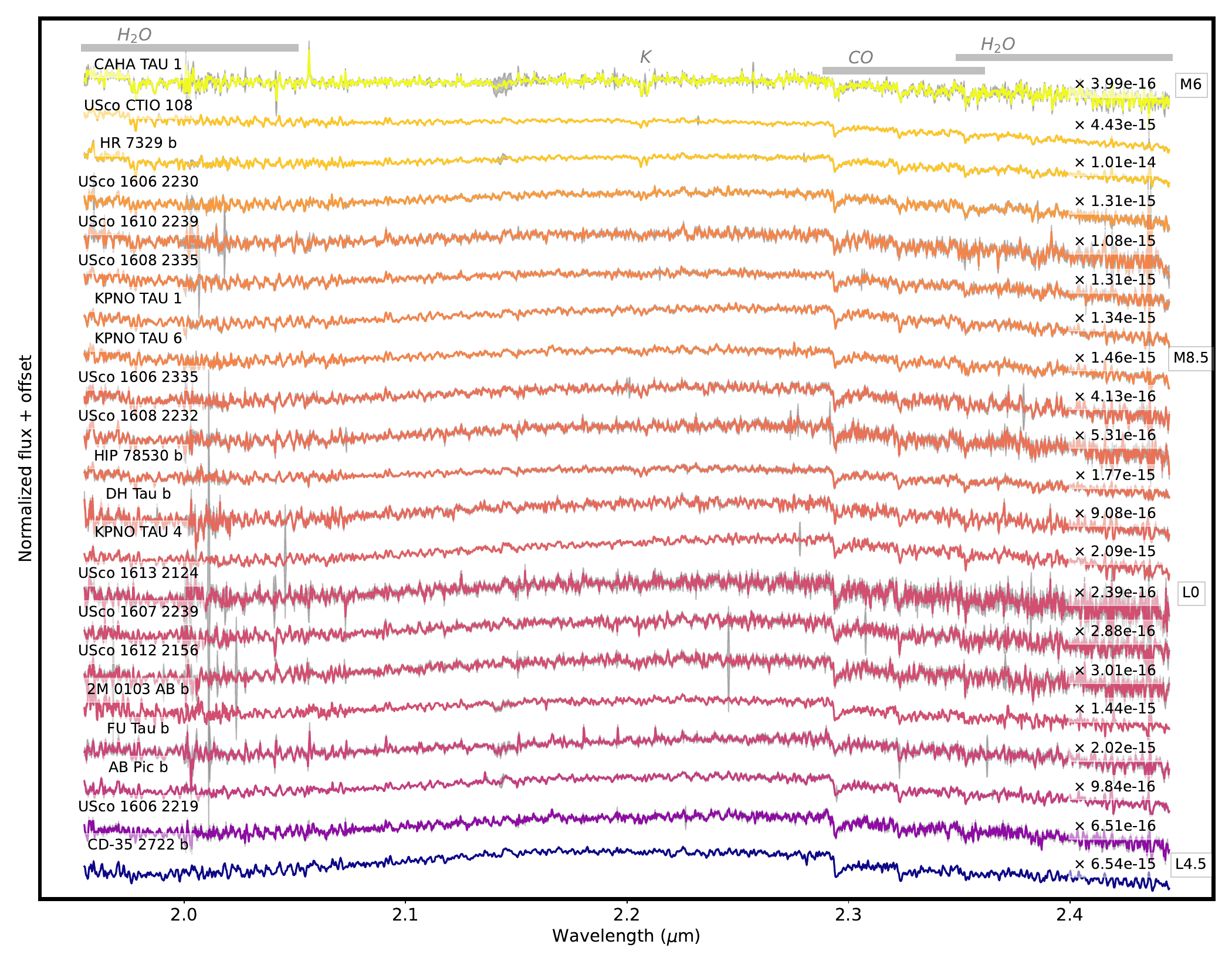}
\caption{Final extracted spectra for each of the 21 targets in the sample, organized and colored by their spectral type reported in the literature (see Table \ref{tab:Targets2}). The errors are reported in gray, and the main absorption features are labeled at the top of the figure, also in gray. To the right side, we have annotated the flux scaling factor for each target in units of $W \, m^{-2}\, \mu m^{-1}$, also reported in \ref{tab:Targets2}.
}
\label{fig:SINFONI}
\end{figure*}

At this stage, we have 12 corrected spectral extractions for each observation of each target. 
We will use AB Pic b as an example target from here onward.
Under ideal conditions, we can assume that the FWHM of the target’s PSF can be approximated by a Gaussian. For AB Pic b, a bright target, the FWHM is approximately 5 pixels, so we could select this aperture radius, as was done in \citet{PalmaBifani2023}. However, the FWHM can vary significantly with the considered plate scale, the seeing, MACAO's setup, and other factors. Therefore, having an objective criterion to select the best aperture radius becomes important to ensure that our extracted spectra for all 21 targets have the best possible quality.
To select the best aperture radius, we proceed as follows: In the wavelength range from 2.05 to 2.2 \mic, we know that M to L spectral type targets exhibit a positive-to-zero slope. 
Therefore, we fit a first-order polynomial to the data and analyze how the slope varies as a function of the aperture radius.
The left panel of Figure \ref{fig:red_radselec} illustrates this. Each color represents a different exposure of AB Pic b. 
In general, we observe that for most observations, the slope exhibits two distinct behaviors: for small aperture radii, it increases rapidly; beyond a critical radius, the rate of change of the slope decreases.

Building on this criterion, we can go further. Specifically, we estimated the signal-to-noise ratio (S/N) in the same spectrum wavelength range for each exposure. In the central panel of Figure \ref{fig:red_radselec}, we observe that the S/N ratio initially increases until it reaches a maximum value and then decreases. This behavior arises because, for small radii, not all of the PSF signal is captured within the aperture; consequently, both the signal and noise remain low. As the aperture radius increases, more planetary flux is included, raising the S/N. However, background contamination becomes dominant when the aperture radius is too large, thereby reducing the S/N ratio.
The central panel of Figure \ref{fig:red_radselec} illustrates this behavior. 
As an example, in the central panel of Figure \ref{fig:red_radselec}, we added a square symbol to mark the aperture radius that maximizes the S/N, which is selected as the final aperture radius. This applies to all targets and datasets.

With the described criterion, we can select the aperture radius, but the quality of the SINFONI observations varies from dataset to dataset. Some datasets exhibit behaviors that are clearly driven by systematic errors, such as wiggles in the spectra or discontinuities in the slope as a function of aperture radius.
In this sense, we also implemented a rejection criterion. We computed the derivative of the slope as a function of the aperture radius, as shown in the right panel of Figure \ref{fig:red_radselec}. 
If this derivative is not monotonically decreasing, we reject the dataset. For AB Pic b, this led to the rejection of the entire second night of observations. For some targets, we relaxed the criterion with a threshold, especially when MACAO was not used.

Finally, we median-combine the accepted spectra to create the final spectrum for each target. We report the corresponding errors as the standard deviation across the included datasets.
For each target, we then calibrated the extracted normalized spectrum in flux units ($W\,m^{-2}\,\mu m^{-1}$) using the K-band magnitude values reported in Table \ref{tab:Targets2} and the VLT/NaCo Ks filter, as previously done in \citet{PalmaBifani2023}. We report the flux factor used to multiply the normalized spectra to recover them in apparent flux units in Table \ref{tab:Targets2}.

\subsection{The SINFONI library}\label{ssec:spectra}

When we refer to the SINFONI library, we mean the collection of 21 reduced and corrected K-band spectra of planetary-mass companions and isolated targets. The entire SINFONI library is shown in Figure \ref{fig:SINFONI}. The library is organized by spectral type, from earlier to later types, as indicated by the color of each spectrum, labeled to the right. The whole collection of spectra is available together with this publication for public use.

From Figure \ref{fig:SINFONI}, we see that some targets appear "noisier," such as most of the Upper Scorpius (USco) targets. This results from observing without MACAO and using the largest plate scale, meaning that the FWHM of the PSF is larger, leading to observations with a lower S/N.
For example, we can also see that FU Tau b's spectrum exhibits many "emission" lines. 
These are not real emission lines from the companion but artifacts from imperfect OH-sky emission line correction, given that the FU Tau b observations are heavily contaminated.
Targets observed with MACAO and in long exposure sequences (such as CD-35 2722 b, observed with an intermediate plate scale and exposures of 150 seconds) exhibit very stable behavior, resulting in a high S/N spectrum.

In addition to the spectra, we have summarized the general properties of the sample in Tables \ref{tab:Targets1} and \ref{tab:Targets2}. These tables include the Gaia DR3 parallaxes \citep{GaiaCollaboration2023}, the corresponding distances in parsecs, and the expected interstellar medium extinction values derived from the 3D maps of \citet{Lallement2022}.
We also provide the age, spectral type, apparent and absolute K-band magnitudes, flux, and all the references used for each target.
To complement this information, we have used the age and absolute magnitudes, along with the BHAC15 evolutionary models \citep{Baraffe2015}, to compute predictions for key properties. Specifically, we employed both the DUSTY and COND versions of the BHAC15 models to estimate the expected effective temperature, surface gravity, radius, and mass, reported in Table \ref{tab:Targets_evolu}.
Having introduced our library, we proceed to further understand the main physical properties of these objects by characterizing the atmospheric properties of this sample.

\begin{table*}[ht!]
\centering
\caption{Comparison of the Different Grids of Atmospheric Models.
Columns 2 to 6 show the parameter ranges explored. Column 7 indicates whether clouds were included in the models, and Column 8 specifies whether the models were calculated at chemical equilibrium ($\rightleftharpoons$) or out of equilibrium, specified by an equality ($=$) or inequality ($\neq$) symbol, respectively. The final column presents the approximate spectral resolution of the grid at the K-band.}
\renewcommand{\arraystretch}{1.5} 
\begin{tabular*}{\linewidth}{@{\extracolsep{\fill}} lrccccccr }
\hline
Name & \Teff\,\,\,\,\,\,\, & \logg & \met & \co & others & Clouds & $\rightleftharpoons$ &  R$_{\lambda}$ \\ 
\hline
\atmo &  800 - 3000 & 2.5 - 5.5 & -0.6 - 0.6 & 0.3 - 0.7 & $\gamma$ $\in$ (1.01 - 1.05) & No & $\neq$ & $\sim$5000 \\
\btex & 1200 - 2900 & 2.5 - 5.5 & 0 & & & Yes & $\neq$ &$\sim$200000\\
\sono & 900 - 2400 & 3.5 - 5.5 & 0 & & \fsed $\in$ (1 - 8) & Yes & $=$ & $\sim$100000\\
\btse & 1400 - 2200 & 3.5 - 5.0 & 0 & 0.27 - 1.09 & & Yes & $\neq$ & $\sim$200000\\
\exor &  400 - 2000 & 3 - 5 & -0.5 - 1.0 & 0.1 - 0.8 & & Yes & $\neq$ & $\sim$5000 \\
\hline
\end{tabular*}
\label{tab:atmomodels}
\end{table*}

\section{Atmospheric modeling}\label{sec:models}

In the community, we typically encounter two approaches to modeling exoplanet atmospheres: the forward problem and the inverse modeling problem. These two methods can be seen as addressing the Bayesian problem from opposite directions, which, due to Bayes' theorem, are perfectly equivalent.
Our work here focuses on the forward modeling problem. 
In other words, we examine existing families of models, which are typically parameterized by a few variables. Some advantages of this approach are that the modeling process remains self-consistent and the computational cost for medium-resolution observations remains reasonable.
We present our modeling framework in Subsection \ref{ssec:formosa}. We then describe the grids of models we implemented in Subsection \ref{ssec:grids} and our fitting strategy for modeling and analyzing the entire SINFONI library in Subsection \ref{ssec:fitting}.

\subsection{\formosa}\label{ssec:formosa}

To model exoplanetary atmospheres and understand their physical and chemical processes, we use \formosa\footnote{\formosa stands for FORward MOdeling tool for Spectral Analysis}, an open-source Python package developed within our group (\formosa Collaboration et al., in prep.).
It has already been extensively used and described in various works, such as \citet{Petrus2021, PalmaBifani2023, Petrus2024, PalmaBifani2024}, as well as in our ReadTheDocs\footnote{\formosa documentation: \url{https://formosa.readthedocs.io}} documentation page.

In its simplest form, \formosa receives as input a set of observations, a specific grid of models, and a configuration file. 
First, \formosa adapts the grid of models to match the wavelength distribution of the observations. 
Then, \formosa performs a nested sampling parameter exploration, for which in this work, we selected the \texttt{Pymultinest}\footnote{\texttt{Pymultinest} documentation: \url{https://johannesbuchner.github.io/PyMultiNest/}} sampler with 500 live points. 
Finally, \formosa saves the outputs and includes a plotting module that allows quick visualizations and provides access to all inputs, parameters, and variables.
In this regard, \formosa also enables exploration of the posterior distribution of a selection of additional parameters, such as the radial velocity, rotational velocity, interstellar medium extinction, and the radius of the target, to better adapt the models to the data.

\subsection{Self-consistent grids of atmospheric models}\label{ssec:grids}

Currently, several efforts are underway in our community to develop atmospheric models. Each of these efforts makes different assumptions and simplifications. While different groups share their models as parameterized grids, these grids often have distinct features and explore varying parameters across different ranges.
Comparing these grids is essential, as demonstrated by \citet{Petrus2024}, who extensively modeled the atmosphere of VHS 1256 b observed with NIRSpec and MIRI on JWST \citep{Miles2023TheB}. The models implemented on that study included \atmo \citep{mod_Tremblin2015, mod_Tremblin2017}, \btse \citep{Allard2012}, \exor \citep{mod_Charnay2018}, \sono \citep{mod_Morley2024}, and \texttt{DRIFT-Phoenix} \citep{mod_Helling2008}.

Here, we propose a similar approach; however, instead of having a broad wavelength range as in the observations of VHS 1256 b, we focus solely on the K-band and model these observations homogeneously for the 21 targets in our sample.
For this, we use five different grids: \atmo, two versions of \btex (with and without exploring the \co ratio), \exor, and \sono. 
A comparison of the parameter space explored by each of these grids is provided in Table \ref{tab:atmomodels}.
The spectral resolutions in the last column were estimated, assuming that the synthetic spectra are Nyquist sampled. Hereafter, we provide a brief description of each grid.

\atmo \citep{mod_Tremblin2015, mod_Tremblin2017}:
One of the main characteristics of the \atmo models, which differentiates them from other grids, is that they do not consider clouds. The authors demonstrate that cloudless atmospheric models, which assume a temperature gradient reduction caused by fingering convection, provide a very good model for reproducing NIR spectra.
This spectral window is dominated by molecular features from H$_2$O, NH$_3$, CH$_4$, and CO.
Therefore, the models focus on non-equilibrium chemistry processes occurring at the L-T and T-Y transitions, including CO/CH$_4$ and N$_2$/NH$_3$, as described in \citet{Venot2012}. Differences in mean molecular weights across atmospheric layers can drive fingering convection (mixing two liquids with different densities), which reduces the temperature gradient between layers. This effect can be modeled using the adiabatic index parameter ($\gamma$), explaining the reddening observed in low-mass objects.
In these models, mixing length theory is used to describe convective mixing. Additionally, the models include Rayleigh scattering from H$_2$ and He, and assume chemical equilibrium among 150 species, including 30 condensates, which are modeled by minimizing the Gibbs free energy.
Finally, the radiative transfer is solved using the line-by-line approach described in \citet{Amundsen2014}, with the correlated-k method employed to enhance computational efficiency.

\btex and \btse \citep{Allard2012, Allard2014}:
The \btex models originated in the stellar modeling community and were later extended to lower temperatures. For this extension, non-equilibrium chemistry was incorporated similarly to \atmo, but without the parametrization for fingering convection.
For the chemistry, solar abundances were adopted from the revised values of \citet{Asplund2009}. The \citet{Rossow1978} cloud model was implemented, incorporating radiation-hydrodynamical simulations of mixing under hydrostatic equilibrium, which allows for supersaturation and mixing processes. This framework enables modeling of the stellar–substellar transition by comparing the timescales of condensation, sedimentation, and mixing (extrapolated from convective velocities into stable layers) while ignoring coalescence and coagulation.
The models assume efficient nucleation for 40 species, and the radiative transfer through the atmosphere is solved using the PHOENIX atmospheric code \citep{Hauschildt1997, Allard2001}.
The most extensive version of the BT-SETTL grid spans a wide temperature range, from 70,000 K for stars to 400 K for young planets. However, the versions used here were truncated to match the \Teff of interest, and \btse also explores the \co ratio as an additional parameter. An observed limitation of \btex is that it does not allow for sufficient dust formation in brown dwarf atmospheres due to a conservative supersaturation threshold. This results in a persistent offset when reproducing the spectra of ultra-cool objects.

\exor \citep{Baudino2015, Baudino2017, mod_Charnay2018, Blain2021}:
The \exor models are based on radiative-convective equilibrium with a simplified treatment of cloud microphysics relatively similar to \btse. They incorporate opacities for collision-induced absorption by H$_2$ and He, as well as nine molecular and atomic species, and solar abundances as defined by \citet{Lodders2010}. For these models, atmospheric vertical mixing is parameterized using a fixed eddy mixing coefficient for a cloud-free environment, with clouds subsequently added on top of it. Similar to \atmo and \btex, \exor includes non-equilibrium chemistry, particularly relevant in cooler and metal-rich atmospheres where vertical mixing dominates.
The \exor models also account for clouds composed of Fe and Mg$_2$SiO$_4$, modeled using a simplified microphysics approach. Cloud particle radii are determined by comparing the timescales for condensation growth, coalescence, vertical mixing, and sedimentation with the shortest timescale process governing the particle size. A single representative radius is assumed for all cloud particles \citep{Rossow1978}, with particle growth limited by removal through sedimentation or mixing.
The primary advantage of \exor lies in its computational efficiency and its ability to explore non-solar metallicities.

\sono \citep{mod_Morley2024}:
The \sono grid is a recent addition to the SONORA family, which also comprises the \texttt{SONORA-Bobcat} \citep{Marley2021}, \texttt{SONORA-Cholla} \citep{Karalidi2021}, and \texttt{SONORA-Elf-Owl} \citep{Mukherjee2024} models.
The \sono models are based on the radiative-convective equilibrium framework described in \citet{MarleyMcKay1999}, with vertical mixing modeled using mixing-length theory. 
\sono includes opacities for 15 molecules and atoms (e.g., silicates, sulfides, salts, Fe, and H$_2$O), as well as collision-induced absorption by H$_2$ and He, with atmospheric chemistry set to solar abundances \citep{Lodders2010}. \sono include clouds, but unlike all previously presented grids, chemical equilibrium is assumed throughout the atmosphere.
Clouds in \sono are parameterized using the approach of \citet{AckermanMarley2001}, where sedimentation efficiency is controlled by the \fsed parameter. Unlike \exor, this cloud parametrization assumes that at each vertical level, particle size evolves by matching the timescale of mixing to the timescale of sedimentation. This parameterization enables the model to span a wide range of cloud conditions, from entirely clouded (\fsed = 1) to nearly cloudless (\fsed = 8) atmospheres.
The \texttt{SONORA} family of models is notable for its high spectral resolution and flexible cloud modeling, allowing for accurate simulations of various atmospheric conditions, particularly for cooler substellar atmospheres. However, the assumption of chemical equilibrium throughout the atmosphere limits its ability to capture non-equilibrium processes, often crucial for atmospheres with substantial vertical mixing.

\begin{table*}[ht!]
\centering
\caption{Summary of processing times and mean $\chi^2_{red}$ for the different model configurations. Each number reported to the left represents the time it took for the 21 targets to converge for that specific model setup in units of minutes. The number reported to the right represents the $\chi^2_{red}$ average value between the 21 targets of our sample, truncated to the first decimal.}
\renewcommand{\arraystretch}{1.5} 
\begin{tabular*}{\linewidth}{@{\extracolsep{\fill}} clccccc }
\hline
Label & Configuration & \atmo & \btex & \btse & \exor & \sono\\ 
 & & 
\scalebox{0.6}{ time | $< \chi^2_{red} >$} & 
\scalebox{0.6}{ time | $< \chi^2_{red} >$} & 
\scalebox{0.6}{ time | $< \chi^2_{red} >$} & 
\scalebox{0.6}{ time | $< \chi^2_{red} >$} & 
\scalebox{0.6}{ time | $< \chi^2_{red} >$} \\ 
\hline
v01 & Grid                      & 260  |  35.3 &  33  |  22.8 & 44  |  22.0   & 375  |  27.8   & 40  |  26.7  \\    
v02 & Grid + RV                 & 381  |  23.2 &  41  |  17.5 & 21  |  18.5   & 514  |  19.8   & 55  |  22.2 \\  
v03 & Grid + RV + $\beta$       & 558  |  15.6 &  72  |  15.3 & 41  |  18.1   & 754  |  15.7   & 101  |  19.6\\ 
v04 & Grid + \Av                & 335  |  29.9 &  72  |  18.5 & 23  |  21.7   & 423  |  26.2   & 60  |  20.6  \\     
v05 & Grid + \Av + RV           & 476  |  18.8 &  70  |  13.2 & 36  |  18.2   & 938  |  19.5   & 74  |  16.9  \\ 
v06 & Grid + \Av + RV + $\beta$ & 789  |  12.8 & 133  |  11.3 & 56  |  17.7  & 1322  |  15.7  & 117  |  15.7  \\ 
v07 & Grid + $A_{\mathrm{v}}^{Gaia}$ + RV + $\beta$  & 876  |  14.6 & 311  |  14.6 & 223  |  17.8 & 1156  |  15.4 & 316  |  19.5 \\
v08 & Same as v07 + $p$(\logg)  & 1440  | 14.9  &   723 | 14.6  & 924  | 20.1  & 2032  | 16.4 & 469 | 19.7  \\
\hline
\end{tabular*}
\label{tab:processing_times}
\end{table*}

As observed, each family of models adopts different assumptions and explores a distinct set of parameters, as illustrated in Table \ref{tab:atmomodels}. 
While all models achieve resolutions equal to or higher than the SINFONI data and consistently explore \Teff and \logg as free parameters, their parameter space coverage varies significantly.
From the \Teff ranges presented in Table \ref{tab:atmomodels}, we note that the lower limits of the grids extend below the expected temperature range for our library (between 1500 K and 2500 K, expected for spectral types from L5 to M5). However, the upper limits of the grids vary: only \atmo and \btse extend sufficiently high to encompass this range fully. The other grids have truncated \Teff ranges, which may not cover the hottest objects in our sample.
This limitation means that we cannot use all grids for all targets despite our aim for a homogeneous analysis. Instead, our fitting approach must account for these discrepancies, as further described in the following section. Moreover, the differences in model assumptions, such as the treatment of clouds, chemical equilibrium, and vertical mixing, further emphasize the need to carefully interpret the results and consider each grid's specific strengths and weaknesses in the context of our data.

\subsection{Fitting strategy}\label{ssec:fitting}

When performing an atmospheric parameter exploration with \formosa, the first thing that becomes evident is that the posterior distributions vary depending on the chosen combination of parameters. For a preliminary exploration of the entire sample and to address questions such as how \Teff and \logg behave and whether the solutions are physically consistent, we repeated the modeling with different setups.
In these setups, we systematically increased the number of free parameters explored. We began with only the grid parameters, then incrementally added parameters such as radial velocity (RV), a spectral line broadening factor ($\beta$), interstellar medium extinction (\Av), and various combinations of these parameters, considering their physical significance. These setups were labeled from v01 to v07 in Table \ref{tab:processing_times}. 
Hereafter, we revisit the meanings of RV, $\beta$, and $\Av$ as additional free parameters.

\begin{figure}[ht!]
\centering
\includegraphics[width=1\linewidth,angle=0]{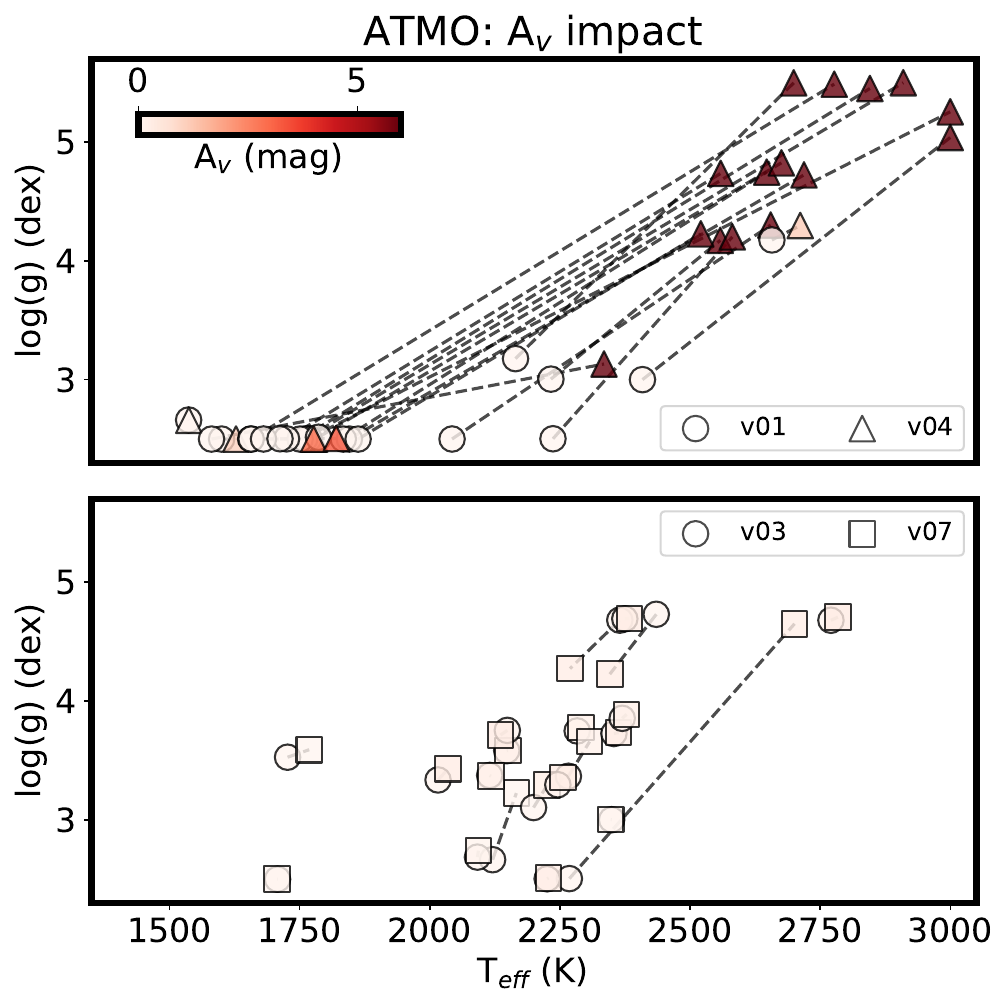}
\caption[]{Exploration of the impact of \Av on the derived posterior values of \Teff and \logg for the \atmo grid of models.
The \Av value is indicated by the color of each symbol. In the top panel, we compare \Teff and \logg from model runs v01 (\Av = 0) and v04 (\Av freely explored). Dashed lines connect the estimated values for a single target across the two runs. This shows that both \Teff and \logg are highly degenerate with \Av. In the lower panel, we perform a similar comparison between model runs v03 (\Av = 0) and v07 (\Av = $A_{\mathrm{v}}^{Gaia}$). Here, the derived parameters exhibit improved physical consistency, underscoring the importance of utilizing fixed interstellar extinction values based on Gaia DR3 data \citep{Lallement2022}.
}
\label{fig:avteff}
\end{figure}

The RV parameter represents a Doppler shift applied to the spectrum, corresponding to the absolute Doppler shift of spectral lines caused by the target's motion relative to the solar system. The derived RV value includes both a system component and a planet for gravitationally bound targets. In our sample, more than half of the objects are isolated but belong to young associations and moving groups. 
We use the RV parameter in our analysis because it is necessary for the models to accurately reproduce the depth of certain spectral lines.

The spectral line broadening factor, denoted as $\beta$, is a parameter introduced to account for broadening effects in spectral lines, such as those caused by a planet's rotation. This parameter can only be explored together with RV. In this context, $\beta$ equals the \vsini parameter used in \citet{PalmaBifani2023}. 
However, the ability to measure the impact of projected rotational velocity on spectral lines depends on the spectral resolution of the observation. The minimum measurable value is inversely proportional to the spectral resolution, given by the relation $\beta_{min} = \nicefrac{c}{R_\lambda}$, where $c$ is the speed of light.
For SINFONI observations, as mentioned earlier, the spectral resolution varies between 4490 and 5950, depending on the plate scale. Additionally, the SINFONI user manual specifies that its detector undersamples the observations. Thus, the true spectral resolution of the data is governed by the Nyquist sampling of the spectral points, which is approximately 4000 for the shortest wavelengths and 5000 for the longest. At a resolution of 4000, $\beta_{min}$ is about 75 km/s. This implies that, for this dataset, we cannot measure any spectral broadening smaller than 75 km/s—an unrealistically high threshold for most planets' rotation velocities.
In \citet{PalmaBifani2023}, the authors reported a \vsini value of 73 km/s for AB Pic b. However, we have now revised this and identified that this value is only an upper limit. 
Despite this limitation, we still fit within \formosa for $\beta$ as a free parameter because, along with the RV parameter, it aids in accurately fitting the depths of the CO bands. This, in turn, enables accurate fitting of the CO features and, consequently, reliable measurements of the \co ratio for some of our targets.
When fitting for $\beta$ in \formosa, we use a fixed limb-darkening coefficient of 0.6, which is reasonable for young, low-mass objects \citet{Claret1998}.

Next, we consider the interstellar medium extinction (\Av).
To explore \Av in \formosa, we use the \texttt{extinction.fm07} package\footnote{\texttt{extinction.fm07} documentation: \url{https://extinction.readthedocs.io/en/latest/api/extinction.fm07.html}}, which implements the \citet{Fitzpatrick2007} extinction model. Since our targets are members of young associations, interstellar dust can cause non-negligible extinction, potentially affecting the spectral slope for some of them. For this reason, we initially explored this parameter freely.
However, we noticed that allowing \Av to vary freely could significantly impact the derived \Teff values by more than 1000 K. This effect is illustrated in Figure \ref{fig:avteff}, which shows results for the \atmo models, but it is worth mentioning that this behavior is not grid dependent. 
In the upper panel of Figure \ref{fig:avteff}, we compare the derived \Teff and \logg values for model runs v01 to v04. The dashed lines connect the estimates from v01 to v04. When \Av is freely explored, the models often select very high values (up to more than 5 magnitudes), leading to substantial biases in the derived \Teff and \logg.
To mitigate this issue—while still accounting for the importance of \Av for some targets—we decided in run v07 to use a fixed interstellar extinction value in \formosa, denoted as $A_{\mathrm{v}}^{Gaia}$. These values are derived from the accurate 3D extinction maps, thanks to the Gaia Data Release 3 \citep{Lallement2022}. We report the estimate for each of our targets in Table \ref{tab:Targets1}. The lower panel of Figure \ref{fig:avteff} compares the results of model runs v03 and v07. This comparison shows that, for most targets, the \Teff values remain stable when comparing no extinction to fixed extinction. However, as expected, the fixed \Av adjusts the posterior distributions for some targets, demonstrating the importance of adequately accounting for interstellar extinction.

\begin{figure}[ht!]
\centering
\includegraphics[width=0.85\linewidth,angle=0]{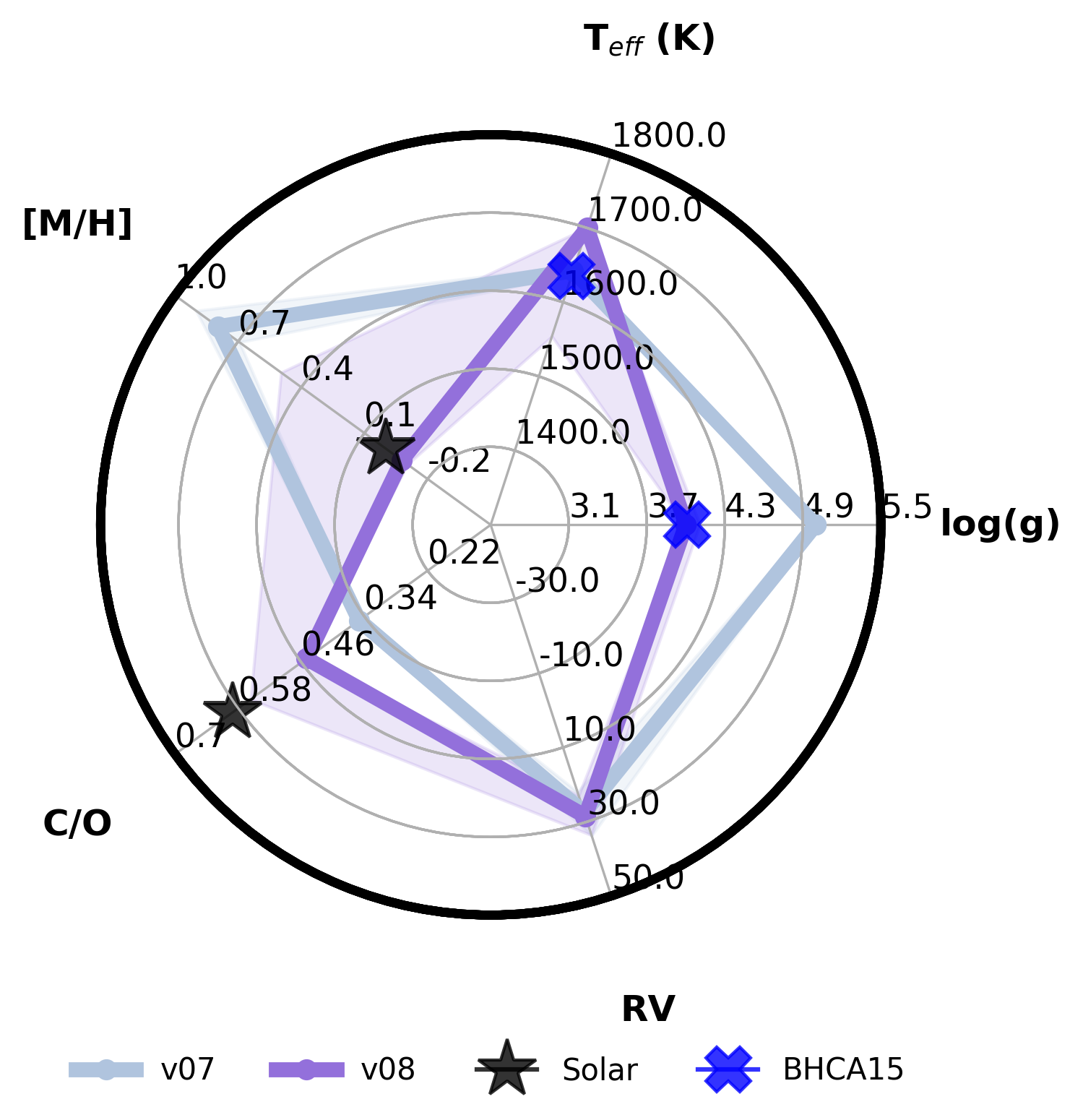}
\caption[]{Comparison of the posterior distributions using a uniform versus Gaussian prior on \logg for AB Pic b (v07 vs. v08 model predictions, shown in light blue and purple, respectively). The Gaussian prior mean value is marked with a blue cross on the \logg axis, based on the BHAC15 and COND03 evolutionary model predictions. We also marked the \Teff predictions with a blue cross, although this was not used as prior information. For comparison, the solar values for \met and the \co ratio are shown as black stars.
}
\label{fig:priorvo7vo8}
\end{figure}

We performed the \formosa exploration using uniform priors across the entire parameter range defined by each grid. We also used uniform priors for the RV and $\beta$, exploring values between -100 and 100 km/s and 0 and 500 km/s, respectively.
As mentioned above, for \Av, we initially applied uniform priors between 0 and 15 magnitudes. However, we later fixed this parameter to the values reported in Table \ref{tab:Targets1}.
Additionally, and specifically for run v08, we employed informed Gaussian priors for the surface gravity. This decision was made because, as demonstrated in \citet{PalmaBifani2024}, using informed priors for the case of AF Lep b helped recover a physically consistent solution and overcome the so-called "radius problem", often observed when modeling the atmospheres of directly imaged companions. 
Here our informed prior was constructed by using the \logg and its associated error, estimated from the BHAC15 COND03 evolutionary models, described above and reported in Table \ref{tab:Targets_evolu}), as the mean and variance for a Gaussian prior ($\mathcal{N}(\mu, \sigma^2)$).
The exception was for \btex, where we loosened the restriction to $10\times \sigma$ for the variance, otherwise, the prior was overly constraining and resulted in extremely long convergence times (over 12 hours for a single target) compared to the previous runs (see times of convergence reported in Table \ref{tab:processing_times}).

In Figure \ref{fig:priorvo7vo8}, we show an example of the impact of this prior using the case of AB Pic b, illustrating how the posterior distributions differ between the v07 and v08 runs. We observe that the v07 run resulted in a very high \logg of $\sim$4.9 dex. This value is unrealistic, as the evolutionary model predictions point to a $\sim$4 dex solution, marked by the blue cross in the figure. 
The impact of adding this prior is particularly evident in the retrieved values for \met and the \co ratio. With the \logg prior, we recover solutions much closer to stellar values (indicated by the black stars) compared to the results without the prior.
This behavior is observed only for HIP 78530 b and USco 1606 2335, in addition to AB Pic b, while the other targets in the sample already have realistic \logg posteriors in the v07 runs. 
However, for consistency and because adding this constraint does not negatively affect the solutions, we report the v08 runs for all targets in the results presented hereafter.

\begin{figure*}[ht]
\centering
\begin{subfigure}[b]{0.61\textwidth}
    \includegraphics[width=\textwidth]{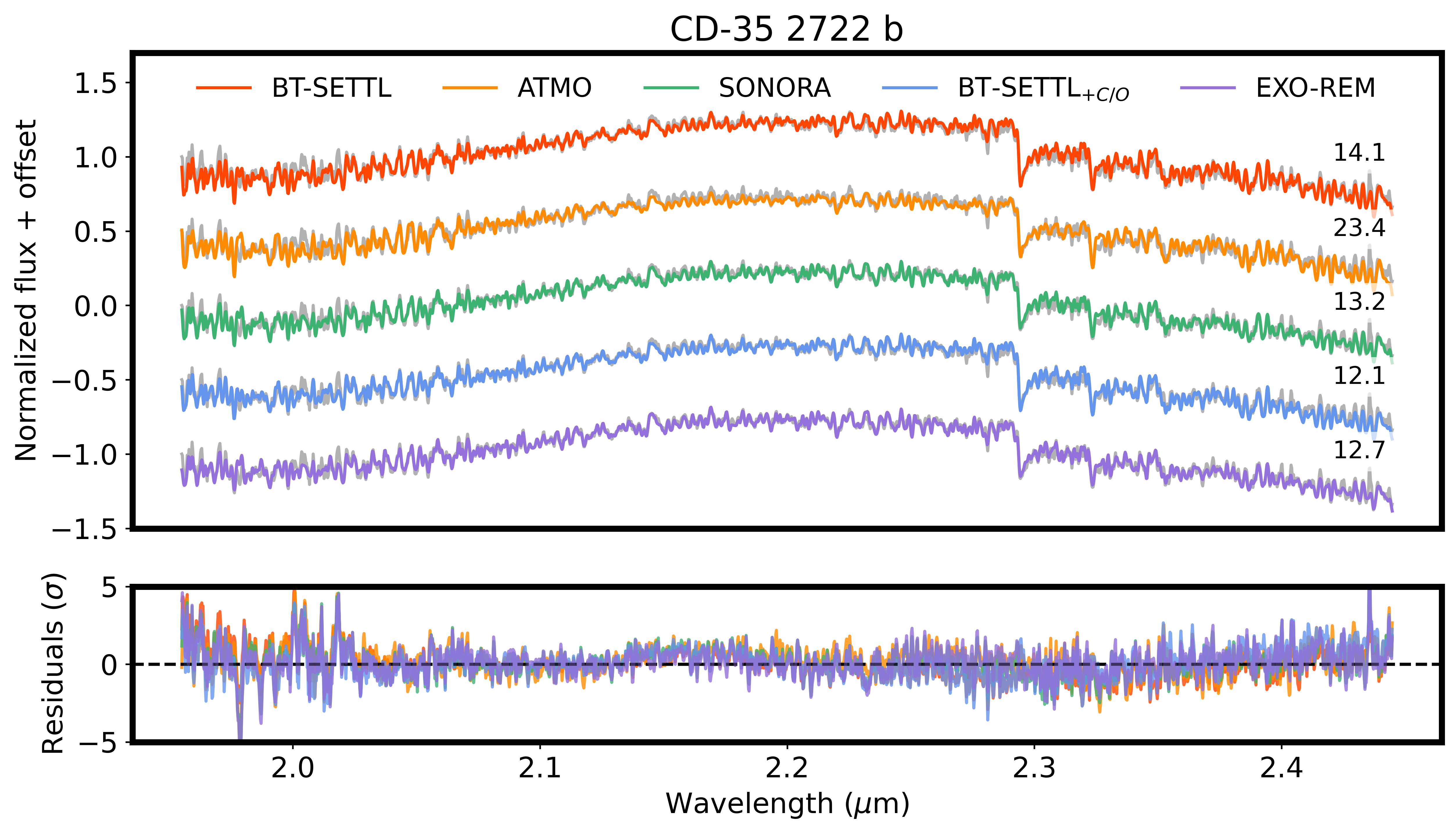}
\end{subfigure}
\hfill
\begin{subfigure}[b]{0.33\textwidth}
    \includegraphics[width=\textwidth]{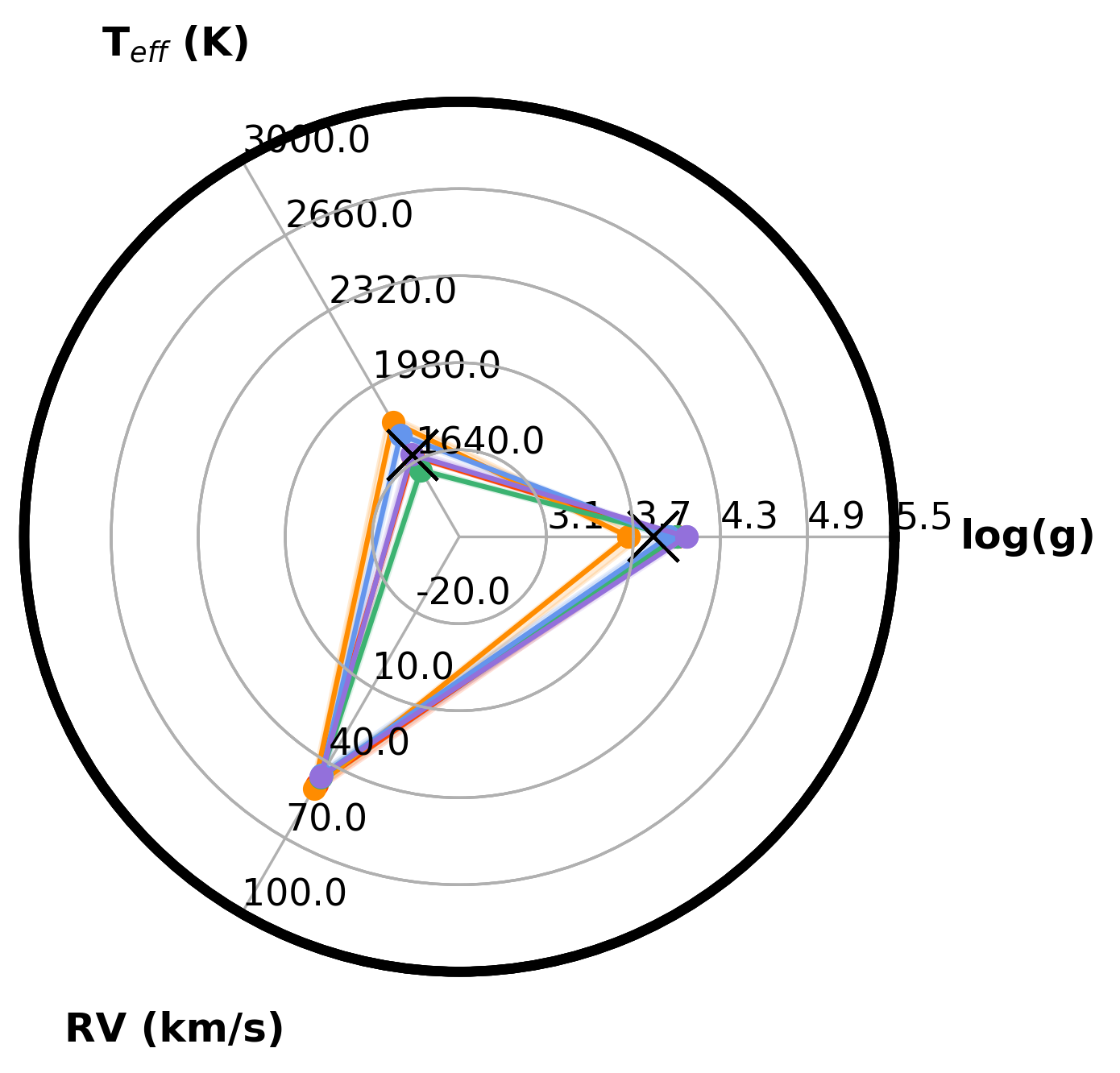}
\end{subfigure}
\caption{Best-fit results from each atmospheric grid using the setup of run v07 for the companion CD-35 2722 b. 
On the left, we display the observed spectrum (black) and the best-fit models from each grid in the different colors, offset vertically for clarity. The $\chi^2_{red}$ values for each fit are shown on the right, on top of each model fit. The bottom-left panel shows the residuals of each fit. 
On the right, a spider plot compares the derived posterior values among the grids, with the light-colored area representing 99.73\% confidence intervals on the posteriors.
For this target, all grids except ATMO perform similarly in terms of $\chi^2_{red}$ and best posterior values. The \Teff values are consistent at approximately 1650 K, while \logg varies between 4.0 and 4.3 dex.}
\label{fig:target_CD-35_2722_b}
\end{figure*}

\begin{table*}[ht]
\centering
\caption{CD-35 2722 b posteriors with $1 \sigma$ asymmetric uncertainties for the models presented in Figure \ref{fig:target_CD-35_2722_b}. The last two columns correspond to the Bayesian evidence (ln(z)) and the $\chi^2_{red}$.}
\renewcommand{\arraystretch}{1.5} 
\resizebox{\textwidth}{!}{%
\begin{tabular}{lcccccccccccc}
\hline
model& \Teff (K) & \logg (dex) & \met & \co & $\gamma$ & \fsed & RV (km/s) & $\beta$ (km/s) & ln(z) & $\chi ^2 _{red}$ \\ 
\hline
\btex&$1664^{+7}_{-7}$&$4.0^{+0.03}_{-0.04}$&&&&&$48.58^{+4.31}_{-4.23}$&$79.74^{+10.55}_{-9.44}$&-14131.7&14\\ 
\atmo&$1819^{+38}_{-13}$&$3.67^{+0.09}_{-0.03}$&$0.35^{+0.04}_{-0.03}$&$0.7^{+0.0}_{-0.01}$&$1.05^{+0.0}_{-0.0}$&&$50.27^{+3.17}_{-2.7}$&$111^{+7}_{-5}$&-23393.0&23\\ 
\sono&$1601^{+16}_{-10}$&$4.0^{+0.02}_{-0.05}$&&&&$1.83^{+0.13}_{-0.13}$&$45.78^{+3.67}_{-3.41}$&$75.05^{+9.03}_{-7.66}$&-13198.2&13\\ 
\btse&$1760^{+20}_{-15}$&$3.94^{+0.03}_{-0.06}$&&$0.54^{+0.02}_{-0.01}$&&&$45.01^{+3.69}_{-4.04}$&$76.45^{+9.12}_{-8.65}$&-12086.4&12\\ 
\exor&$1671^{+29}_{-14}$&$4.07^{+0.03}_{-0.07}$&$-0.0^{+0.01}_{-0.1}$&$0.5^{+0.01}_{-0.0}$&&&$45.38^{+2.6}_{-2.65}$&$87.49^{+5.67}_{-4.95}$&-12694.3&13\\ 
\hline
\end{tabular}}
\label{tab:target_CD-35_2722_b}
\end{table*}

\subsection{Atmospheric modeling results}\label{sec:results}

\begin{figure*}[ht!]
\centering
\includegraphics[width=0.95\hsize]{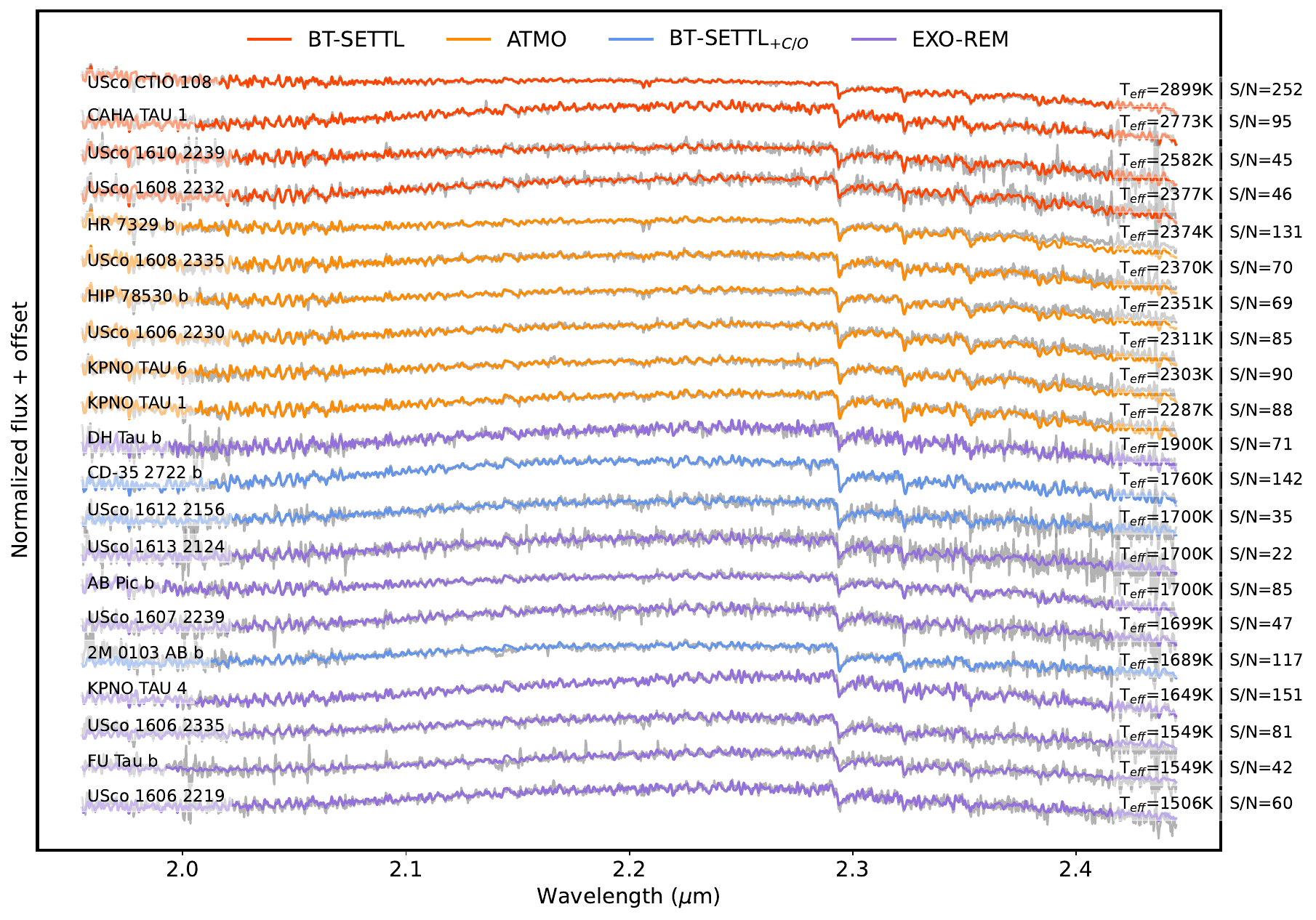}
\caption{Best spectral fit for each target. Here, the targets are organized by their derived \Teff, from hottest at the top to coldest at the bottom. The \Teff and S/N values are shown on the right side of each spectrum. The models are color-coded to indicate the grid that provided the best match to the data, selected based on the lowest $\chi^2_{red}$ value.}
\label{fig:full_res_teff}
\end{figure*}

\begin{figure}[ht!]
\centering
\includegraphics[width=0.95\linewidth,angle=0]{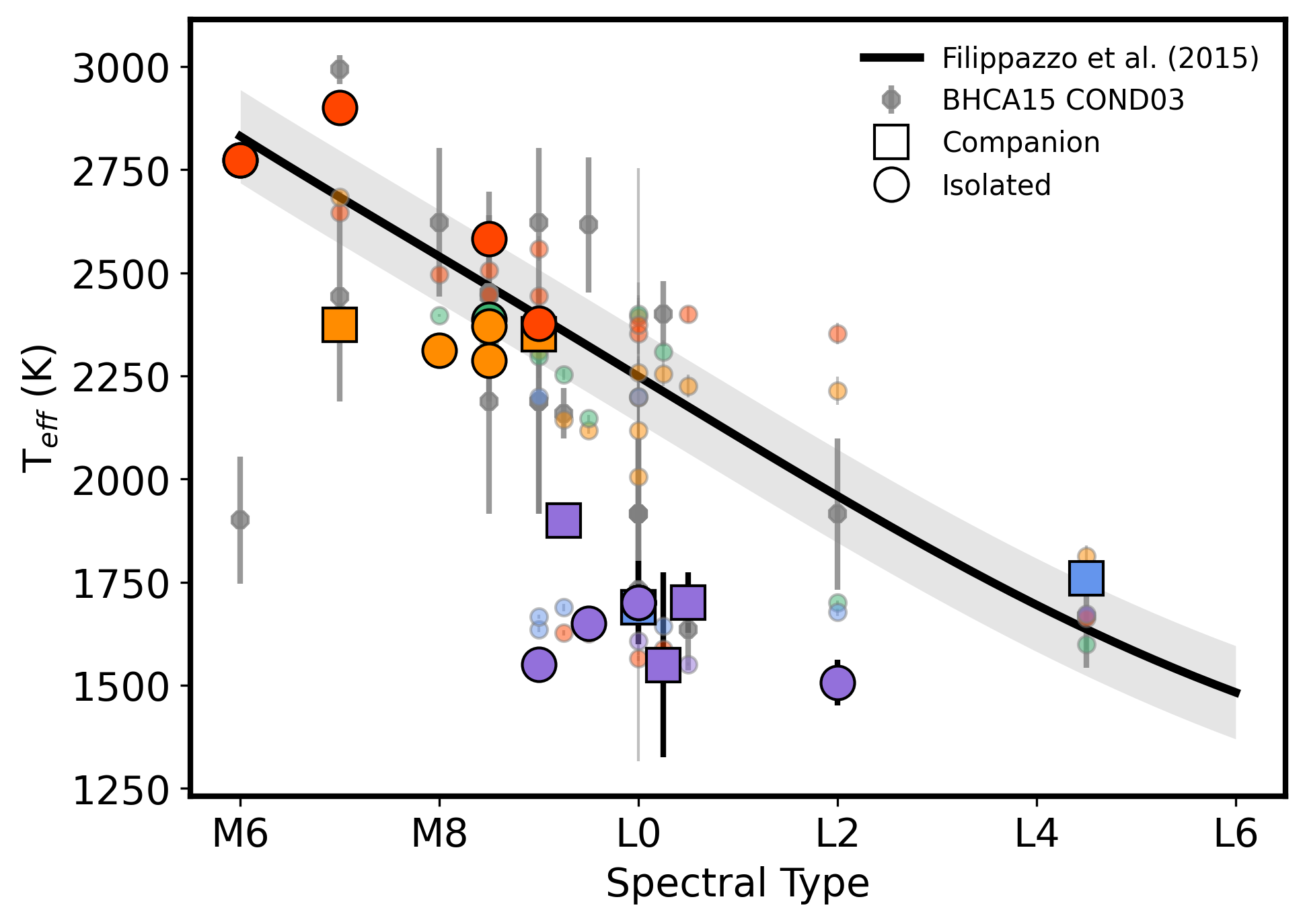}
\caption[]{Spectral type vs. \Teff comparison. Large markers represent the best-model \Teff values, presented in Figure \ref{fig:full_res_teff}. Smaller markers indicate the \Teff solutions from all other grids, and gray points show the \Teff predictions from evolutionary models. Companions in our sample are highlighted with square markers, and isolated targets are marked with circles. The black curve and gray area represent the relationship determined by \citet{Filippazzo2015} for field brown dwarfs.
}
\label{fig:teff_sptype}
\end{figure}

To summarize, Table \ref{tab:processing_times} provides an overview of all the model runs performed to explore the parameter space of the atmospheric models for our sample. For each run involving looping over the 21 targets, we report the time it took to complete in minutes and the mean value of the reduced chi-squared ($\chi^2_{red}$) for those models. This provides us with a statistical measure to evaluate how well a model fits the entire dataset.
We observe that the mean value of $\chi^2_{red}$ is lower (indicating a better fit) for the runs between v05, v06, v07, and v08, depending on the family of models. However, we have already mentioned the issue of freely exploring \Av as well as the problem of not retrieving realistic \logg solutions, so we report the results from runs v08 only hereafter.

Here, we first present our best run per target, as illustrated in Figure \ref{fig:target_CD-35_2722_b} and Table \ref{tab:target_CD-35_2722_b} for the example of the target CD-35 2722 b. In Figure \ref{fig:target_CD-35_2722_b}, we compare the five models, one for each grid. The left panel displays the best-fit model for each grid, along with the residuals. Each best fit includes the $\chi^2_{red}$ value, displayed on the right side of the spectrum, to assess the quality of the fit.
In the same figure, the right panel displays a spider plot comparing the best-fit parameters across the different grids as a function of \Teff, \logg, and RV. The light-colored area in the spider plot represents the $99.73\%$ confidence intervals (3$\sigma$).
For CD-35 2722 b, all grids perform remarkably well, except for ATMO, which exhibits a higher $\chi^2_{red}$ value and a poorer fit when analyzing the spectral features.
The derived values for each explored parameter from each grid are reported in Table \ref{tab:target_CD-35_2722_b}. The errors are asymmetric $1\sigma$ intervals, assuming Gaussian distributions for the posteriors. These values do not account for additional systematics and should therefore be considered purely statistical, as already identified in previous works with \formosa \citep{Petrus2021, PalmaBifani2023}.

Although all models perform similarly for CD-35 2722 b, this is not true for other targets.
First, we observe that several of our targets exhibit high S/N values ($>$80), as reported on the right side of each spectrum in Figure \ref{fig:full_res_teff}. 
The targets with high S/N include USco CTIO 108 A, CAHA Tau 1, HR 7329 b, KPNO Tau 6, KPNO Tau 1, CD-35 2722 b, AB Pic b, 2M 0103 AB b, and KPNO Tau 4. The remaining targets display lower S/N ratios, necessitating extra caution when interpreting the derived parameters.
In addition, not all targets can be modeled with all grids, because they are too hot for some of our grids, as discussed in Section \ref{sec:models}.
Therefore, we made an initial guess of the \Teff of our targets based on the spectral type. 
This means in practice that, even though we have used all grids for all targets, for targets with spectral types between M5 and M7, we report results only with \btex and \atmo. For targets with spectral types extending to M8, we included \sono. Finally, for targets down to L4, we utilized all five grids.
Figures similar to Figure \ref{fig:target_CD-35_2722_b} and tables comparable to Table \ref{tab:target_CD-35_2722_b} are provided in Appendix \ref{sec_a:pertarget} for all other targets in our library.

Upon analyzing the outcomes, it becomes clear that there is no definitive "best grid." 
However, observing the best-derived atmospheric parameters for each target, we explored which grid performed better for each case. 
To this end, we selected the model with the lowest $\chi^2_{red}$ for each target and organized the spectra by the value of the derived \Teff, observable in Figure \ref{fig:full_res_teff}. 
We excluded the \sono grid from this analysis because only KPNO Tau 6 favored this grid, and, as shown in Figure \ref{fig:target_KPNO_Tau_6}, the preference is negligible, and the derived parameters with the other grids are very similar.

In Figure \ref{fig:full_res_teff}, we observe that our sample can essentially be divided into two domains. Below 2000 K, where all grids fall within the parameter space, the best fits are consistently provided by \exor or \btse. Among the 13 targets with spectral types below M9, 11 have \Teff values below 2000 K.
For targets with \Teff above 2000 K, only \atmo and \btex remain within the parameter space. In this regime, \btex is preferred for the four hottest targets, offering excellent agreement with the spectral features, particularly for the two hottest targets, USco CTIO 108 A and CAHA Tau 1. For the remaining targets where \atmo is favored, additional challenges arise. Despite being the preferred grid, \atmo exhibits discrepancies, especially in reproducing the depth of the K lines at 2.2 $\mu$m and the overall shape of the continuum at longer wavelengths.
The limitations of the grids and the reliability of our derived parameters are further addressed in the discussion section.

\section{Discussion}\label{sec:discussion}

From our results, several important points emerge that can be grouped into two categories.
First, the observed temperature-domain difference raises the question of whether the split has a physical explanation or whether it is a model-dependent feature related to the physics included in the different atmospheric models.
Second, for the low-temperature targets, we measured the \co ratio, a parameter often referred to as a formation tracer \citep{oberg2011}. Here, we present these measurements alongside the possible implications for the formation histories of these targets.

\subsection{Assessing the temperature-domain discontinuity}

\begin{figure*}[ht!]
\centering
\includegraphics[width=0.95\hsize]{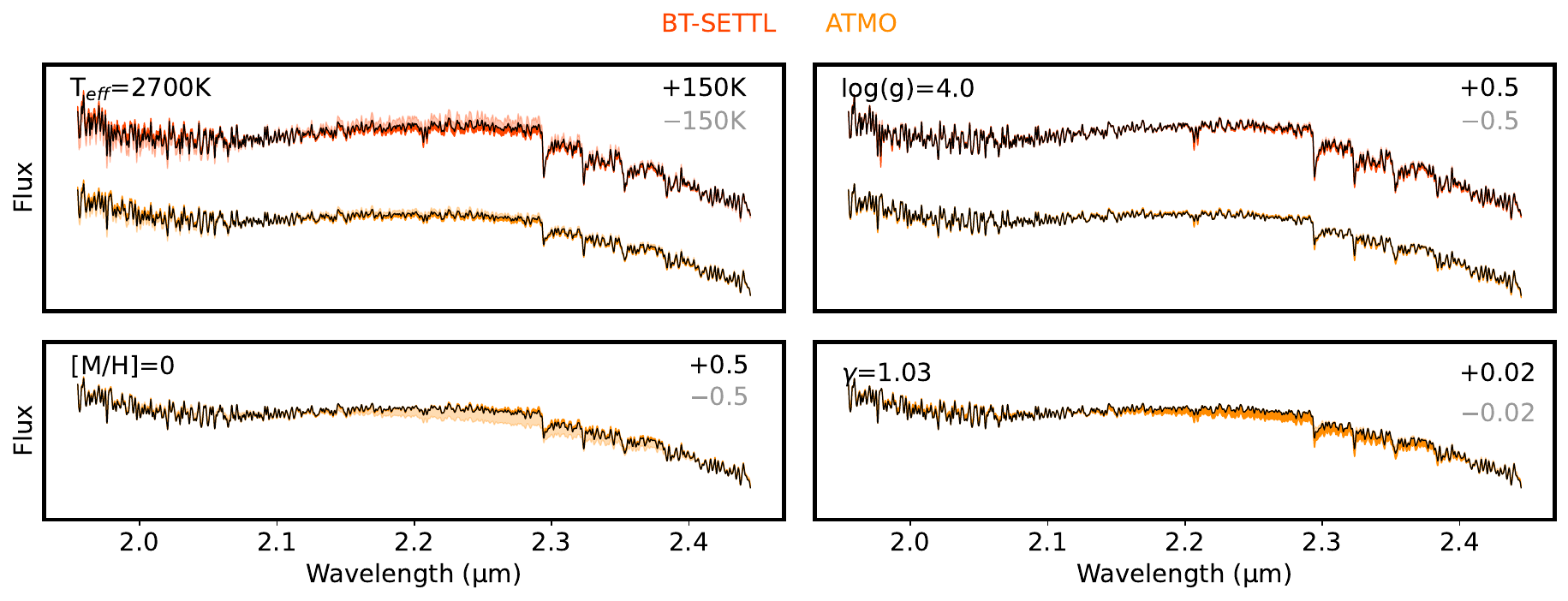}
\caption{Visualization of the two grids that reach the \Teff domain of the hottest targets in our sample. Here, we compare models from both grids extracted at \Teff = 2700 K, \logg = 4 dex, \met = 0, and $\gamma = 1.03$, in black. In each panel, we illustrate the spectral variations caused by altering one parameter by the values listed in the upper right corner. The vivid color represents an increase, while the lighter (more transparent) color indicates a negative variation.}
\label{fig:grids_variations_comp_2700}
\end{figure*}

\begin{figure*}[ht!]
\centering
\includegraphics[width=0.95\hsize]{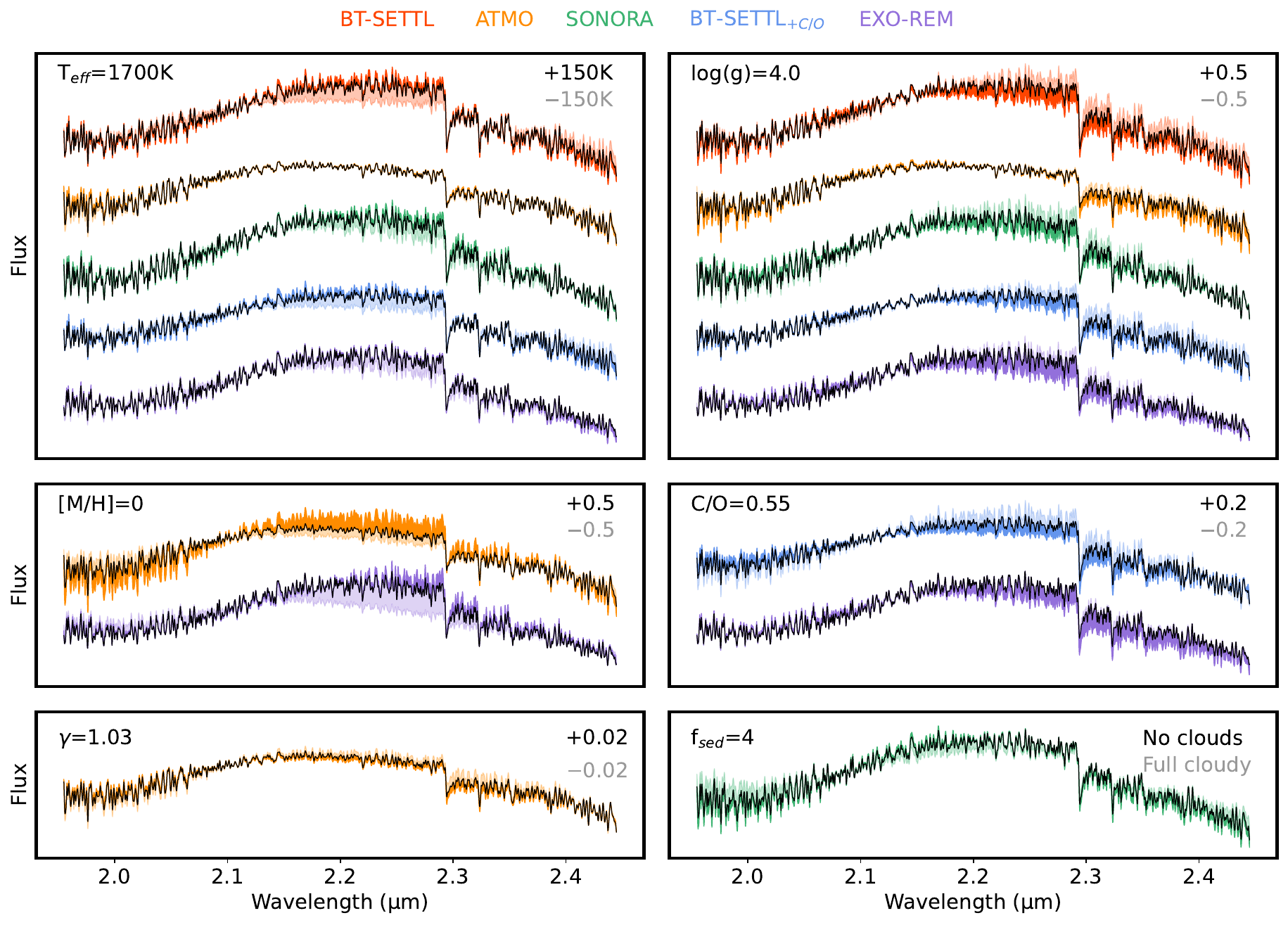}
\caption{Same as Figure \ref{fig:grids_variations_comp_2700}, but comparing the five grids at parameters compatible with the coldest targets in our sample. Here the models presented in black are at \Teff = 1700 K, \logg = 4 dex, \met = 0, \co = 0.55, $\gamma = 1.03$, and \fsed = 4.}
\label{fig:grids_variations_comp_1700}
\end{figure*}

We begin by comparing the derived \Teff values of our sample to their spectral types, shown in Figure \ref{fig:teff_sptype}. 
This figure includes several layers of information, which we will break down for clarity hereafter.
First, the primary data points, represented by the large markers outlined in black, are the \Teff results for the selected model among all five grids for each target. These points are color-coded depending on the best fit presented in Figure \ref{fig:full_res_teff}. 
This comparison shows that the hottest targets, which we modeled using \atmo and \btex, closely follow the sixth-order polynomial fit from \citet{Filippazzo2015}, which describes the relationship between \Teff and spectral type for field brown dwarfs. However, for targets near spectral type M9, where \exor and \btse yield the best solutions, we notice a particular shift in the trend toward lower \Teff values.
This behavior for young planetary-mass objects has been reported in previous studies, such as \citet{Filippazzo2015} and \citet{Bonnefoy2014s}, and is likely linked to the physical properties of the atmospheres of objects near the M/L transition. 
According to \citet{Filippazzo2015}, young objects exhibit similar or slightly higher bolometric luminosities compared to field-age objects of the same spectral type. 
However, since they are still contracting, their larger radii require cooler photospheres to maintain the same luminosity. As a result, a lower \logg in young dwarfs could lead to reddening and a higher apparent spectral type.
Furthermore, the sensitivity of radius to age in young $M8 – L0$ dwarfs creates a significant dispersion in \Teff, up to $\sim$ 500 K, at the M/L transition. This dispersion could explain the deviations observed in our results for targets near M9, as these objects fall into the age and temperature regime where such variability is expected.

In addition to the large markers representing the best solutions (those with the lowest $\chi^2_{red}$), Figure \ref{fig:teff_sptype} includes smaller circles indicating the \Teff values derived using each grid for each target, consistent with the respective \Teff limits.
From the smaller colored points in Figure \ref{fig:teff_sptype}, we observe that different grids behave distinctly. 
The \exor grid, shown in purple, has an upper \Teff limit of 2000 K and consistently yields solutions below the \citet{Filippazzo2015} fit for all our targets within its range.
Next, \btse and \btex, in blue and red, respectively, share similar underlying physics, but \btse explores an additional free parameter, accounting for changes in the atmospheric composition. Notably, at spectral type M9, \btex shows a drop in the \Teff solutions, aligning with the \btse and \exor results, except for the L2 target, USco 1606-2219, which has a spectral type uncertainty of $\pm$1.
For \btex, the same discontinuity was observed previously by \citep{Sanghi2023}, and they attributed it to the fact that, since cloud opacity increases as \Teff decreases, \btex does not incorporate sufficient dust to accurately mimic the atmospheric behavior at these temperatures, leading to an underestimated \Teff.
For the \sono and \atmo model grids, we do not observe the same \Teff drop. Their \Teff solutions as a function of spectral type remain closely aligned with the trend observed for field dwarfs. 
This behavior is particularly interesting, as \atmo does not include clouds, and \sono assumes equilibrium chemistry only, meaning that this different behavior can potentially be explained by a lack of physics in these models to accurately represent the M/L transition.

To further investigate the similarities and differences among the grids, we present Figures \ref{fig:grids_variations_comp_2700} and \ref{fig:grids_variations_comp_1700}, which illustrate how the spectral features in the K-band vary as different parameters are adjusted. 
Specifically, we selected two example spectra representing the two temperature domains of our sample: one at \Teff=2700 K and one at \Teff=1700 K, both at \logg=4 dex and solar \met and \co.
In Figure \ref{fig:grids_variations_comp_2700}, we display the case for the hotter spectrum at 2700 K, which lies within the parameter space explored by \atmo and \btex. The figure contains four sub-panels, each demonstrating the impact of varying one of the grid parameters. 
Interestingly, we observe that the continuum shape of the \btex spectrum changes significantly when \Teff is altered by $\pm$150 K. In contrast, the same variation in \atmo produces a much less noticeable effect.
For \logg, neither grid shows substantial variations when adjusting the value by $\pm$0.5 dex. However, we note that the K-band lines at 2.2 $\mu$m become deeper with increasing \logg.
To explore fingering convection, \atmo incorporates three additional parameters compared to \btex: \met, \co, and $\gamma$. 
Varying \met and $\gamma$ results in noticeable changes to the continuum shape, pointing towards a possible inverse proportionality among them. Here we did not explore the behavior of \co due to the limited sampling of this parameter in the grid (only 3 points from 0.3 to 0.7), which, as evidenced by our fits always preferring values at 0.7 (see Tables \ref{tab:target_2M0103AB_b} to \ref{tab:target_UScoCTIO108}), is incapable of reliable exploring this dimension.

For the colder targets in our sample, we have all five grids available, allowing us to make a comparison similar to the one in Figure \ref{fig:grids_variations_comp_2700} for models from all grids at \Teff=1700 K, which we present in Figure \ref{fig:grids_variations_comp_1700}. 
This time, the figure comprises six sub-panels as we explore additional dimensions, including \co and \fsed, along with the previous ones.
In the first panel, we show \Teff variations of $\pm$150 K. We observe that \btex, \sono, \btse, and \exor all show similar behavior: increasing \Teff leads to a slight increase in the central bump while decreasing \Teff results in a decrease, flattening the spectrum. 
More interestingly, \atmo exhibits a very different behavior overall. The continuum shape of the selected model differs significantly, and the \Teff variations have almost no impact on the observed features. This raises questions about \atmo's ability to represent the physics of these objects at these temperatures.
For \logg, we observe similar, strong effects across all grids, with \atmo again showing the least noticeable impact. 
Interestingly, \logg strongly affects the depth of the CO overtone. We can compare this to the fourth panel, which explores the variations due to changes in \co for the \exor and \btse grids, where we observe a similar impact. However, the negative \co variations are more noticeable in \btse than in \exor, potentially indicating a limitation in exploring sub-solar \co values with this grid, at least when the \met is set to solar.
For the \met, only \atmo and \exor explore non-solar values, showing different variations in the spectral features, making comparisons between these grids more challenging. 
For \atmo and \sono, we also demonstrate how $\gamma$ and \fsed affect the spectral features; however, we do not explore this further, as the effects are more pronounced over a wider wavelength range.

The K-band is a rich spectral region that exhibits a strong correlation between \Teff and spectral type. As discussed, we observe a drop in \Teff of around 500 K at the M-L transition.
The available grids currently limit our ability to fully understand this transition, as they cannot be compared homogeneously across this critical point. 
One possible step to overcome these limitations would be to extend grids such as \exor to higher \Teff ranges, which would require the addition of missing opacity sources.
Additionally, multi-band comparisons could provide a better understanding of the parameters that shape the spectra of these objects. However, both of these ideas are beyond the scope of the present work.

\subsection{Reliability of our posteriors}\label{sec:reliability}

To understand and discuss the reliability of our posteriors, we categorized the different behaviors observed in our sample. As shown in Figure \ref{fig:full_res_teff}, our sample exhibits various S/N values and spans a broad range of \Teff. This diversity allows us to assess the performance of the atmospheric grids under various conditions.
To this end, we will make a broad distinction between the hot ($M5-M9$) and the cold targets of our sample ($M9-L5$).

Among the hot targets in our sample, several distinct behaviors emerge. KPNO Tau 1, KPNO Tau 6, USco 1606-2230, USco 1608-2335, and USco 1610-2239 demonstrate excellent alignment with predictions from evolutionary models and spectral type estimates, suggesting high confidence in the derived atmospheric parameters for these objects. 
In contrast, USco 1608-2232 and USco CTIO 108 A yield \Teff solutions that are significantly hotter than expected from evolutionary models or spectral type relations but with very low $\chi^2_{red}$ and a close match to the spectral features, especially for USco CTIO 108, indicating we can reliably report the physical properties (\Teff and \logg) of these targets. 
CAHA Tau 1 and HR 7329 b converged towards solutions differing from evolutionary models' \logg predictions, but providing overall a great spectral match.
We urge special caution when referencing the parameters of USco 1610-2239 and USco 1608-2232, as the S/N for these observations is below 50, which, although still high, is among the lowest values in our sample. Finally, for the targets where \atmo provides the best match, as the continuum level is not well fitted for the longest wavelengths, the \Teff could potentially be biased.

Among the cold targets in our sample, distinct behaviors can be identified. CD-35 2722 b and 2M 0103 AB b show strong agreement across all grids, indicating high confidence in the derived parameters for these objects. AB Pic b stands out as a low-\Teff target with non-stellar metallicity values that align well with evolutionary model predictions, demonstrating the potential of grids to capture formation-tracing properties. 
However, several targets, including DH Tau b, FU Tau b, HIP 78530 b, KPNO Tau 4, USco 1606-2219, USco 1606-2335, and USco 1607-2239, exhibit poor agreement between atmospheric models and evolutionary model predictions for \Teff. These discrepancies are likely influenced by the generally low S/N of their observations. 
Finally, USco 1612-2156 and USco 1613-2124 have data that is too noisy to yield reliable estimates of atmospheric parameters.

\subsection{Formation-tracers}

To discuss the possible formation scenarios of our sample, we can explore the values of the so-called formation tracers, as well as the large-scale behavior of atmospheric properties. 
In broad terms, as expected, we observe no distinction in \Teff and \logg behavior as a function of spectral type between isolated targets and companions. For example, in Figure \ref{fig:teff_sptype}, we have labeled the companions with square symbols and the isolated young brown dwarfs with circles. We observe that both groups follow the same trend, potentially suggesting a shared origin of formation for all these targets.

\begin{figure}[ht!]
\centering
\includegraphics[width=1\linewidth,angle=0]{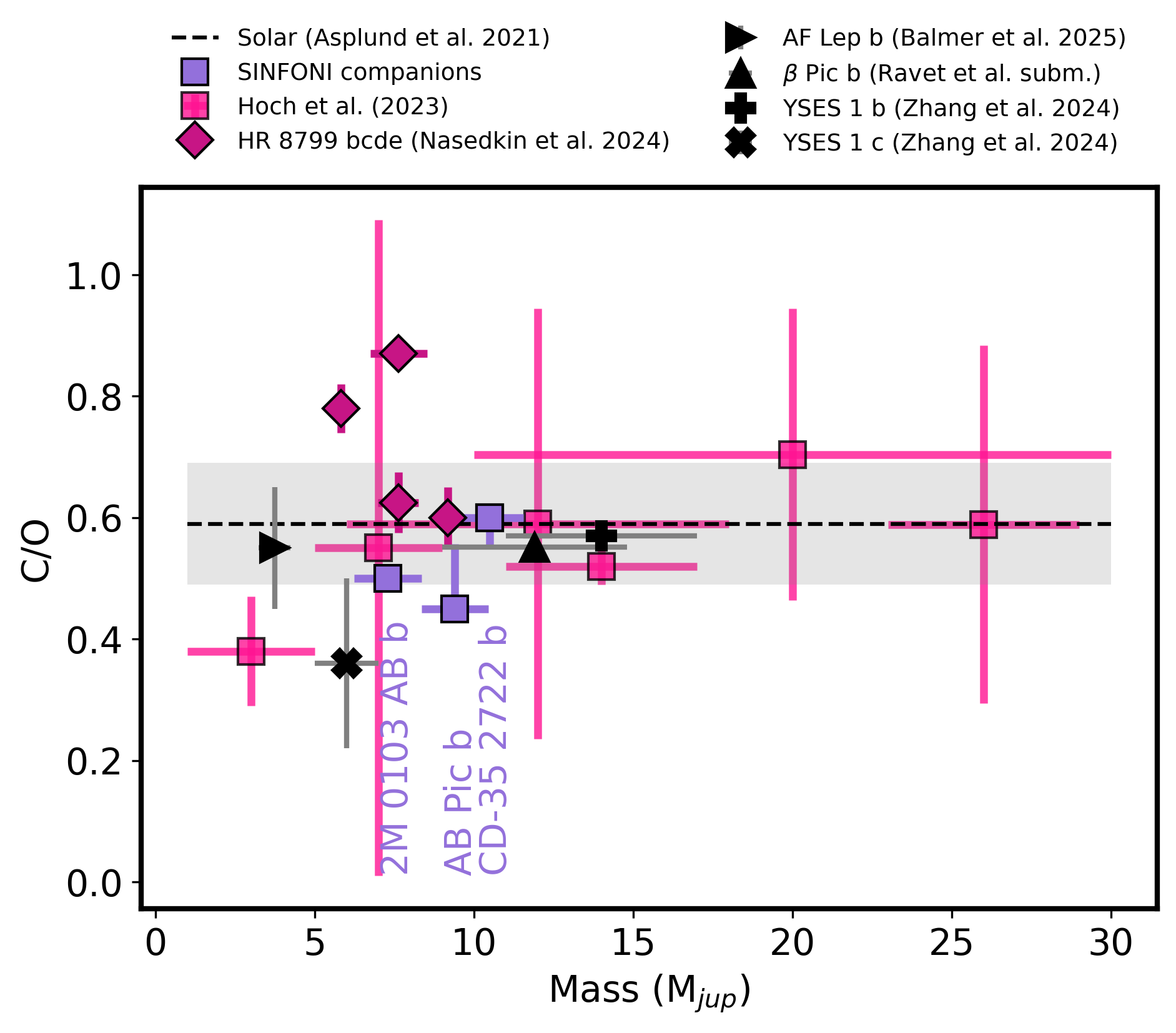}
\caption[]{
Mass vs. \co ratio for directly imaged companions, building on the work of \citet{Hoch2023}. The magenta points represent the original sample of directly imaged planets, while the purple points indicate the three new measurements provided by this work. Recently published values have been added in black, including updated measurements for HR 8799 bcde \citep{Nasedkin2024}, AF Lep \citep{Balmer2025}, $\beta$ Pic b (Ravet et al., in prep.), and YSES 1 bc \citep{Zhang2024_YSES}. The dashed line and gray shaded area represent the solar \co ratio with uncertainties (0.59 $\pm$ 0.1) as provided by \citet{Asplund2021}.
}
\label{fig:mass_co}
\end{figure}

Regarding the formation tracers, with \formosa we only fit and constrain for the \co ratio with \exor, as it is the only grid that samples this dimension densely enough.
In addition, because of the \Teff range of \exor and the S/N of our sample, we reliably measured \co for a limited subset of our sample. As detailed in \ref{sec:reliability}, these targets include the companions 2M0103 AB b, AB Pic b, and CD-35 2722 b. 
In Figure \ref{fig:mass_co}, we place our measured \co ratios in the context of known values for other companions in the field. For the three targets in our sample, the \co ratios are consistent with solar values. While this figure does not allow for definitive conclusions about formation histories, our targets fall well within the observed distribution, broadly consistent with the solar value revisited by \citet{Asplund2021}. 
However, the \co ratios for the reported targets in Figure \ref{fig:mass_co} were measured differently across studies, complicating their comparison and interpretation.
For now, we expect that as more observations are added to this plot, it will eventually enable us to draw population-based conclusions, including connections to other formation tracers, such as isotopologue ratios like D/H and $^{12}$C/$^{13}$C \citep{molliere2019, Zhang2022}.

\section{Conclusions}\label{sec:conclusions}

Our work highlights the importance of analyzing large datasets homogeneously to gain a deeper understanding of planet formation, evolution, and the atmospheric imprints of these processes. 
Using \formosa, we have modeled a sample of 21 young substellar companions observed with VLT/SINFONI.
This homogeneous analysis reveals several key conclusions.

First, we have observed that under a forward modeling approach, informing the priors is crucial. Allowing parameters such as interstellar extinction and surface gravity to vary freely, without physical constraints, can lead to nonphysical or degenerate solutions. With the release of Gaia DR3, we now have access to high-precision distance measurements and 3D dust maps. This is one example of how invaluable the prior information can be, as it provides robust insights on some physical properties from unrelated measurements.

Next, we confirm that the K-band spectrum is a reliable proxy for estimating effective temperature.
Our findings align with the \Teff versus spectral type behavior identified by \citet{Bonnefoy2014s} and \citet{Filippazzo2015} for young planetary-mass objects. The observed discontinuity in \Teff occurs precisely at the M/L transition. 
However, current atmospheric models still fail to adequately capture the full range of relevant physical conditions across this transition, as previously identified by \citet{Sanghi2023} for \btex.
Therefore, there is a pressing need to further explore the discrepancies between self-consistent atmospheric model grids, which currently sample the parameter space unevenly and can produce diverging results.

Third, while we performed a homogeneous analysis of this unique library of SINFONI data, each target holds deeper mysteries. Many have also been observed with other instruments, which could provide valuable insights into their individual properties.
Additionally, there are more SINFONI archival observations of companions, which were excluded from this study due to heavy stellar speckle contamination but could potentially be revisited using advanced molecular mapping techniques, as was the case for $\beta$ Pic b by \citet{Kiefer2024}.

Finally, the lack of strong spectral differences in the K-band between wide-orbit planetary companions and isolated brown dwarfs may suggest that they have similar formation mechanisms. However, this remains an open question that requires further investigation through both observational and theoretical studies.
We encourage the community to revisit and reanalyze archival datasets—not only those from SINFONI, but also those from legacy instruments. Such efforts may reveal insights that have long remained hidden.

\begin{acknowledgements}
To the memory of Dr. France Allard and her contributions to the domain of atmospheric physics of stars, brown dwarfs, and exoplanets.
We thank Carine Babusiaux for providing us with the Gaia measurements of interstellar extinction, derived using the 3D extinction map from \citet{Lallement2022}.
This publication utilized the SIMBAD and VizieR databases, operated by the CDS in Strasbourg, France. This work has utilized data from the European Space Agency (ESA) mission Gaia \url{https://www.cosmos.esa.int/gaia}, processed by the Gaia Data Processing and Analysis Consortium (DPAC, \url{https://www.cosmos.esa.int/web/gaia/dpac/consortium}). Funding for the DPAC has been provided by national institutions, particularly those participating in the Gaia Multilateral Agreement.
We acknowledge support in France from the French National Research Agency (ANR) through project grants ANR-20-CE31-0012 (FRAME) and ANR-23-CE31-0006 (MIRAGES), as well as the Programmes Nationaux de Planetologie et de Physique Stellaire (PNP and PNPS). This project is supported in part by the European Research Council (ERC) under the European Union’s Horizon
2020 research and innovation program (COBREX; grant agreement n° 885593).
P.P.-B., M.B., and G.C. received funding from the French Programme National de Planétologie (PNP) and de Physique Stellaire (PNPS) of CNRS (INSU).
S.P.'s research is supported by an appointment to the NASA Postdoctoral Program at the NASA–Goddard Space Flight Center, administered by Oak Ridge Associated Universities under contract with NASA.
\end{acknowledgements}

\bibliographystyle{aa}
\bibliography{aa54894-25}

\begin{appendix}

\onecolumn
\section{Background information for all targets}\label{sec_a:bigtables}

This appendix summarizes each target's properties and key information in the format of tables to facilitate easy comparison. 
Table \ref{tab:Obslog} provides details of the observing runs. Tables \ref{tab:Targets1} and \ref{tab:Targets2} list the targets general properties, and their references.
Table \ref{tab:Obslog} provides an overview of each sample target's observing conditions, including program IDs, principal investigators, and key observational setup parameters.
The reported values of this table were extracted from the original FITS file headers.
We observe that the observations were conducted under varying conditions, with some targets observed using the MACAO adaptive optics module while others were observed without.
The seeing values for the runs are also indicated. In some cases, marked with a star symbol, this information was missing and later retrieved from the ESO Ambient Conditions Database\footnote{ESO Ambient Conditions Database: \url{https://archive.eso.org/asm/ambient-server?night=18+june+2014&site=paranal}}.

\begin{table*}[ht]
\centering
\caption[Observation Log of the SINFONI Library]{Observation log of the SINFONI Library. All the information was taken from the headers of the raw data cubes.
For the seeing, the values accompanied by a star symbol ($^{\star}$) refer to the lower or upper limit of the night as reported by the ESO Ambient Conditions Database.}
\renewcommand{\arraystretch}{1.5} 
\resizebox{\textwidth}{!}{%
\begin{tabular}{lllccccccc}
\hline
PI \& & Name & Date & MACAO & Platescale & Exposure & ndit & nexp & Airmass & Seeing   \\
Program &  &  &  & (arcsec) & (s) & & & & (FWHM) \\
\hline
Kopytova & CD-35 2722 b      & 26-11-2013 & on & 0.1    & 150 & 4 & 8 & 1.055 - 1.354      & 1.04 - 1.66  \\
092.C-0803(A) & 2M 0103 AB b   & 18-07-2014 & on & 0.1    & 70  & 8 & 9  & 1.159 - 1.183   & 0.5$^{\star}$ - 1.53    \\
093.C-0829(A\&B) &          & 11-08-2014 & on & 0.1    & 70  & 16 & 9  & 1.126 - 1.186     & 0.7$^{\star}$  - 1.32    \\
 &                          & 15-08-2014 & on & 0.1    & 70  & 8 & 9   & 1.145 - 1.202     & 1.02 - 1.74  \\
 & HR 7329 b                & 05-06-2014 & on & 0.025  & 60  & 8 & 9   & 1.155 - 1.182     & 0.7$^{\star}$  - 1.19    \\
 &                          & 25-04-2014 & on & 0.025  & 60  & 8 & 9   & 1.152 - 1.170     & 0.8$^{\star}$  - 2.14    \\
 &                          & 10-07-2014 & on & 0.025  & 60  & 9 & 9   & 1.152 - 1.175     & 1.31 - 1.87  \\
\hdashline
Radigan & USco 1606-2219         & 18-05-2014 & off & 0.25  & 300  & 4 & 4 & 1.016 - 1.030  & 0.81 - 1.08   \\
093.C-0502(A) & USco 1606-2230   & 02-06-2014 & off & 0.25  & 150  & 4 & 4 & 1.261 - 1.299  & 0.82 - 0.90   \\
 & USco 1606-2335                & 21-05-2014 & off & 0.25  & 600  & 3 & 3 & 1.006 - 1.019  & 0.86 - 0.98   \\
 & USco 1607-2239                & 21-05-2014 & off & 0.25  & 600  & 4 & 2 & 1.005 - 1.108  & 0.74 - 0.99   \\
 &                              & 28-05-2014 & off & 0.25  & 600  & 2 & 2 & 1.005 - 1.01    & 0.74 - 0.99   \\
 & USco 1608-2232                & 21-05-2014 & off & 0.25  & 300  & 4 & 4 & 1.188 - 1.244  & 0.99 - 1.56   \\
 & USco 1608-2335                & 19-05-2014 & off & 0.25  & 150  & 4 & 4 & 1.057 - 1.071  & 1.21 - 1.38   \\
 & USco 1610-2239                & 27-04-2014 & off & 0.25  & 150  & 4 & 4 & 1.366 - 1.414  & 1.36 - 1.79   \\
 & USco 1612-2156                & 18-05-2014 & off & 0.25  & 600  & 4 & 4 & 1.059 - 1.117  & 0.74 - 1.91   \\
 &                              & 18-05-2014 & off & 0.25  & 600  & 3 & 3 & 1.219 - 1.299   & 1.31 - 1.37   \\
 & USco 1613-2124                & 19-05-2014 & off & 0.25  & 600  & 4 & 4 & 1.008 - 1.032  & 1.20 - 1.55   \\
 &                              & 20-05-2014 & off & 0.25  & 600  & 3 & 3 & 1.179 - 1.248   & 0.9$^{\star}$ - 2$^{\star}$            \\
 & USco CTIO 108 AB                & 17-06-2014 & off & 0.25  & 300  & 5 & 5 & 1.010 - 1.023 & 1$^{\star}$ - 3$^{\star}$       \\
 & HIP 78530 b                    & 28-05-2014 & on & 0.25   & 600  & 4 & 4 & 1.049 - 1.102  & 1.09 - 3$^{\star}$    \\
 &                              & 28-05-2014 & on & 0.25   & 600  & 4 & 4 & 1.112 - 1.194  & 1.03 - 1.22  \\
092.C-0535(A) & CAHA Tau 1      & 10-10-2013 & off & 0.25  & 600  & 3 & 5 & 1.516 - 1.556  & 0.68 - 0.88  \\
 &                              & 11-10-2013 & off & 0.25  & 60.  & 4 & 5 & 1.581 - 1.677  & 0.61 - 0.75  \\
 & DH Tau b                     & 09-11-2013 & on & 0.025  & 900  & 2 & 3 & 1              & 0.6$^{\star}$ - 1   \\
 &                              & 09-11-2013 & on & 0.025  & 900  & 2 & 3 & 1              & 0.6$^{\star}$ - 1   \\
 &                              & 03-12-2013 & on & 0.025  & 900  & 2 & 3 & 1              & 1 - 3$^{\star}$\\
 & FU Tau b                     & 09-10-2013 & off & 0.25  & 100  & 6 & 9 & 1.645 - 1.770  & 0.6$^{\star}$ - 1.24  \\
 & KPNO Tau 1                   & 07-10-2013 & off & 0.25  & 300  & 6 & 9 & 1.646 - 1.663  & 0.94 - 1.24  \\
 & KPNO Tau 4                   & 08-10-2013 & off & 0.25  & 150  & 6 & 9 & 1.590 - 1.623  & 0.63 - 3$^{\star}$   \\
 & KPNO Tau 6                   & 11-10-2013 & off & 0.25  & 300  & 6 & 9 & 1.579 - 1.595  & 0.62 - 0.79  \\
\hdashline
Patience & AB Pic b             & 01-12-2014 & on & 0.1    & 300  & 5 & 6 & 1.198 - 1.205  & 0.67 - 1.08  \\
092.C-0809(A) &                 & 15-12-2013 & on & 0.1    & 300  & 5 & 6 & 1.245 - 1.283  & 0.66 - 1.02  \\
\hline
\end{tabular}}
\label{tab:Obslog}
\end{table*}

\begin{table*}[ht]
\centering
\caption[Sample information]{
General information about the SINFONI Library sample.
The right ascension (RA) and declination (Dec) are provided in ICRS J2000 format, queried from \href{https://simbad.cds.unistra.fr/simbad/}{SIMBAD} \citep{Wenger2000_SIMBAD} for each target. Parallaxes, in milliarcseconds, were obtained from the Gaia DR3 data release and converted to distances in parsecs. For targets without reported parallaxes, we used the median distance of their associated group, with a conservative uncertainty of 50 parsecs.
The expected interstellar extinction in the visible (\Av) was measured and kindly provided by Carine Babusiaux using the 3D extinction map from \citet{Lallement2022} and is rounded to two decimal places or listed as an upper limit in this table.
}
\renewcommand{\arraystretch}{1.5} 
\centering
\resizebox{\textwidth}{!}{%
\begin{tabular}{ lllcccc}
\hline
Name & RA, Dec &  Association & Parallax (mas) & Distance (pc) & Age (Myr) & $A_{\mathrm{v}}^{Gaia}$ (mag)\\
\hline
2M 0103 AB b   & 01 03 35.66 -55 15 56.24 & Tucana-Ho. & 21.18$\pm$1.37  &  47.21$\pm$3.00 & $30\pm1$     & $<$0.004 \\
AB Pic b       & 06 19 12.91 -58 03 15.52 & Carina     & 19.95$\pm$0.01  &  50.14$\pm$0.03 & $13.3\pm1.1$ & $<$0.004 \\
CAHA Tau 1     & 04 36 09.00 +24 08 21.20 & Taurus     & -               & 140$\pm$50      & $3\pm2$      & 0.23$\pm$0.28 \\
CD-35 2722 b   & 06 09 19.21 -35 49 31.06 & AB Doradus & 44.72$\pm$0.01  &  22.36$\pm$0.01 & $133\pm20$    & $<$0.003 \\
DH Tau b       & 04 29 41.66 +26 32 56.50 & Taurus     &  7.49$\pm$0.03  & 133.45$\pm$0.53 & $3\pm2$      & 0.18$\pm$0.01 \\
FU Tau b       & 04 23 35.75 +25 02 59.64 & Taurus     &  7.48$\pm$1.12  & 133.78$\pm$17.4 & $3\pm2$      & 0.18$\pm$0.15 \\
HIP 78530 b    & 16 01 55.65 -21 58 53.01 & Upper Sco. &  7.42$\pm$0.03  & 134.70$\pm$0.54 & $10\pm1$     & 0.23$\pm$0.01 \\
HR 7329 b      & 19 22 51.36 -54 25 31.56 & $\beta$ Pictoris & 20.60$\pm$0.10  &  48.54$\pm$0.23 & $20\pm4$     & $<$0.006 \\
KPNO Tau 1     & 04 15 14.72 +28 00 09.38 & Taurus     &  7.77$\pm$0.61  & 128.72$\pm$9.37 & $5\pm4$      & 0.13$\pm$0.07 \\
KPNO Tau 4     & 04 27 27.99 +26 12 05.08 & Taurus     &  7.49$\pm$1.02  & 133.43$\pm$16.0 & $5\pm4$      & 0.18$\pm$0.14 \\
KPNO Tau 6     & 04 30 07.24 +26 08 20.71 & Taurus     &  8.15$\pm$0.38  & 122.72$\pm$5.47 & $5\pm4$      & 0.09$\pm$0.05 \\
USco 1606-2219 & 16 06 03.76 -22 19 29.95 & Upper Sco. & 10.16$\pm$2.41  &  98.39$\pm$19.0 & $ 8\pm1$     & 0.06$\pm$0.09 \\
USco 1606-2230 & 16 06 48.18 -22 30 40.02 & Upper Sco. &  6.49$\pm$0.49  & 154.17$\pm$11.0 & $11\pm1$     & 0.30$\pm$0.03 \\
USco 1606-2335 & 16 06 06.29 -23 35 13.36 & Upper Sco. & -               & 145$\pm$50      & $11\pm1$     & 0.27$\pm$0.12 \\
USco 1607-2239 & 16 07 27.82 -22 39 04.06 & Upper Sco. & -               & 145$\pm$50      & $11\pm1$     & 0.28$\pm$0.12 \\
USco 1608-2232 & 16 08 18.43 -22 32 24.82 & Upper Sco. & -               & 145$\pm$50      & $11\pm1$     & 0.28$\pm$0.12 \\
USco 1608-2335 & 16 08 30.49 -23 35 10.98 & Upper Sco. &  7.14$\pm$0.42  & 104.04$\pm$7.78 & $11\pm1$     & 0.26$\pm$0.03 \\
USco 1610-2239 & 16 10 47.13 -22 39 49.32 & Upper Sco. &  8.08$\pm$0.59  & 123.81$\pm$8.42 & $11\pm1$     & 0.19$\pm$0.05 \\
USco 1612-2156 & 16 12 27.64 -21 56 40.71 & Upper Sco. & -               & 145$\pm$50      & $11\pm1$     & 0.30$\pm$0.13 \\
USco 1613-2124 & 16 13 02.32 -21 24 28.37 & Upper Sco. & -               & 145$\pm$50      & $11\pm1$     & 0.31$\pm$0.14 \\
USco CTIO 108 A  & 16 05 54.07 -18 18 44.37 & Upper Sco. &  6.85$\pm$0.13  & 146.02$\pm$2.72 & $11\pm1$     & 0.31$\pm$0.01 \\
\hline
\end{tabular}}
\label{tab:Targets1}
\end{table*}

As mentioned above, in Tables \ref{tab:Targets1} and \ref{tab:Targets2}, we provide an overview of the general properties of each object, organized in alphabetical order. The information is split into two main tables for clarity.
Table \ref{tab:Targets1} lists the spatial coordinates, including right ascension and declination, the parallax in mas, and the corresponding distance in parsec. The last two columns provide details on the distance and age of each target.
Table \ref{tab:Targets2} continues with the spectral type of each target, its apparent and absolute magnitudes in the K band, and the flux calibration factor for the SINFONI spectra in units of $W \, m^{-2} \, \mu m ^{-1}$. In the last column of this table, we also provide all the references used to gather the information about the targets.

In addition, Table \ref{tab:Targets_companions} provides various properties of the companion objects in our sample. Specifically, it includes information on the projected distance to their host stars, the spectral type of the host stars, and, when available, the metallicity of the host stars. For comparison, we also include the metallicity and \co ratio measurements from our work for the targets where such data are available. 
The chemical composition of the Sun can be found in \citet{Asplund2021}. The metallicity for the Taurus-Auriga association is provided by \citet{Dorazi2011}. More metallicities of young stellar associations can be extracted from \citet{VianaAlmeida2009} and \citet{Torres2006}.
Additionally, a column is provided for specific comments or annotations regarding our measurements.

The last table included in this appendix summarizes all the properties derived from the evolutionary models. We have used the BHAC15 COND03 and DUSTY00 models from \citep{Baraffe2015} to provide estimated values for the mass (\MJup), \Teff (K), \logg (dex), and radius (\RJup). Specifically, we used the \logg values from the COND03 predictions as priors for our atmospheric models. The predictions presented here were based on the SPHERE IRDIS K12,2 filter. Since some of our targets have distances derived from their mean association distances, the evolutionary predictions for some targets should be interpreted with caution.

\begin{table*}[ht]
\caption[SINFONI Library information, part 2]{In this table, the second part of the general information for the SINFONI Library is provided. Specifically, we listed the spectral types of our objects, along with their K-band apparent and absolute magnitudes, and the flux scaling factor. The corresponding references are found in the right-hand column. 
}
\renewcommand{\arraystretch}{1.5} 
\centering
\resizebox{0.9\textwidth}{!}{%
\begin{tabular}{lccccc}
\hline
Name & Spectral type & $m_{K}$ (mag) & $M_{K}$ (mag) & Flux ($W\,m^{-2}\,\mu m^{-1}$) & References \\
\hline
2M 0103 AB b   & $L0 \pm 0.5$    & $13.690\pm0.022$ & $10.32 \pm 0.07$ & $1.4361\times10^{-15}$ & 1,2,3,4,11,24 \\
AB Pic b       & $L0.5 \pm 0.5$  & $14.090\pm0.08$  & $10.59 \pm 0.08$ & $9.8374\times10^{-16}$ & 1,4,5,6,7,8,9,10,11 \\
CAHA Tau 1     & $M6 \pm 5$      & $15.110\pm0.09$  & $9.38 \pm 0.11$  & $3.9881\times10^{-16}$ & 1,11,12,13 \\
CD-35 2722 b   & $L4.5 \pm 5$    & $12.01\pm0.16$   & $10.26 \pm 0.30$ & $6.5393\times10^{-15}$ & 1,4,11,14,25 \\
DH Tau b       & $M9.25 \pm 0.25$& $14.19\pm0.02$   & $8.56 \pm 0.04$  & $9.0759\times10^{-16}$ & 1,7,9,11,15 \\
FU Tau b       & $L0.25 \pm 0.25$& $13.329\pm0.098$ & $7.70 \pm 0.11$  & $2.0223\times10^{-15}$ & 1,11,12,15 \\
HIP 78530 b    & $M7 \pm 1$      & $13.491\pm0.003$ & $7.84 \pm 0.02$  & $1.7653\times10^{-15}$ & 1,11,17,19 \\
HR 7329 b      & $M7 \pm 1$      & $11.6\pm0.1$     & $8.17 \pm 0.17$  & $1.0106\times10^{-14}$ & 1,7,11,20 \\
KPNO Tau 1     & $M8.5 \pm 0.5$  & $13.772\pm0.035$ & $8.22 \pm 0.05$  & $1.3410\times10^{-15}$ & 1,11,21 \\
KPNO Tau 4     & $M9.5 \pm 0.5$  & $13.28\pm0.03$   & $7.65 \pm 0.05$  & $2.0882\times10^{-15}$ & 1,7,11,12,21 \\
KPNO Tau 6     & $M8.5 \pm 0.5$  & $13.689\pm0.037$ & $8.25 \pm 0.06$  & $1.4592\times10^{-15}$ & 1,11,21,22 \\
USco 1606-2219 & $L2 \pm 1$      & $14.550\pm0.086$ & $9.59 \pm 0.45$  & $6.5125\times10^{-16}$ & 1,11,23 \\
USco 1606-2230 & $M8 \pm 0.5$    & $13.806\pm0.042$ & $7.87 \pm 0.26$  & $1.3052\times10^{-15}$ & 1,11,23 \\
USco 1606-2335 & $M9 \pm 0.5$    & $15.052\pm0.161$ & $9.25 \pm 0.40$  & $4.1317\times10^{-16}$ & 1,11,23 \\
USco 1607-2239 & $L0 \pm 0.5 $   & $15.436\pm0.019$ & $9.63 \pm 0.26$  & $2.8820\times10^{-16}$ & 1,11,23 \\
USco 1608-2232 & $M9 \pm 0.5$    & $14.775\pm0.113$ & $8.97 \pm 0.35$  & $5.3114\times10^{-16}$ & 1,11,23 \\
USco 1608-2335 & $M8.5 \pm 0.5$  & $13.814\pm0.054$ & $8.73 \pm 0.39$  & $1.3102\times10^{-15}$ & 1,11,23 \\
USco 1610-2239 & $M8.5 \pm 0.5$  & $14.025\pm0.058$ & $8.56 \pm 0.34$  & $1.0767\times10^{-15}$ & 1,11,23 \\
USco 1612-2156 & $L0 \pm 1$      & $15.395\pm0.166$ & $9.59 \pm 0.40$  & $3.0080\times10^{-16}$ & 1,11,23 \\
USco 1613-2124 & $L0 \pm 1$      & $15.652\pm0.022$ & $9.85 \pm 0.26$  & $2.3869\times10^{-16}$ & 1,11,23 \\
USco CTIO 108 A& $M7 \pm 0.5$    & $12.51\pm0.03$   & $6.69 \pm 0.27$  & $4.4330\times10^{-15}$ & 1,7,11 \\
\hline
\end{tabular}}
\tablebib{
(1) \citet{Wenger2000_SIMBAD}; 
(2) \citet{Liu2013}; 
(3) \citet{Kraus2014}; 
(4) \citet{Stone2016}; 
(5) \citet{chauvin2005}; 
(6) \citet{Bonnefoy2010}; 
(7) \citet{Bonnefoy2014s}; 
(8) \citet{Patience2012s}; 
(9) \citet{Martinez2022}; 
(10) \citet{Booth2021}; 
(11) \citet{GaiaCollaboration2021}; 
(12) \citet{Quanz2010}; 
(13) \citet{Luhman2010}; 
(14) \citet{Wahhaj2011}; 
(15) \citet{Wu2020}; 
(16) \citet{Cheetham2015}; 
(17) \citet{Lachapelle2015}; 
(18) \citet{Ireland2011}; 
(19) \citet{Petrus2020s}; 
(20) \citet{Lowrance2000}; 
(21) \citet{Briceno2002}; 
(22) \citet{Akeson2019}; 
(23) \citet{Lodieu2011}; 
(24) \citet{Delorme2013};
(25) \citet{Gagne2018}.
}
\label{tab:Targets2}
\end{table*}

\begin{table*}[ht]
\caption{Table \ref{tab:Targets_companions} lists the properties of the companion objects in our sample. Host star metallicities were taken from \citet{Swastik2021}.
Additionally, we have included the \co ratio and metallicity measurements from our work, along with specific annotations where relevant. 
}
\renewcommand{\arraystretch}{1.5} 
\centering
\resizebox{\textwidth}{!}{%
\begin{tabular}{lcccccc}
\hline
Name & Distance host (au) & Spectral type host & \met host  & \met companion & \co companion & comment\\
\hline
2M 0103 AB b   &  $84 \pm 1$     & $M6$     & - & $0.15^{+0.2}_{-0.05}$&$0.6^{+0.0}_{-0.05}$ & - \\
AB Pic b       &  $273 \pm 1$    & $K2$     & $0.04\pm0.02$& $-0.08^{+0.57}_{-0.03}$&$0.45^{+0.11}_{-0.0}$ & - \\
CD-35 2722 b   &  $67 \pm 1$     & $M1$     & - & $-0.0^{+0.01}_{-0.1}$&$0.5^{+0.01}_{-0.0}$ & - \\
DH Tau b       &  $330 \pm 5$    & $M0.5$   & - &  $-0.11^{+0.04}_{-0.02}$&$0.4^{+0.0}_{-0.0}$ & S/N$<80$ \\
FU Tau b       &  $800 \pm 10$   & $M7.25$  & - &  $0.61^{+0.13}_{-0.09}$&$0.65^{+0.01}_{-0.35}$ & S/N$<80$ \\
HIP 78530 b    &  $623 \pm 8$    & $B9$     & $-0.5\pm0.03$ & - & - & \Teff$>2000K$\\
HR 7329 b      & $136 \pm 10 $   & $A0$      & -  & - & - & \Teff$>2000K$\\
\hline
\end{tabular}}
\label{tab:Targets_companions}
\end{table*}

\begin{table*}[ht]
\caption{We report the masses, effective temperatures, surface gravities, and radii for all our targets using the BHAC15 COND03 and DUSTY00 models \citep{Baraffe2015}. The predictions are based on the SPHERE IRDIS K12\,2 filter. 
For these calculations we used the ages, which are taken from Table \ref{tab:Targets1}, and the absolute magnitudes taken from Table \ref{tab:Targets2}, which were derived using the distances from Table \ref{tab:Targets1} and the apparent magnitudes from Table \ref{tab:Targets2}.
}
\renewcommand{\arraystretch}{1.5} 
\centering
\resizebox{\textwidth}{!}{%
\begin{tabular}{lcccccccc}
\hline
Name  & \multicolumn{2}{c}{Mass (\MJup)} & \multicolumn{2}{c}{\Teff (K)} & \multicolumn{2}{c}{\logg (dex)} & \multicolumn{2}{c}{Radius (\RJup)} \\
    & cond & dusty & cond & dusty & cond & dusty & cond & dusty \\
\hline
2M 0103 AB b     & 10.5 $\pm$ 1.1 & 12.6 $\pm$ 2.1 & 1731 $\pm$ 96  & 1801 $\pm$ 193 & 4.04 $\pm$ 0.03 & 4.07 $\pm$ 0.05 & 1.54 $\pm$ 0.03 & 1.63 $\pm$ 0.05 \\
AB Pic b         & 9.4  $\pm$ 1.1 & 10.5 $\pm$ 2.1 & 1635 $\pm$ 99  & 1608 $\pm$ 72  & 4.01 $\pm$ 0.03 & 4.02 $\pm$ 0.04 & 1.51 $\pm$ 0.03 & 1.58 $\pm$ 0.10 \\
CAHA Tau 1       & 5.2  $\pm$ 1.1 & 12.6 $\pm$ 2.1 & 1901 $\pm$ 154 & 2036 $\pm$ 164 & 3.47 $\pm$ 0.03 & 3.96 $\pm$ 0.05 & 2.09 $\pm$ 0.15 & 1.85 $\pm$ 0.07 \\
CD-35 2722 b     & 7.3  $\pm$ 1.1 & 7.3  $\pm$ 1.1 & 1668 $\pm$ 125 & 1668 $\pm$ 125 & 3.84 $\pm$ 0.04 & 3.84 $\pm$ 0.04 & 1.62 $\pm$ 0.05 & 1.62 $\pm$ 0.05 \\
DH Tau b         & 8.4  $\pm$ 1.1 & 8.4  $\pm$ 1.1 & 2159 $\pm$ 61  & 2159 $\pm$ 61  & 3.52 $\pm$ 0.00 & 3.52 $\pm$ 0.01 & 2.51 $\pm$ 0.16 & 2.51 $\pm$ 0.16 \\
FU Tau b         & 15.7 $\pm$ 3.1 & 21.0 $\pm$ 5.2 & 2400 $\pm$ 79  & 2544 $\pm$ 94  & 3.54 $\pm$ 0.00 & 3.79 $\pm$ 0.16 & 3.36 $\pm$ 0.37 & 2.89 $\pm$ 0.12 \\
HIP 78530 b      & 31.4 $\pm$ 10.5 & 31.4 $\pm$ 10.5 & 2622 $\pm$ 180 & 2667 $\pm$ 180 & 3.97 $\pm$ 0.03 & 4.05 $\pm$ 0.08 & 2.88 $\pm$ 0.46 & 2.63 $\pm$ 0.26 \\
HR 7329 b        & 21.0 $\pm$ 5.2 & 21.0 $\pm$ 5.2 & 2442 $\pm$ 254 & 2487 $\pm$ 319 & 3.94 $\pm$ 0.10 & 3.97 $\pm$ 0.08 & 2.42 $\pm$ 0.55 & 2.36 $\pm$ 0.50 \\
KPNO Tau 1      & 21.0 $\pm$ 5.2 & 21.0 $\pm$ 5.2 & 2452 $\pm$ 188 & 2504 $\pm$ 250 & 3.90 $\pm$ 0.06 & 3.93 $\pm$ 0.05 & 2.54 $\pm$ 0.48 & 2.48 $\pm$ 0.47 \\
KPNO Tau 4      & 31.4 $\pm$ 10.5 & 21.0 $\pm$ 5.2 & 2616 $\pm$ 164 & 2504 $\pm$ 250 & 3.80 $\pm$ 0.11 & 3.93 $\pm$ 0.05 & 3.53 $\pm$ 0.99 & 2.48 $\pm$ 0.47 \\
KPNO Tau 6       & 21.0 $\pm$ 5.2 & 21.0 $\pm$ 5.2 & 2452 $\pm$ 188 & 2504 $\pm$ 250 & 3.90 $\pm$ 0.06 & 3.93 $\pm$ 0.05 & 2.54 $\pm$ 0.48 & 2.48 $\pm$ 0.47 \\
USco 1606 2219   & 12.6 $\pm$ 2.1 & 15.7 $\pm$ 3.1 & 1915 $\pm$ 184 & 2168 $\pm$ 367 & 4.08 $\pm$ 0.04 & 4.05 $\pm$ 0.02 & 1.61 $\pm$ 0.07 & 1.87 $\pm$ 0.24 \\
USco 1606 2230  & 31.4 $\pm$ 10.5 & 31.4 $\pm$ 10.5 & 2622 $\pm$ 180 & 2667 $\pm$ 180 & 3.97 $\pm$ 0.03 & 4.05 $\pm$ 0.08 & 2.88 $\pm$ 0.46 & 2.63 $\pm$ 0.26 \\
USco 1606 2335   & 15.7 $\pm$ 3.1 & 15.7 $\pm$ 3.1 & 2188 $\pm$ 273 & 2168 $\pm$ 367 & 4.05 $\pm$ 0.04 & 4.05 $\pm$ 0.02 & 1.87 $\pm$ 0.26 & 1.87 $\pm$ 0.24 \\
USco 1607 2239   & 12.6 $\pm$ 2.1 & 12.6 $\pm$ 2.1 & 1915 $\pm$ 184 & 1801 $\pm$ 193 & 4.08 $\pm$ 0.04 & 4.07 $\pm$ 0.05 & 1.61 $\pm$ 0.07 & 1.63 $\pm$ 0.24 \\
USco 1608 2232  & 15.7 $\pm$ 3.1 & 15.7 $\pm$ 3.1 & 2188 $\pm$ 273 & 2168 $\pm$ 367 & 4.05 $\pm$ 0.04 & 4.05 $\pm$ 0.02 & 1.87 $\pm$ 0.26 & 1.87 $\pm$ 0.24 \\
USco 1608 2335   & 15.7 $\pm$ 3.1 & 15.7 $\pm$ 3.1 & 2188 $\pm$ 273 & 2168 $\pm$ 367 & 4.05 $\pm$ 0.04 & 4.05 $\pm$ 0.02 & 1.87 $\pm$ 0.26 & 1.87 $\pm$ 0.24 \\
USco 1610 2239   & 21.0 $\pm$ 5.2 & 21.0 $\pm$ 5.2 & 2442 $\pm$ 254 & 2487 $\pm$ 319 & 3.94 $\pm$ 0.10 & 3.97 $\pm$ 0.08 & 2.42 $\pm$ 0.55 & 2.36 $\pm$ 0.50 \\
USco 1612 2156   & 12.6 $\pm$ 2.1 & 12.6 $\pm$ 2.1 & 1915 $\pm$ 184 & 1801 $\pm$ 193 & 4.08 $\pm$ 0.04 & 4.07 $\pm$ 0.05 & 1.61 $\pm$ 0.07 & 1.63 $\pm$ 0.24 \\
USco 1613 2124  & 12.6 $\pm$ 2.1 & 12.6 $\pm$ 2.1 & 1915 $\pm$ 184 & 1801 $\pm$ 193 & 4.08 $\pm$ 0.04 & 4.07 $\pm$ 0.05 & 1.61 $\pm$ 0.07 & 1.63 $\pm$ 0.24 \\
USco CTIO 108 A  & 12.6 $\pm$ 2.1 & 12.6 $\pm$ 2.1 & 1915 $\pm$ 184 & 1801 $\pm$ 193 & 4.08 $\pm$ 0.04 & 4.07 $\pm$ 0.05 & 1.61 $\pm$ 0.07 & 1.63 $\pm$ 0.24 \\
\hline
\end{tabular}}
\label{tab:Targets_evolu}
\end{table*}

\newpage
\clearpage
\section{Result figures per target}\label{sec_a:pertarget}

In this appendix, we present an overview of each target's previous findings as well as the best-fit models and posteriors derived for each grid with \formosa. For the posteriors, we include spider plots that provide a comparative view of the models' predictions in terms of \Teff, \logg, and RV, with the shaded area representing 99.73\% confidence intervals ($3\sigma$). The accompanying tables list the exact values for each posterior, with asymmetric uncertainty ranges representing a $1\sigma$ confidence interval.

\begin{figure*}[ht]
\centering
\begin{subfigure}[b]{0.61\textwidth}
    \includegraphics[width=\textwidth]{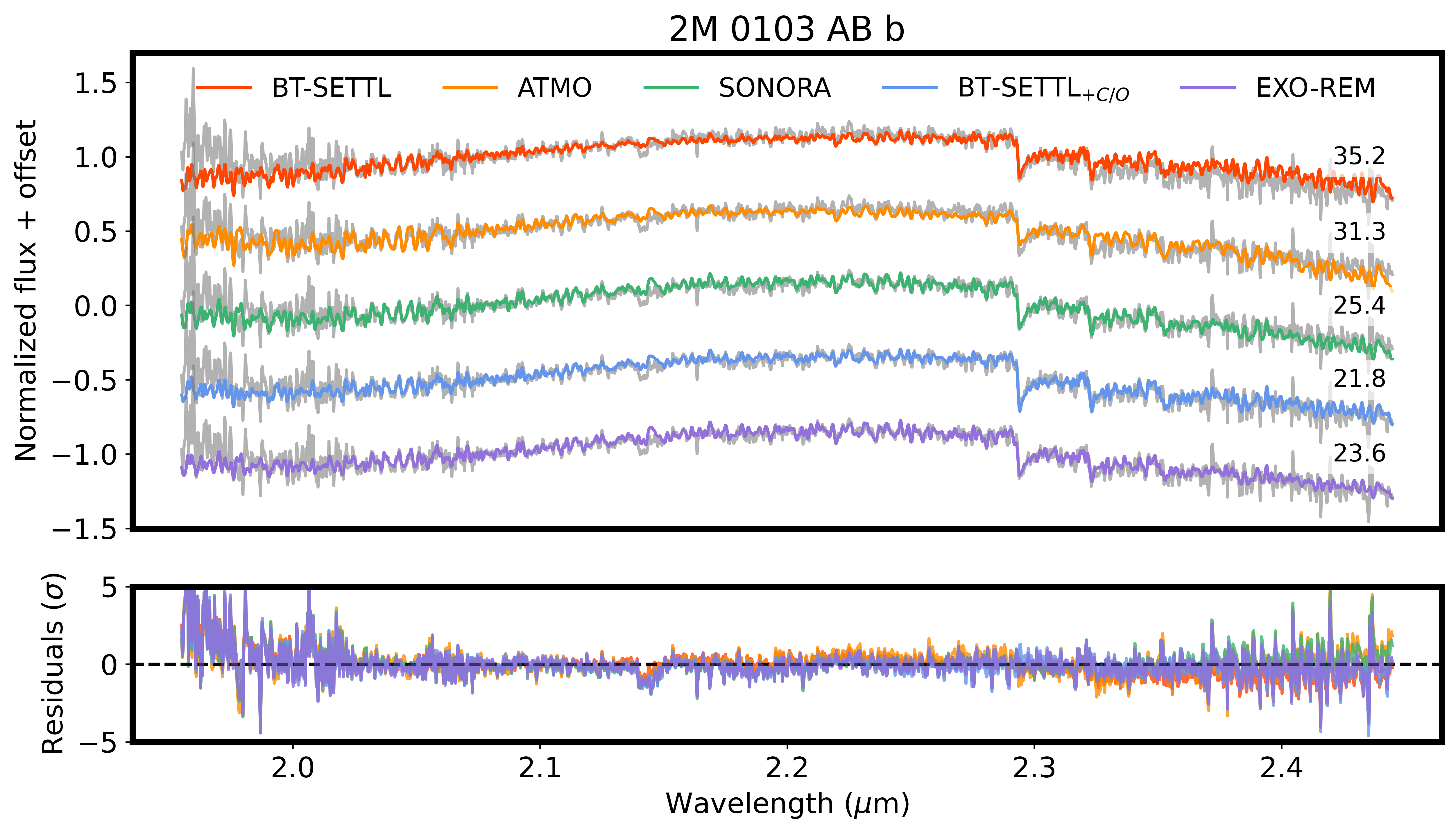}
\end{subfigure}
\hfill
\begin{subfigure}[b]{0.33\textwidth}
    \includegraphics[width=\textwidth]{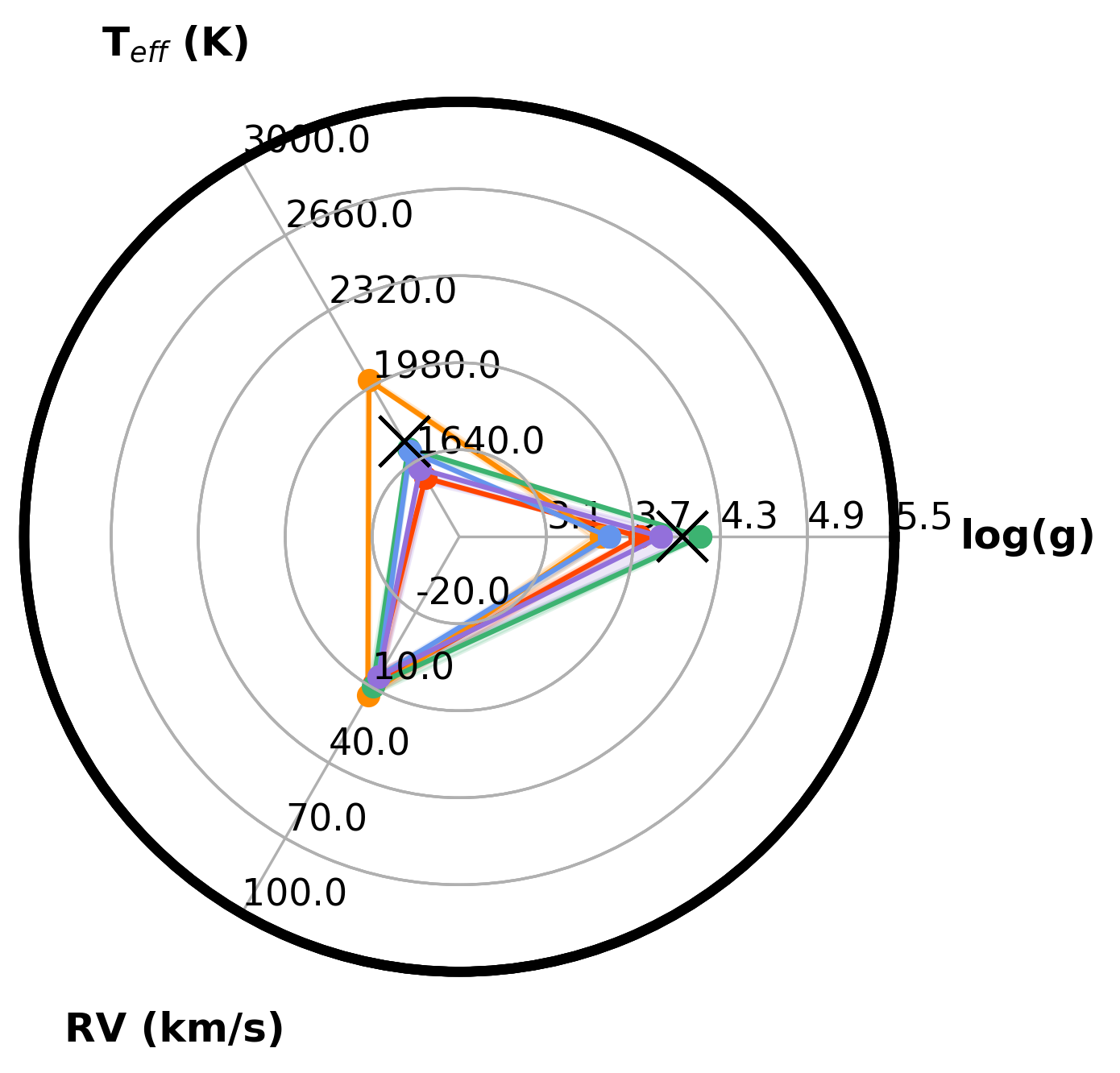}
\end{subfigure}
\caption{Same as Figure \ref{fig:target_CD-35_2722_b} but for 2M 0103 AB b}
\label{fig:target_2M0103AB_b}
\end{figure*}

2M 0103 AB b, also known as Delorme 1 AB b or SCR J0103–5515C, is a first-of-its-kind companion. Its discovery by \citet{Delorme2013} marked the first direct imaging detection of a planetary-mass companion orbiting a very low-mass stellar binary at a wide separation. Since then, it has been followed up with multiple instruments. Preliminary orbital constraints and stability analyses for the system have been presented by \citet{Blunt2017A}. In addition, both optical (H$\alpha$/H$\beta$) and near-infrared (Br$_{\gamma}$) observations have confirmed that the companion is still actively accreting material at an age of approximately 40 Myr, indicating unusually long-lived accretion in a low-mass substellar object \citep{Eriksson2020, Betti2022}. Interestingly, as shown in Figure \ref{fig:target_2M0103AB_b}, we do not detect Br$_{\gamma}$ in these observations. However, we note an odd behavior in our data around $\sim$2.15 $\mu$m, which could be due to systematics.

\begin{table*}[ht]
\centering
\caption{Same as Table \ref{tab:target_CD-35_2722_b} but for 2M 0103 AB b}
\renewcommand{\arraystretch}{1.5} 
\resizebox{\textwidth}{!}{%
\begin{tabular}{lcccccccccccc}
\hline
model& \Teff (K) & \logg (dex) & \met & \co & $\gamma$ & \fsed & RV (km/s) & $\beta$ (km/s) & ln(z) & $\chi ^2 _{red}$ \\ 
\hline
\btex&$1565^{+6}_{-6}$&$3.75^{+0.03}_{-0.02}$&&&&&$8.65^{+5.82}_{-6.04}$&$70.04^{+14.74}_{-14.62}$&-35168.6&35\\ 
\atmo&$2006^{+20}_{-2}$&$3.48^{+0.09}_{-0.07}$&$-0.13^{+0.02}_{-0.02}$&$0.7^{+0.0}_{-0.0}$&$1.03^{+0.0}_{-0.0}$&&$13.03^{+1.18}_{-1.79}$&$118^{+2}_{-3}$&-31471.4&31\\ 
\sono&$1699^{+2}_{-20}$&$4.16^{+0.02}_{-0.04}$&&&&$1.0^{+0.02}_{-0.0}$&$9.44^{+4.44}_{-3.5}$&$82.61^{+10.22}_{-7.84}$&-25422.1&25\\ 
\btse&$1686^{+5}_{-7}$&$3.53^{+0.05}_{-0.03}$&&$0.7^{+0.01}_{-0.02}$&&&$5.69^{+2.26}_{-2.2}$&$77.57^{+4.87}_{-5.77}$&-21978.8&22\\ 
\exor&$1606^{+5}_{-64}$&$3.89^{+0.21}_{-0.05}$&$0.15^{+0.2}_{-0.05}$&$0.6^{+0.0}_{-0.05}$&&&$5.63^{+3.36}_{-3.49}$&$88.09^{+5.42}_{-7.73}$&-23532.1&24\\ 

\hline
\end{tabular}}
\label{tab:target_2M0103AB_b}
\end{table*}

AB Pic b, is a substellar companion discovered by \citet{chauvin2005} via direct imaging with VLT/NaCo. Follow-up spectroscopic observations with VLT/SINFONI confirmed its nature as a young L0–L1 dwarf with intermediate surface gravity, estimating an effective temperature of approximately 2000 K and \logg around 4.0 \citep{Bonnefoy2010}. More recent studies have combined archival near-infrared spectroscopy and multi-band photometry to further constrain its atmospheric properties, suggesting a cooler temperature of $\sim$1700 K, \logg $\sim$ 4.5, and a solar-like \co ratio of 0.58$\pm$0.08 \citep{PalmaBifani2023}. In the same work an orbital analysis was performed using astrometry from NaCo and SPHERE, which suggests a semi-major axis of ~190 au and a high inclination of $\sim$90 degrees, though the eccentricity remains uncertain. In \citet{PalmaBifani2023} and \citet{Gandhi2025}, the authors discovered a spin–orbit misalignment, with an obliquity around $\sim$90 degrees, suggesting that AB Pic b is tilted as Uranus in our solar system. In terms of the architecture of the system, Gaia-Hipparcos proper motion anomalies hint at a potential inner companion down to $\sim$2 times the mass of Jupiter beyond 10 au \citep{PalmaBifani2023, Lagrange2025}. While there is no evidence of ongoing accretion, in the SINFONI observations presented here, we do observe an emission peak at 2.16 $\mu$m, which might be Br$_{\gamma}$, and is calling for further analysis. Finally, the wide separation and system architecture continue to raise questions about its formation, with a planet-like formation mechanism potentially more likely.

\begin{figure*}[ht]
\centering
\begin{subfigure}[b]{0.61\textwidth}
    \includegraphics[width=\textwidth]{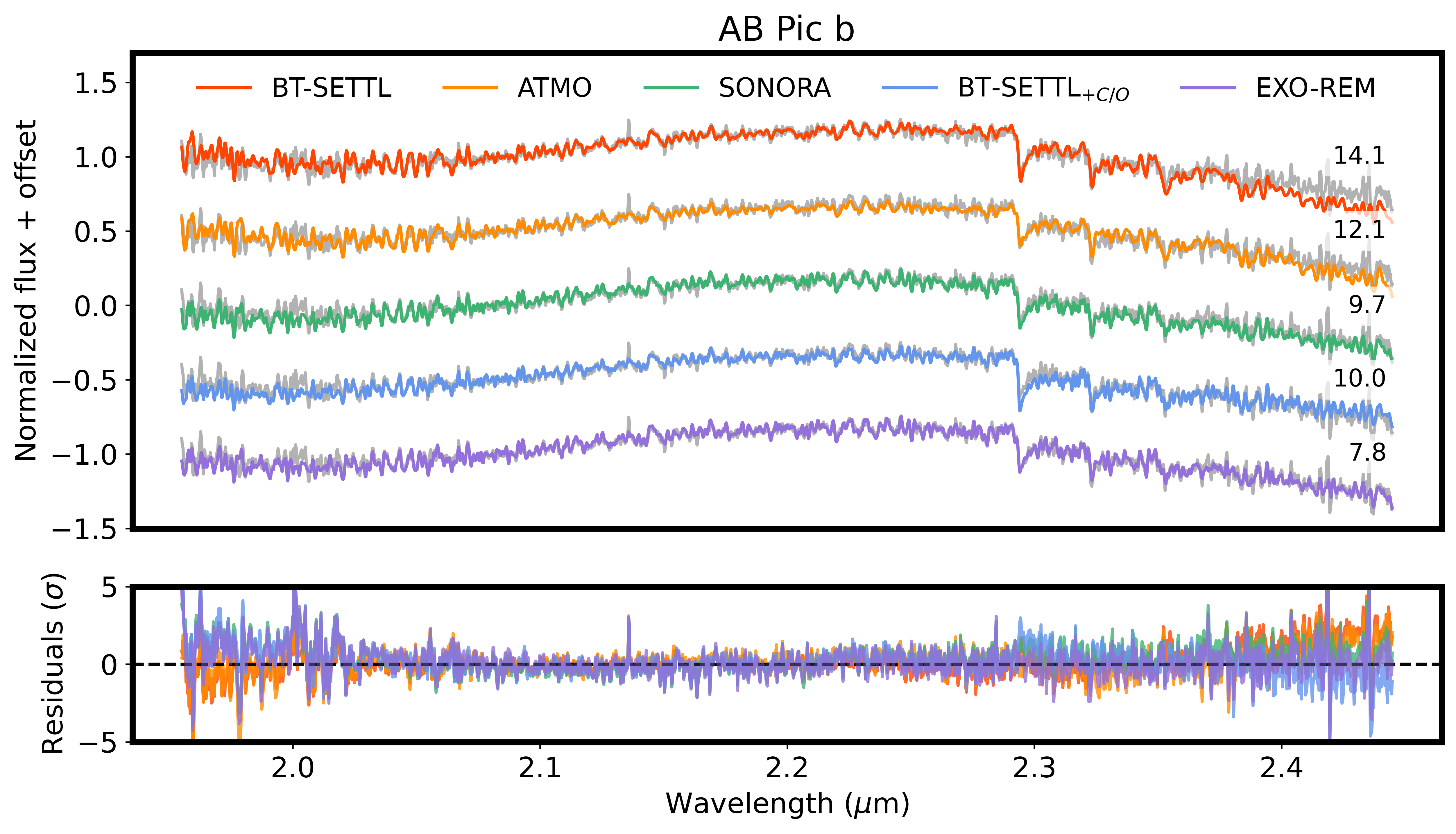}
\end{subfigure}
\hfill
\begin{subfigure}[b]{0.33\textwidth}
    \includegraphics[width=\textwidth]{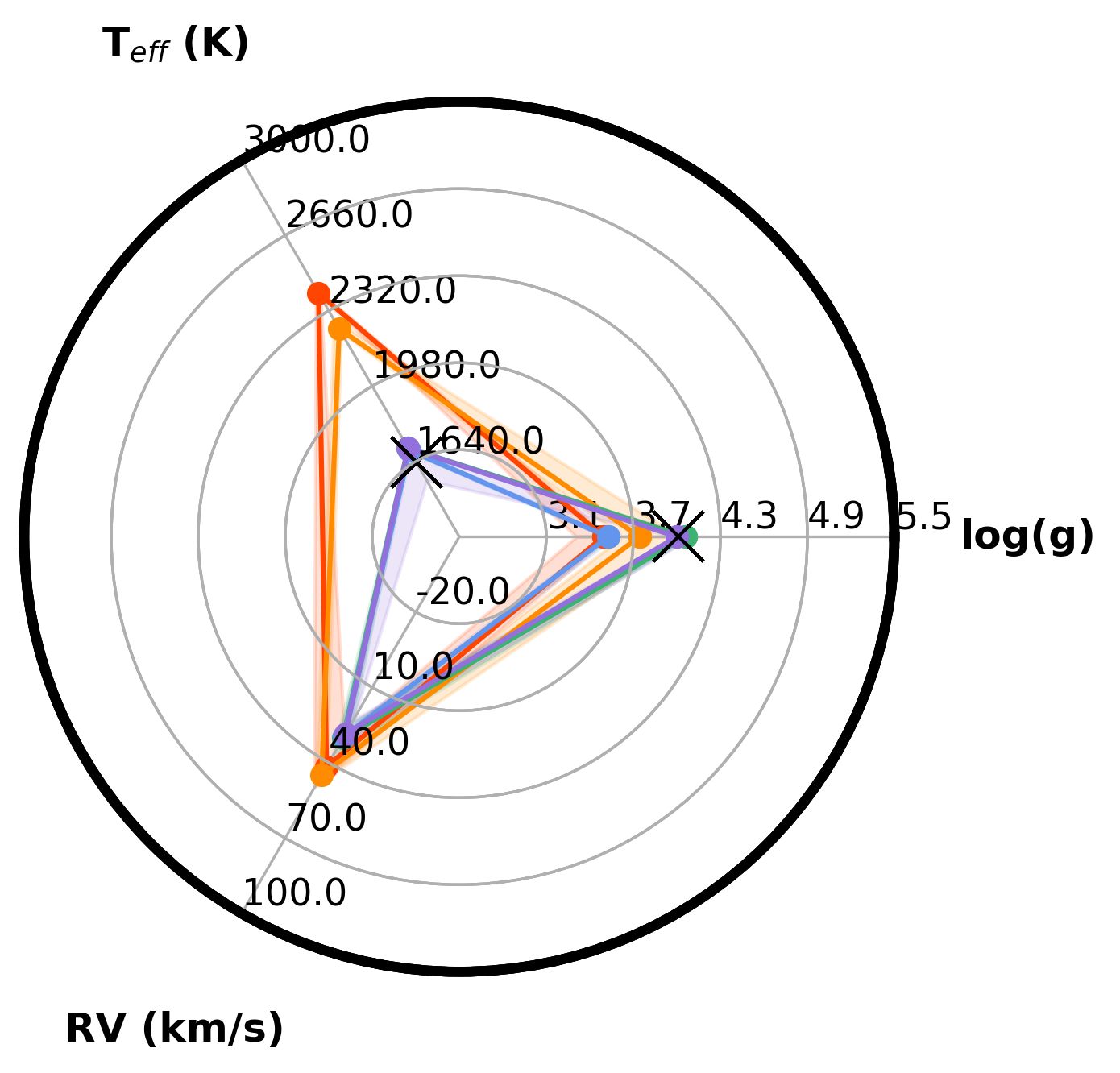}
\end{subfigure}
\caption{Same as Figure \ref{fig:target_CD-35_2722_b} but for AB Pic b}
\label{fig:target_AB_Pic_b}
\end{figure*}

\begin{table*}[ht]
\centering
\caption{Same as Table \ref{tab:target_CD-35_2722_b} but for AB Pic b}
\renewcommand{\arraystretch}{1.5} 
\resizebox{\textwidth}{!}{%
\begin{tabular}{lcccccccccccc}
\hline
model& \Teff (K) & \logg (dex) & \met & \co & $\gamma$ & \fsed & RV (km/s) & $\beta$ (km/s) & ln(z) & $\chi ^2 _{red}$ \\ 
\hline
\btex&$2399^{+21}_{-15}$&$3.49^{+0.04}_{-0.19}$&&&&&$41.53^{+8.55}_{-12.33}$&$124.09^{+18.46}_{-15.45}$&-14087.3&14\\ 
\atmo&$2238^{+41}_{-15}$&$3.74^{+0.28}_{-0.12}$&$0.56^{+0.04}_{-0.05}$&$0.63^{+0.07}_{-0.33}$&$1.04^{+0.0}_{-0.01}$&&$44.9^{+4.11}_{-3.57}$&$129^{+9}_{-7}$&-12119.7&12\\ 
\sono&$1700^{+1}_{-6}$&$4.06^{+0.03}_{-0.05}$&&&&$1.0^{+0.02}_{-0.0}$&$30.21^{+6.75}_{-6.25}$&$74.3^{+15.02}_{-14.7}$&-9702.8&10\\ 
\btse&$1690^{+6}_{-8}$&$3.53^{+0.05}_{-0.03}$&&$0.67^{+0.01}_{-0.02}$&&&$29.56^{+3.09}_{-3.15}$&$62.28^{+7.94}_{-8.96}$&-10147.3&10\\ 
\exor&$1700^{+1}_{-146}$&$4.0^{+0.11}_{-0.01}$&$-0.08^{+0.57}_{-0.03}$&$0.45^{+0.11}_{-0.0}$&&&$28.53^{+5.12}_{-5.6}$&$79.55^{+10.51}_{-19.74}$&-7823.7&8\\ 
\hline
\end{tabular}}
\label{tab:target_AB_Pic_b}
\end{table*}

CAHA Tau 1 is a young, free-floating substellar object identified in the Taurus star-forming region. It was initially discovered by \citet{Quanz2010} during a deep imaging survey targeting planetary-mass candidates, and it was later spectroscopically confirmed as a member of the Taurus population by \citet{Luhman2010}. Its spectral features and luminosity are consistent with those of a very low-mass brown dwarf or a planetary-mass object, placing it near the deuterium-burning limit. 

\begin{figure*}[ht]
\centering
\begin{subfigure}[b]{0.61\textwidth}
    \includegraphics[width=\textwidth]{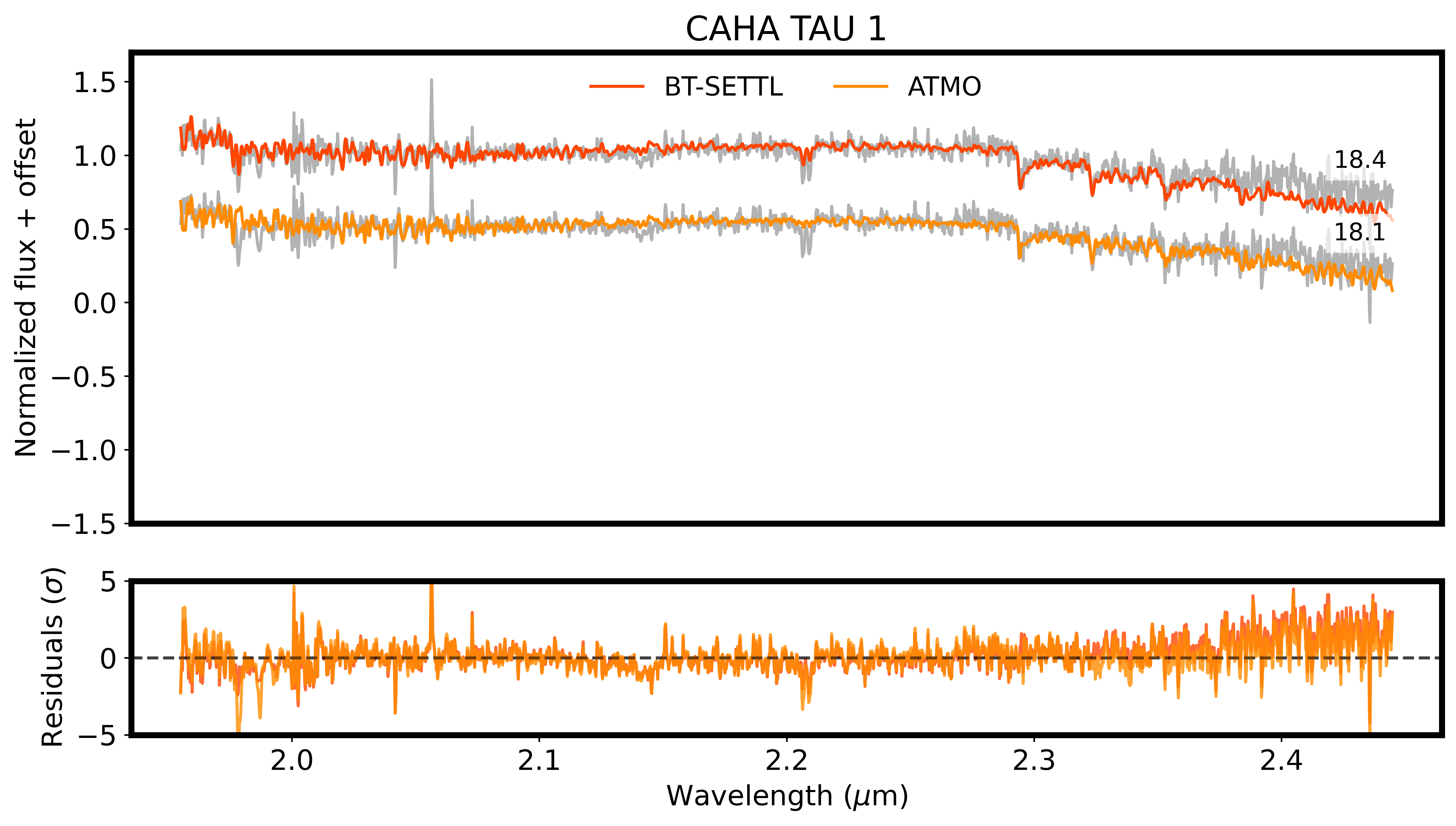}
\end{subfigure}
\hfill
\begin{subfigure}[b]{0.33\textwidth}
    \includegraphics[width=\textwidth]{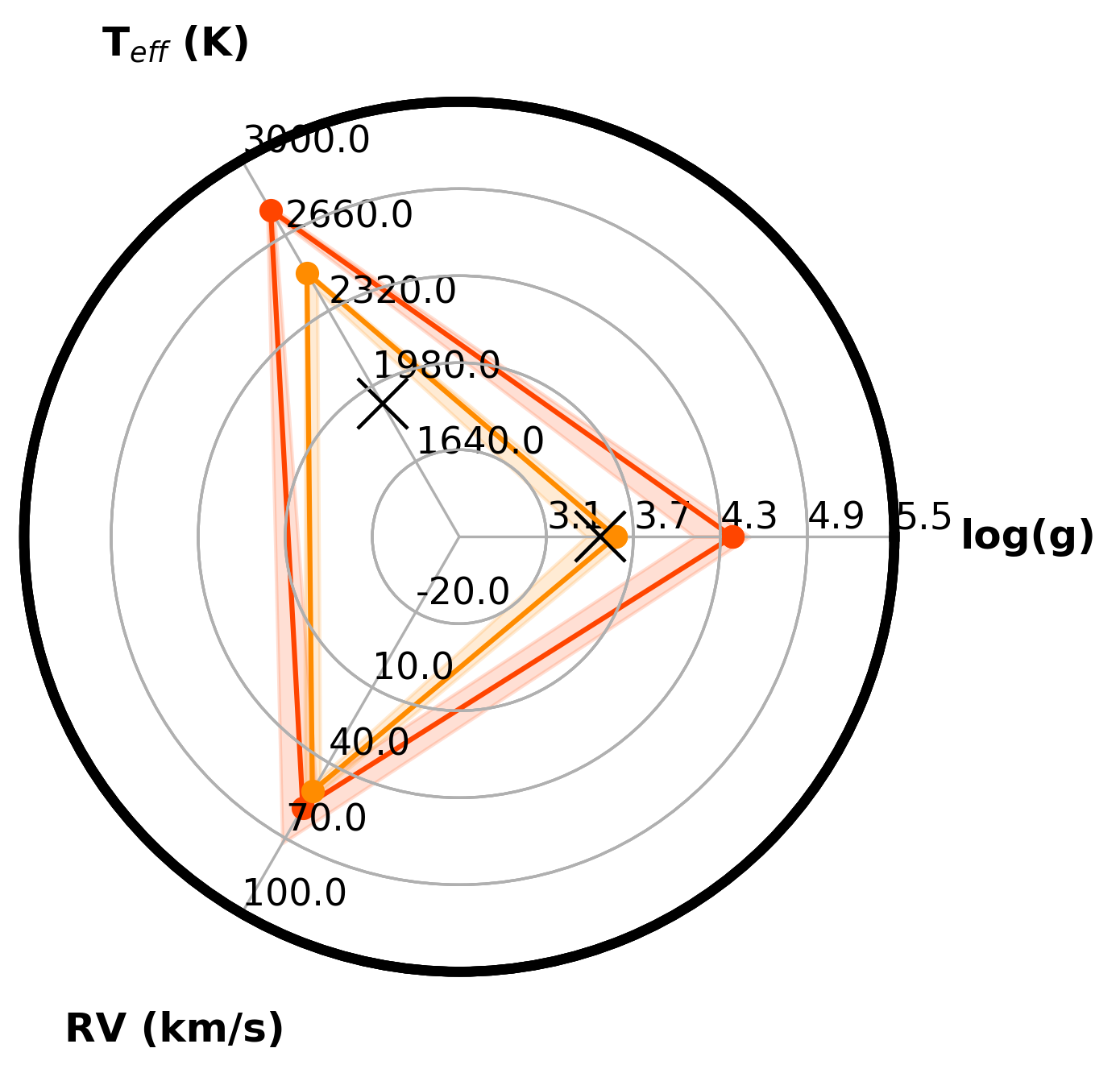}
\end{subfigure}
\caption{Same as Figure \ref{fig:target_CD-35_2722_b} but for CAHA Tau 1}
\label{fig:target_CAHA_Tau_1}
\end{figure*}

\begin{table*}[ht]
\centering
\caption{Same as Table \ref{tab:target_CD-35_2722_b} but for CAHA Tau 1}
\renewcommand{\arraystretch}{1.5} 
\resizebox{\textwidth}{!}{%
\begin{tabular}{lcccccccccccc}
\hline
model& \Teff (K) & \logg (dex) & \met & \co & $\gamma$ & \fsed & RV (km/s) & $\beta$ (km/s) & ln(z) & $\chi ^2 _{red}$ \\ 
\hline
\btex&$2774^{+28}_{-29}$&$4.38^{+0.12}_{-0.26}$&&&&&$57.9^{+14.12}_{-9.07}$&$86.48^{+30.28}_{-26.07}$&-18391.3&18\\ 
\atmo&$2490^{+13}_{-79}$&$3.58^{+0.09}_{-0.2}$&$0.07^{+0.05}_{-0.35}$&$0.7^{+0.0}_{-0.03}$&$1.05^{+0.0}_{-0.01}$&&$51.32^{+5.28}_{-5.5}$&$65^{+13}_{-14}$&-18061.8&18\\  
\hline
\end{tabular}}
\label{tab:target_CAHA_Tau_1}
\end{table*}

CD‑35 2722 b was discovered by \citet{Wahhaj2011} as part of the Gemini NICI Planet-Finding Campaign via direct imaging at a projected separation of $\sim$67 au, and it was confirmed as co-moving over two epochs. Near-IR spectroscopy determined it to be an L4 dwarf with moderately low gravity, with atmospheric model fits yielding a \Teff of 1700–1900 K and mass estimates around 31 M$_\mathrm{Jup}$ based on bolometric luminosity and evolutionary models. Furthermore, CD‑35 2722 b appears over-luminous in the near-IR bands compared to field dwarfs, consistent with its youth. Studies of substellar companion populations (e.g., \citet{Bowler2020}, \citet{Bonavita2016}) place CD‑35 2722 b as a rare, well-characterized young L-type companion, useful for constraining formation models and testing atmospheric evolution. The corresponding Figure \ref{fig:target_CD-35_2722_b} and Table \ref{tab:target_CD-35_2722_b} are located in the main text, as we also use this target to benchmark our methods.

DH Tau b is a companion located at a wide separation ($\sim$330 au) from the T Tauri star DH Tauri. Discovered via direct near-infrared imaging \citep{Itoh2005}, its companionship was confirmed through common proper motion and spectroscopy, which revealed CO and H$_2$O signatures and an effective temperature in the 2500 K range \citep{Itoh2005, Patience2012, Xuan2024}, notably higher than the values found in this work. High-resolution spectroscopy using Keck/NIRSPEC measured a projected rotational velocity of \vsini $= 9.6\pm0.7$ km/s, which is about 10–15\% of the breakup speed, and is consistent with late-stage disk-regulated angular momentum transfer \citep{Xuan2020}. Polarimetric observations with VLT/SPHERE-IRDIS revealed significant H-band linear polarization, attributed to a highly inclined circumplanetary disk misaligned with the circumstellar disk, providing the first direct detection of a circumplanetary disk around a directly imaged companion \citep{vanHolstein2021, Martinez2022}. Additional surveys using SPHERE and Spitzer/IRAC probed for further companions or disks, setting constraints on disk mass and geometry \citep{Wolff2017, Lazzoni2020, Martinez2022}. However, even though there is a circumplanetary disk, we do not see any spectral features suggesting ongoing accretion. These findings collectively suggest that DH Tau b formed early in the system's evolution, likely through disk fragmentation, retains a substantial circumplanetary disk, and exhibits slow rotation, perhaps due to magnetic coupling.

\begin{figure*}[ht]
\centering
\begin{subfigure}[b]{0.61\textwidth}
    \includegraphics[width=\textwidth]{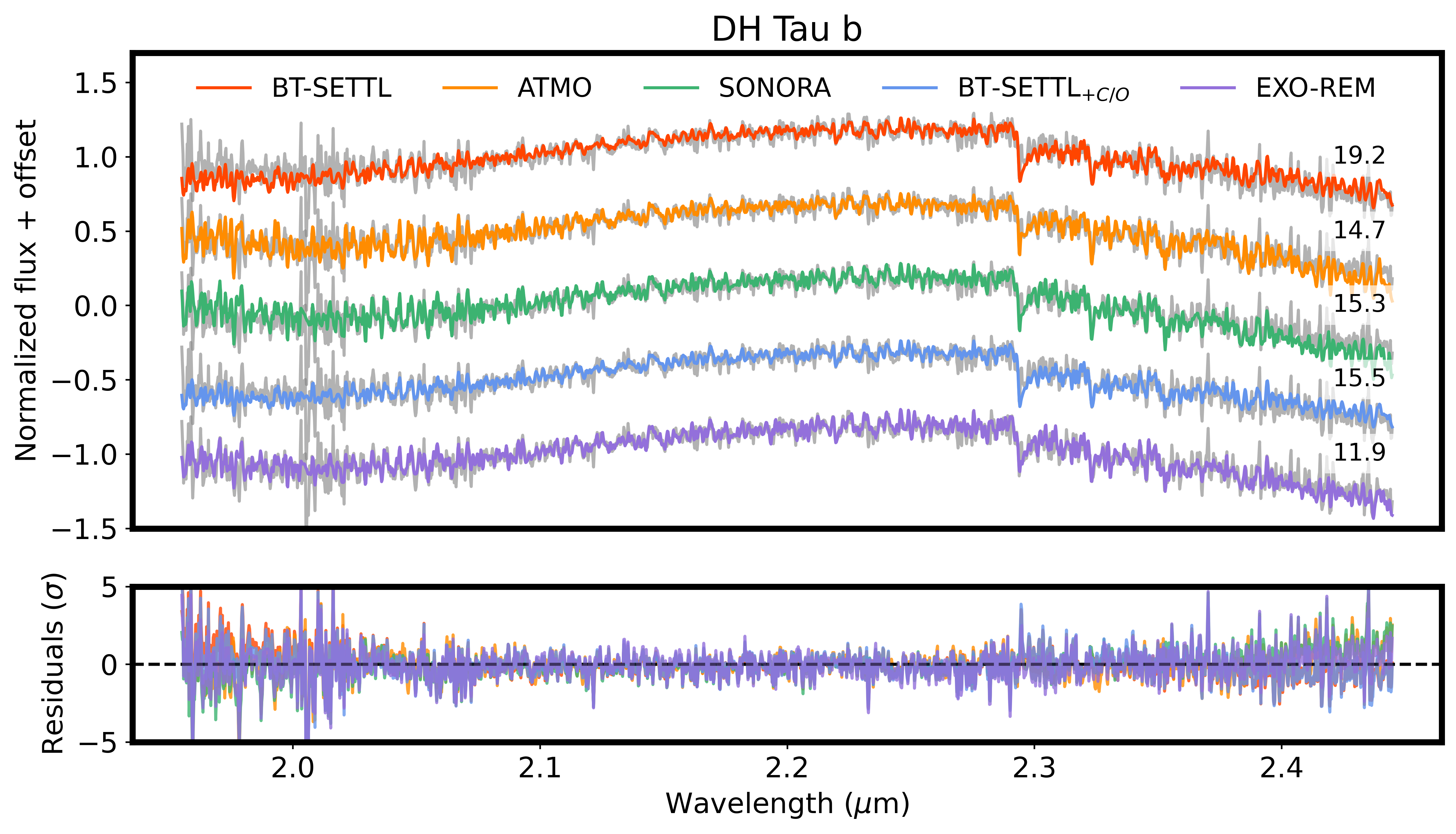}
\end{subfigure}
\hfill
\begin{subfigure}[b]{0.33\textwidth}
    \includegraphics[width=\textwidth]{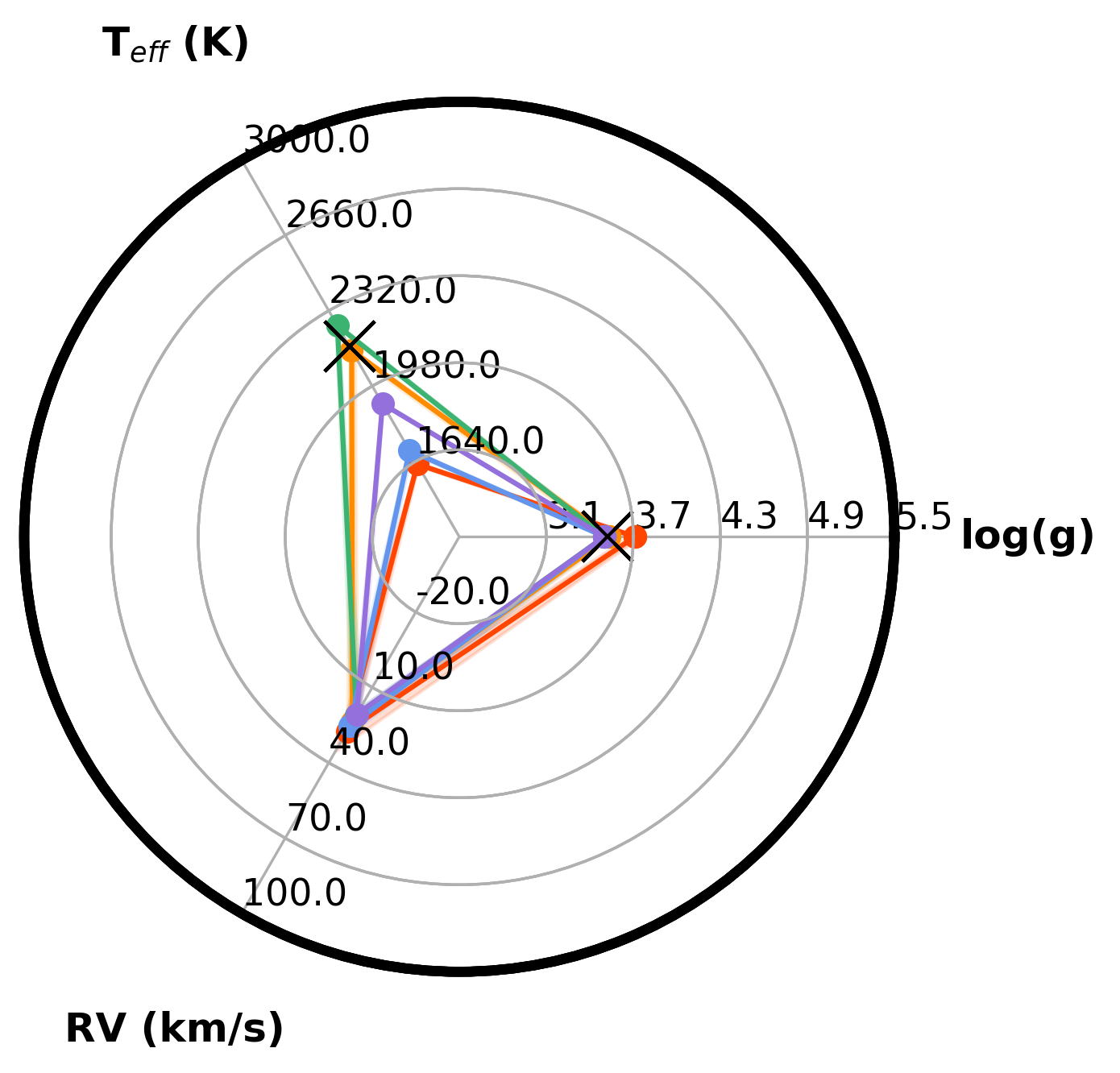}
\end{subfigure}
\caption{Same as Figure \ref{fig:target_CD-35_2722_b} but for DH Tau b}
\label{fig:target_DH_Tau_b}
\end{figure*}

\begin{table*}[ht]
\centering
\caption{Same as Table \ref{tab:target_CD-35_2722_b} but for DH Tau b}
\renewcommand{\arraystretch}{1.5} 
\resizebox{\textwidth}{!}{%
\begin{tabular}{lcccccccccccc}
\hline
model& \Teff (K) & \logg (dex) & \met & \co & $\gamma$ & \fsed & RV (km/s) & $\beta$ (km/s) & ln(z) & $\chi ^2 _{red}$ \\ 
\hline
\btex&$1627^{+6}_{-6}$&$3.71^{+0.04}_{-0.04}$&&&&&$27.37^{+7.95}_{-6.51}$&$63.28^{+14.41}_{-17.05}$&-19217.0&19\\ 
\atmo&$2141^{+12}_{-26}$&$3.53^{+0.03}_{-0.04}$&$0.59^{+0.01}_{-0.12}$&$0.7^{+0.0}_{-0.09}$&$1.04^{+0.0}_{-0.01}$&&$23.68^{+3.51}_{-3.27}$&$61^{+10}_{-8}$&-14666.3&15\\ 
\sono&$2252^{+14}_{-11}$&$3.5^{+0.02}_{-0.0}$&&&&$4.16^{+3.71}_{-0.17}$&$20.97^{+3.07}_{-2.85}$&$13.34^{+18.49}_{-12.93}$&-15269.0&15\\ 
\btse&$1689^{+8}_{-11}$&$3.5^{+0.04}_{-0.0}$&&$0.58^{+0.02}_{-0.02}$&&&$25.34^{+5.51}_{-5.45}$&$50.77^{+16.13}_{-18.47}$&-15472.0&15\\ 
\exor&$1900^{+2}_{-2}$&$3.5^{+0.02}_{-0.01}$&$-0.11^{+0.04}_{-0.02}$&$0.4^{+0.0}_{-0.0}$&&&$20.74^{+2.96}_{-2.72}$&$52.9^{+8.76}_{-8.91}$&-11951.5&12\\ 
\hline
\end{tabular}}
\label{tab:target_DH_Tau_b}
\end{table*}

FU Tau B was initially discovered as a co-moving companion via near-infrared imaging \citep{Luhman2009}. Recent monitoring of its H$_{\alpha}$ emission over six consecutive nights \citep{Wu2023} revealed mild variability but no burst-like accretion events, indicating ongoing, stable mass accretion. ALMA Band-7 (0.88 mm) continuum observations, however, did not detect any disk emission above $\sim$120–210 $\mu$Jy, suggesting either a low-mass circumplanetary disk or rapid radial dust drift \citep{Wu2020}. In the SINFONI K-band observations, where we do not observe evidence of ongoing accretion, however, the dataset is relatively noisy, and some residuals of sky emission lines might still be present. Altogether, FU Tau B might represent a rare example of an isolated, wide-separation, young substellar companion, useful for probing early accretion behavior and potentially the rapid dissipation of circumplanetary disks around low-mass companions.

\begin{figure*}[ht]
\centering
\begin{subfigure}[b]{0.61\textwidth}
    \includegraphics[width=\textwidth]{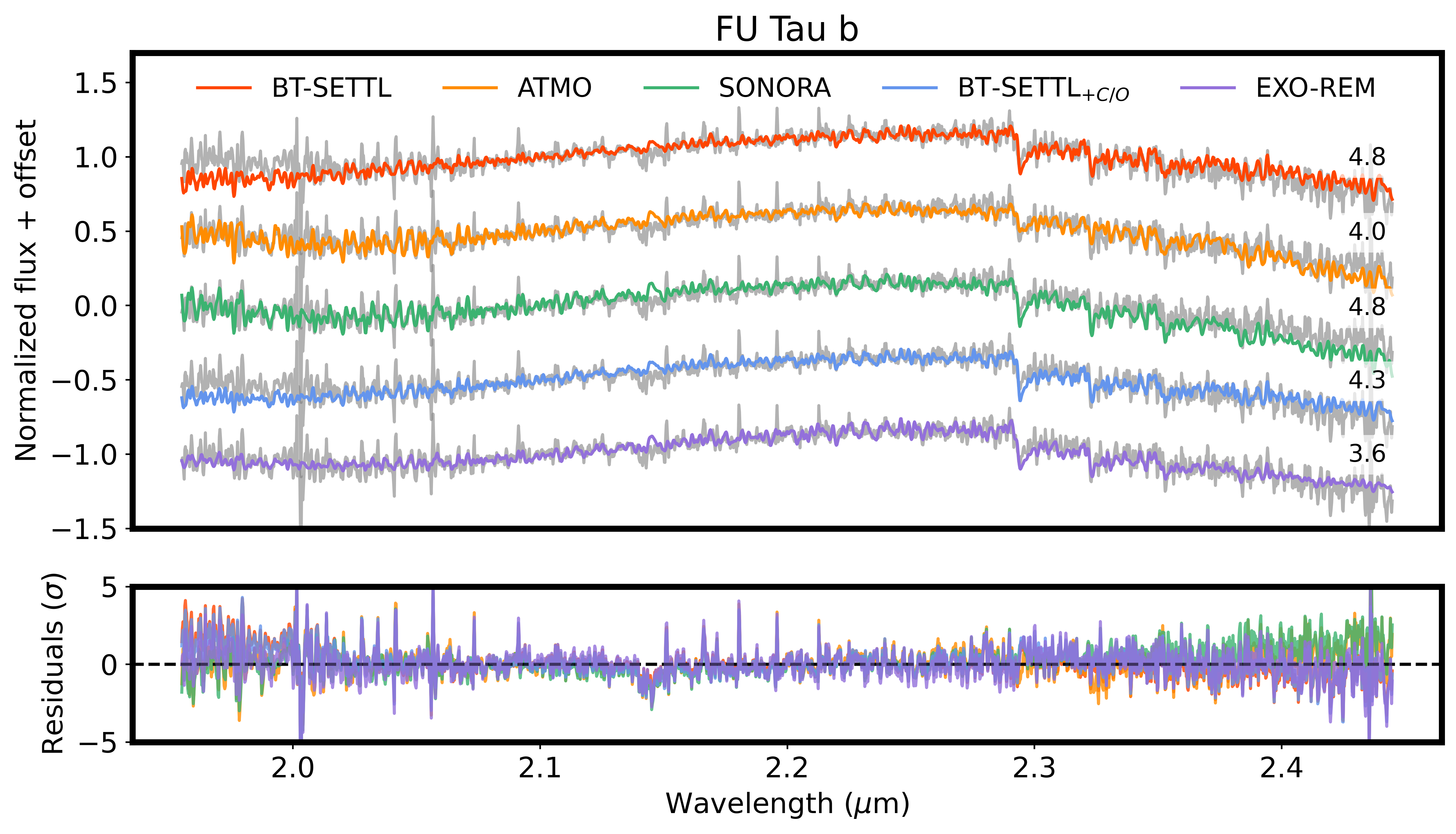}
\end{subfigure}
\hfill
\begin{subfigure}[b]{0.33\textwidth}
    \includegraphics[width=\textwidth]{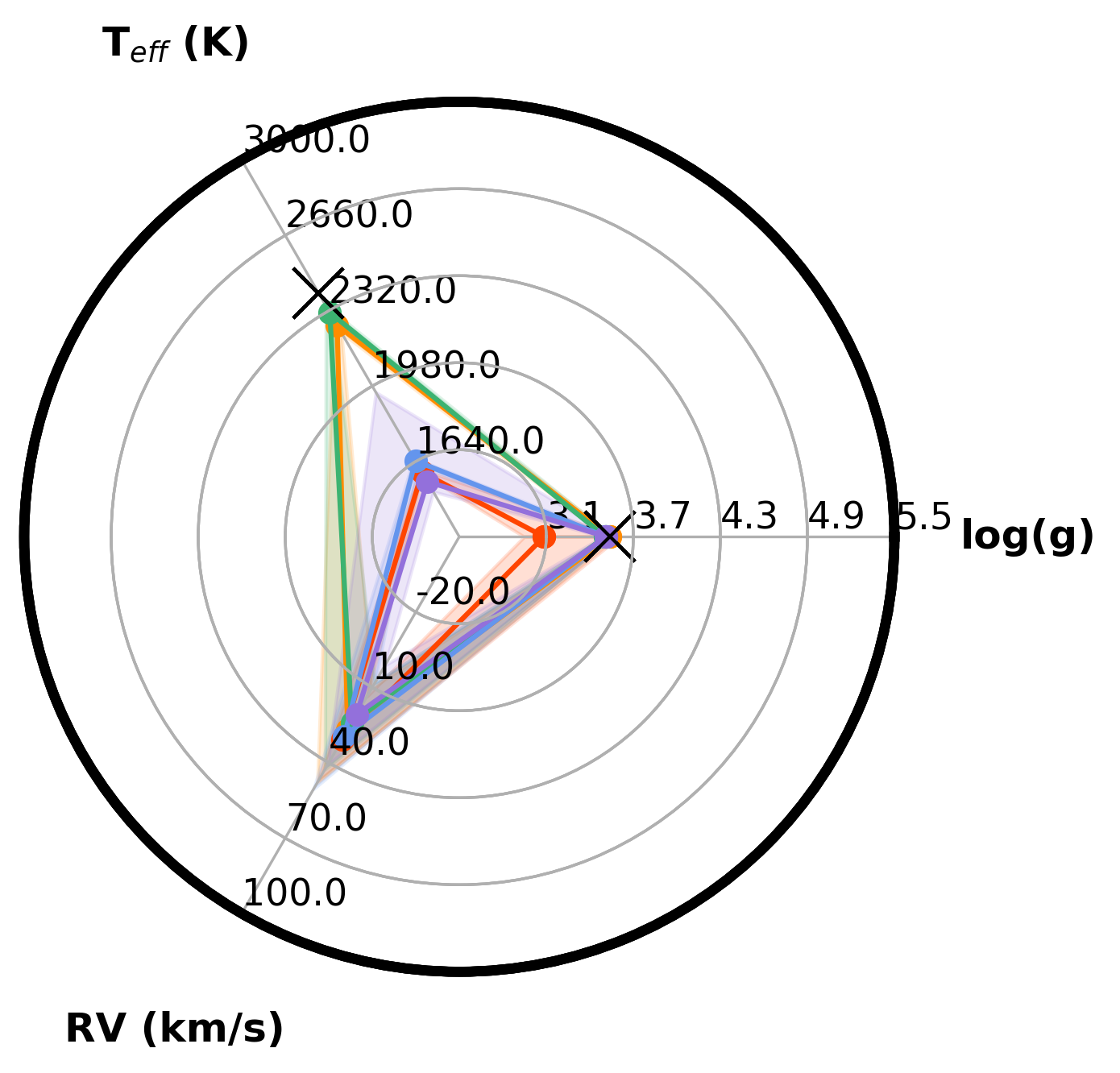}
\end{subfigure}
\caption{Same as Figure \ref{fig:target_CD-35_2722_b} but for FU Tau b}
\label{fig:target_FU_Tau_b}
\end{figure*}

\begin{table*}[ht]
\centering
\caption{Same as Table \ref{tab:target_CD-35_2722_b} but for FU Tau b}
\renewcommand{\arraystretch}{1.5} 
\resizebox{\textwidth}{!}{%
\begin{tabular}{lcccccccccccc}
\hline
model& \Teff (K) & \logg (dex) & \met & \co & $\gamma$ & \fsed & RV (km/s) & $\beta$ (km/s) & ln(z) & $\chi ^2 _{red}$ \\ 
\hline
\btex&$1591^{+14}_{-10}$&$3.08^{+0.52}_{-0.14}$&&&&&$30.8^{+16.35}_{-17.29}$&$83.61^{+31.87}_{-38.97}$&-4849.9&5\\ 
\atmo&$2254^{+24}_{-33}$&$3.54^{+0.03}_{-0.03}$&$0.59^{+0.01}_{-0.2}$&$0.67^{+0.03}_{-0.36}$&$1.01^{+0.01}_{-0.0}$&&$26.98^{+20.87}_{-16.87}$&$104^{+51}_{-29}$&-4048.9&4\\ 
\sono&$2312^{+31}_{-23}$&$3.51^{+0.05}_{-0.01}$&&&&$4.41^{+3.55}_{-0.49}$&$24.12^{+18.46}_{-16.02}$&$87.84^{+40.73}_{-33.59}$&-4799.4&5\\ 
\btse&$1643^{+13}_{-20}$&$3.51^{+0.05}_{-0.01}$&&$0.58^{+0.03}_{-0.05}$&&&$29.3^{+21.13}_{-21.38}$&$77.45^{+52.8}_{-44.9}$&-4281.9&4\\ 
\exor&$1549^{+403}_{-46}$&$3.51^{+0.05}_{-0.02}$&$0.61^{+0.13}_{-0.09}$&$0.65^{+0.01}_{-0.35}$&&&$20.72^{+20.08}_{-19.54}$&$96.28^{+49.45}_{-24.89}$&-3603.7&4\\ 
\hline
\end{tabular}}
\label{tab:target_FU_Tau_b}
\end{table*}

HIP 78530 b was first identified in the Upper Scorpius association by \citet{Lafreniere2011}. The early observations confirmed common proper motion and obtained near-IR (JHK) spectroscopy, revealing a low surface gravity and a \Teff of $\sim$2800 K, consistent with an M8 spectral type \citep{Lafreniere2011}. Positioned at a projected separation of $\sim$710 au, HIP 78530 b ranks among the lowest mass ratio wide-orbit companions known ($\sim$0.009), with an estimated orbital period of $\sim$12000 years \citep{Lafreniere2011}. Subsequent thermal infrared imaging (3–5 $\mu$m) and near-IR spectroscopy characterized its atmospheric properties further (e.g., \citet{Petrus2020s}), yielding a consistent \Teff and a mass between 21–25 M$_{Jup}$. In these studies, no IR excess from circumplanetary material was found.

\begin{figure*}[ht]
\centering
\begin{subfigure}[b]{0.61\textwidth}
    \includegraphics[width=\textwidth]{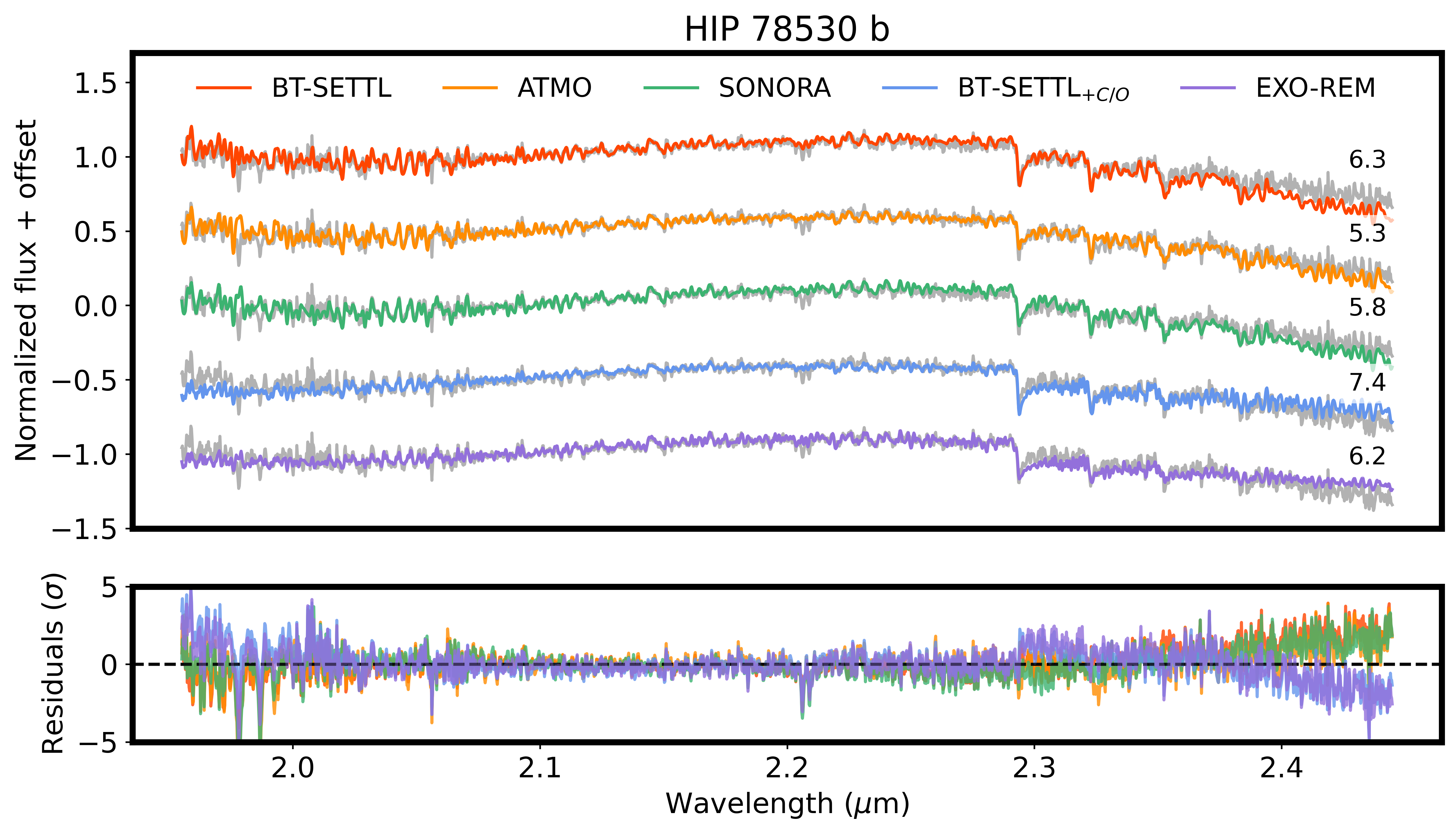}
\end{subfigure}
\hfill
\begin{subfigure}[b]{0.33\textwidth}
    \includegraphics[width=\textwidth]{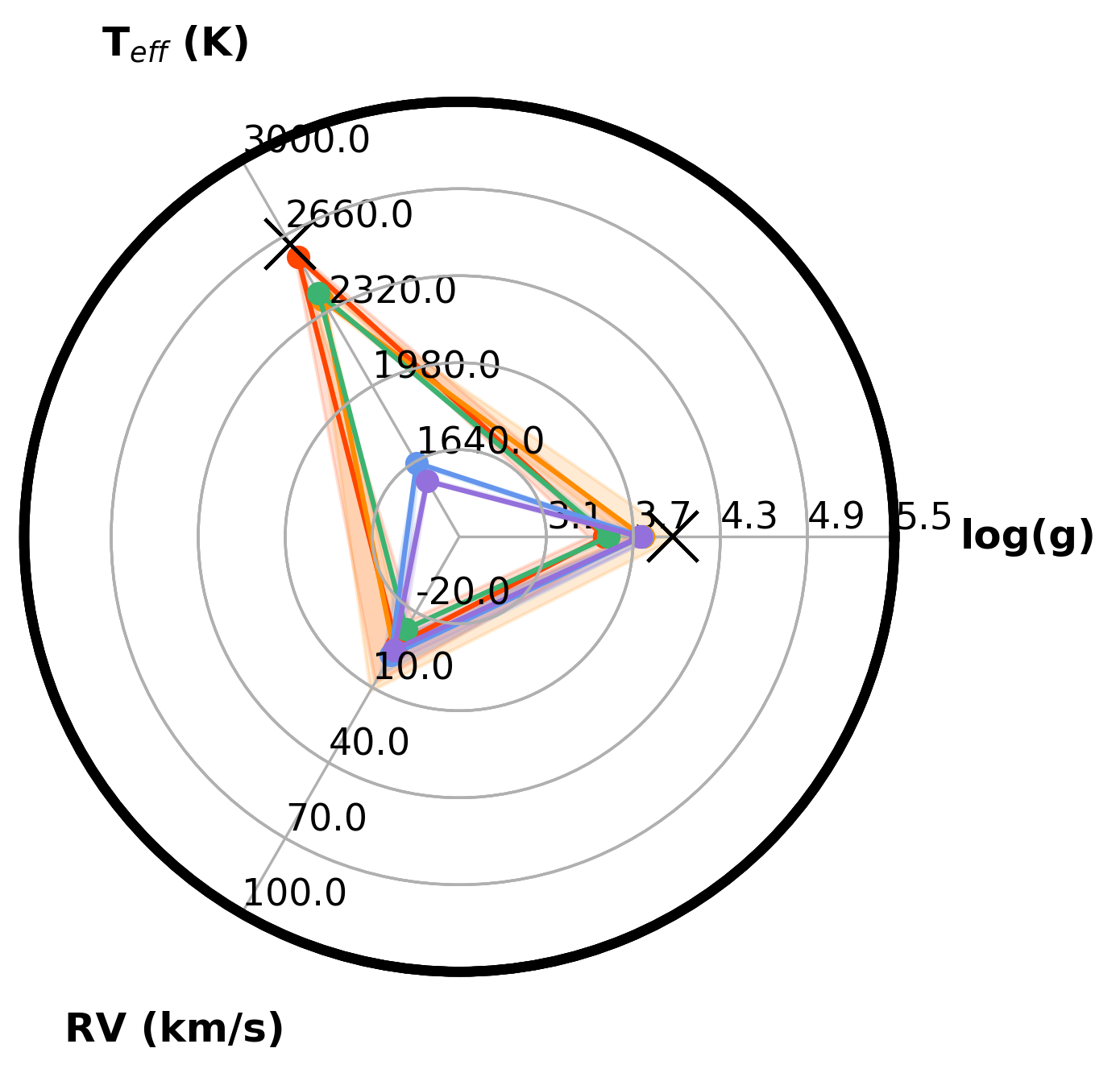}
\end{subfigure}
\caption{Same as Figure \ref{fig:target_CD-35_2722_b} but for HIP 78530 b}
\label{fig:target_HIP78530_b}
\end{figure*}

\begin{table*}[ht]
\centering
\caption{Same as Table \ref{tab:target_CD-35_2722_b} but for HIP 78530 b}
\renewcommand{\arraystretch}{1.5} 
\resizebox{\textwidth}{!}{%
\begin{tabular}{lcccccccccccc}
\hline
model& \Teff (K) & \logg (dex) & \met & \co & $\gamma$ & \fsed & RV (km/s) & $\beta$ (km/s) & ln(z) & $\chi ^2 _{red}$ \\ 
\hline
\btex&$2561^{+33}_{-28}$&$3.5^{+0.21}_{-0.11}$&&&&&$-6.82^{+14.8}_{-10.6}$&$101.97^{+25.61}_{-21.79}$&-6282.3&6\\ 
\atmo&$2368^{+65}_{-19}$&$3.76^{+0.24}_{-0.11}$&$0.59^{+0.01}_{-0.69}$&$0.32^{+0.37}_{-0.02}$&$1.04^{+0.0}_{-0.03}$&&$-4.92^{+16.32}_{-5.25}$&$104^{+20}_{-10}$&-5368.6&5\\ 
\sono&$2400^{+0}_{-0}$&$3.53^{+0.04}_{-0.03}$&&&&$5.75^{+0.0}_{-0.0}$&$-13.28^{+0.11}_{-0.14}$&$88.04^{+0.02}_{-0.01}$&-5944.5&6\\ 
\btse&$1633^{+4}_{-12}$&$3.76^{+0.08}_{-0.04}$&&$0.72^{+0.01}_{-0.03}$&&&$-2.79^{+7.47}_{-7.21}$&$58.14^{+16.66}_{-18.99}$&-7440.0&7\\ 
\exor&$1551^{+11}_{-3}$&$3.75^{+0.1}_{-0.06}$&$0.33^{+0.07}_{-0.04}$&$0.8^{+0.0}_{-0.01}$&&&$-5.01^{+6.52}_{-6.34}$&$62.84^{+12.7}_{-14.02}$&-6268.1&6\\ 
\hline
\end{tabular}}
\label{tab:target_HIP78530_b}
\end{table*}

\clearpage

HR 7329 b, also known as $\eta$ Tel b and HD 181327 b, is a directly imaged substellar companion orbiting a young A-type star part of the $\beta$ Pictoris moving group. It was first detected by \citet{Lowrance2000} and confirmed as a co-moving companion at a projected separation of $\sim$192 au \citep{Neuhauser2011}. HR 7329 A hosts a debris disk detected in mid-infrared, detected by \citet{Lebreton2012} and recently observed with JWST/MIRI by \citet{Chai2024}, where the orbit of planet b and the architecture of the system were studied.

\begin{figure*}[ht]
\centering
\begin{subfigure}[b]{0.61\textwidth}
    \includegraphics[width=\textwidth]{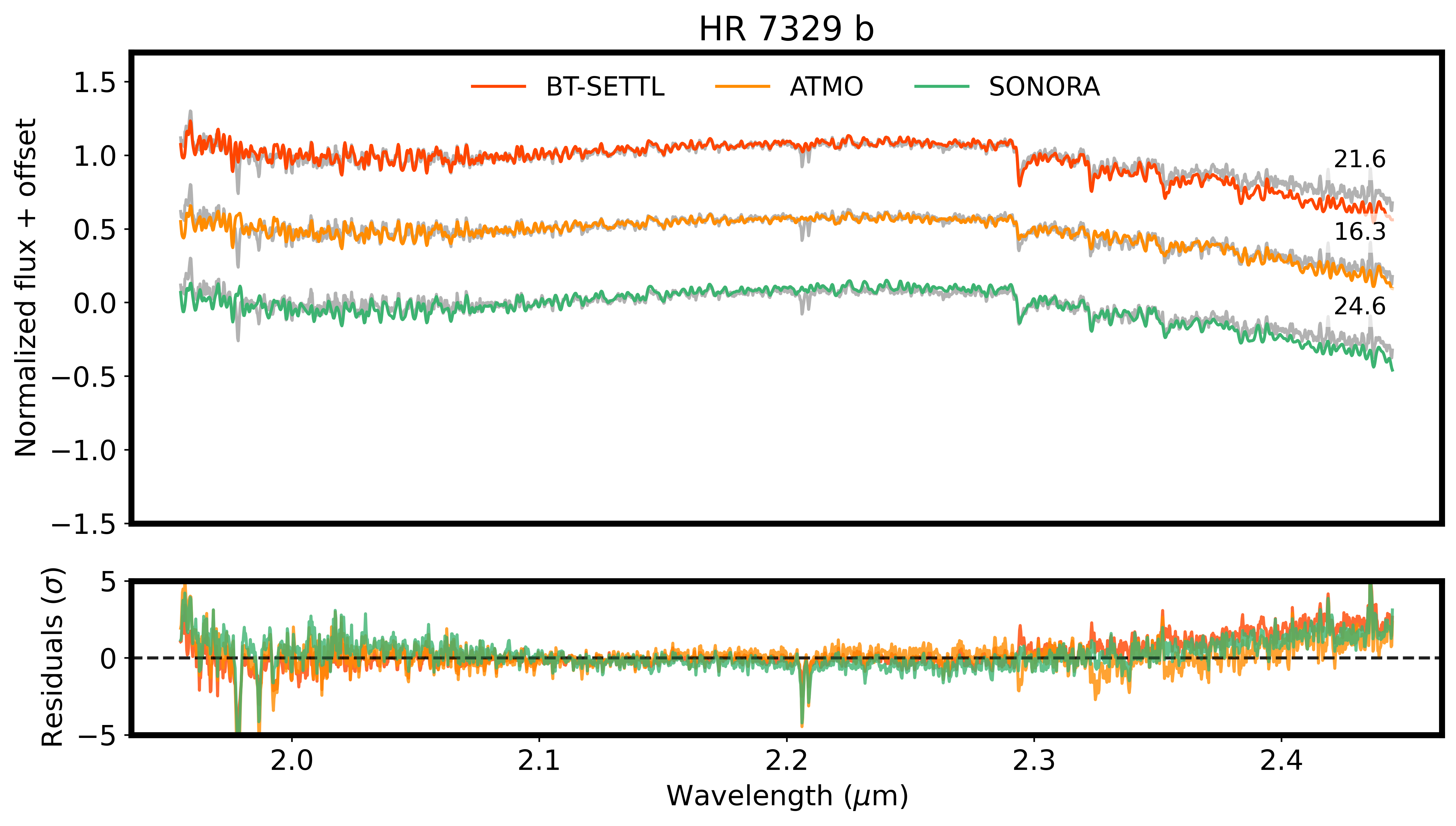}
\end{subfigure}
\hfill
\begin{subfigure}[b]{0.33\textwidth}
    \includegraphics[width=\textwidth]{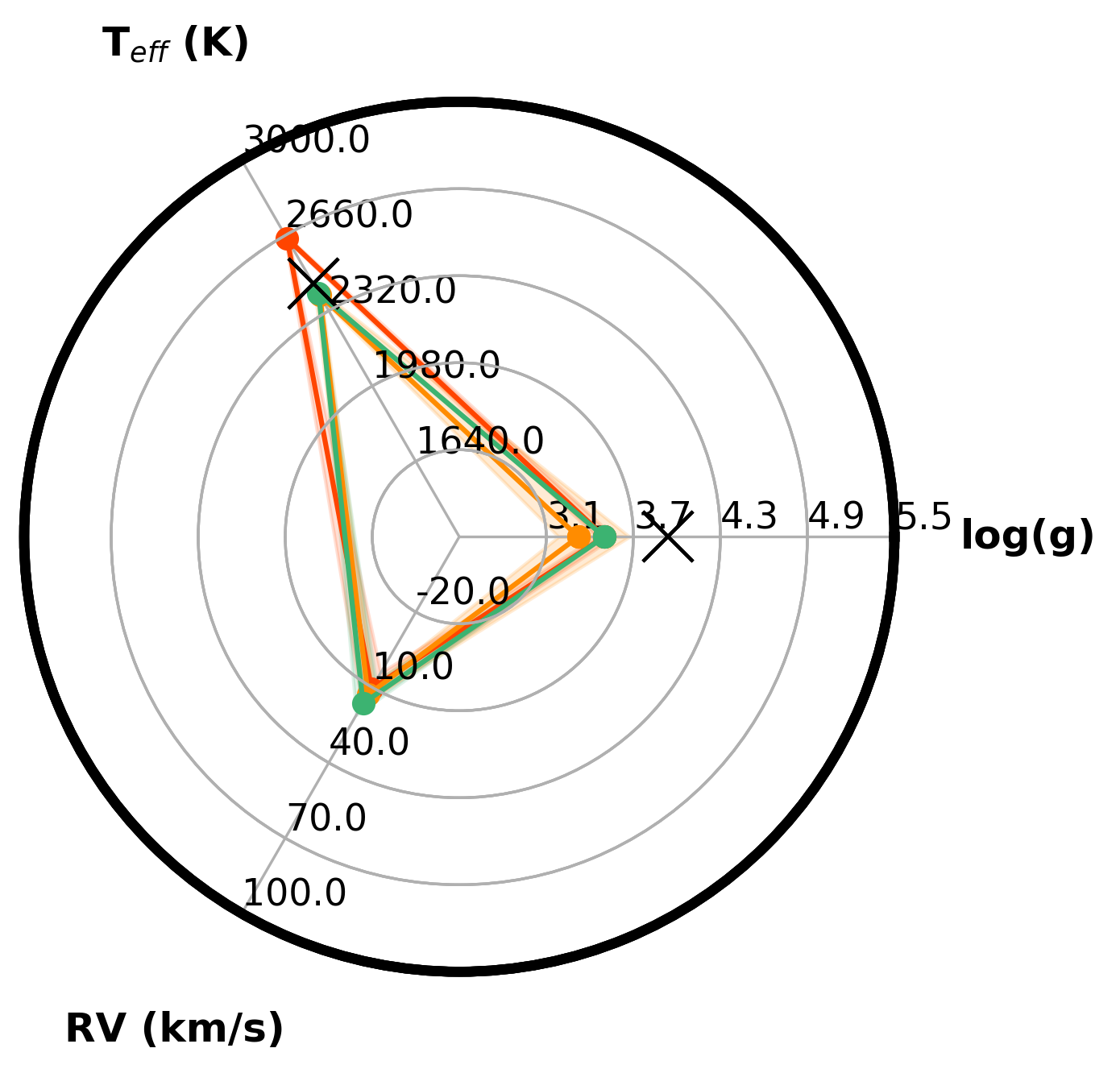}
\end{subfigure}
\caption{Same as Figure \ref{fig:target_CD-35_2722_b} but for HR 7329 b}
\label{fig:target_HR_7329_b}
\end{figure*}

\begin{table*}[ht]
\centering
\caption{Same as Table \ref{tab:target_CD-35_2722_b} but for HR 7329 b}
\renewcommand{\arraystretch}{1.5} 
\resizebox{\textwidth}{!}{%
\begin{tabular}{lcccccccccccc}
\hline
model& \Teff (K) & \logg (dex) & \met & \co & $\gamma$ & \fsed & RV (km/s) & $\beta$ (km/s) & ln(z) & $\chi ^2 _{red}$ \\ 
\hline
\btex&$2645^{+20}_{-17}$&$3.5^{+0.07}_{-0.05}$&&&&&$10.52^{+6.52}_{-6.32}$&$91.15^{+16.04}_{-14.13}$&-21598.8&22\\ 
\atmo&$2390^{+43}_{-15}$&$3.32^{+0.35}_{-0.14}$&$0.6^{+0.0}_{-0.02}$&$0.63^{+0.07}_{-0.33}$&$1.02^{+0.0}_{-0.01}$&&$12.92^{+3.58}_{-3.54}$&$99^{+8}_{-7}$&-16345.4&16\\ 
\sono&$2400^{+0}_{-1}$&$3.5^{+0.0}_{-0.0}$&&&&$5.85^{+0.35}_{-1.69}$&$16.37^{+4.65}_{-8.84}$&$105.58^{+12.59}_{-15.95}$&-24589.2&25\\ 
\hline
\end{tabular}}
\label{tab:target_HR_7329_b}
\end{table*}

\begin{figure*}[ht]
\centering
\begin{subfigure}[b]{0.61\textwidth}
    \includegraphics[width=\textwidth]{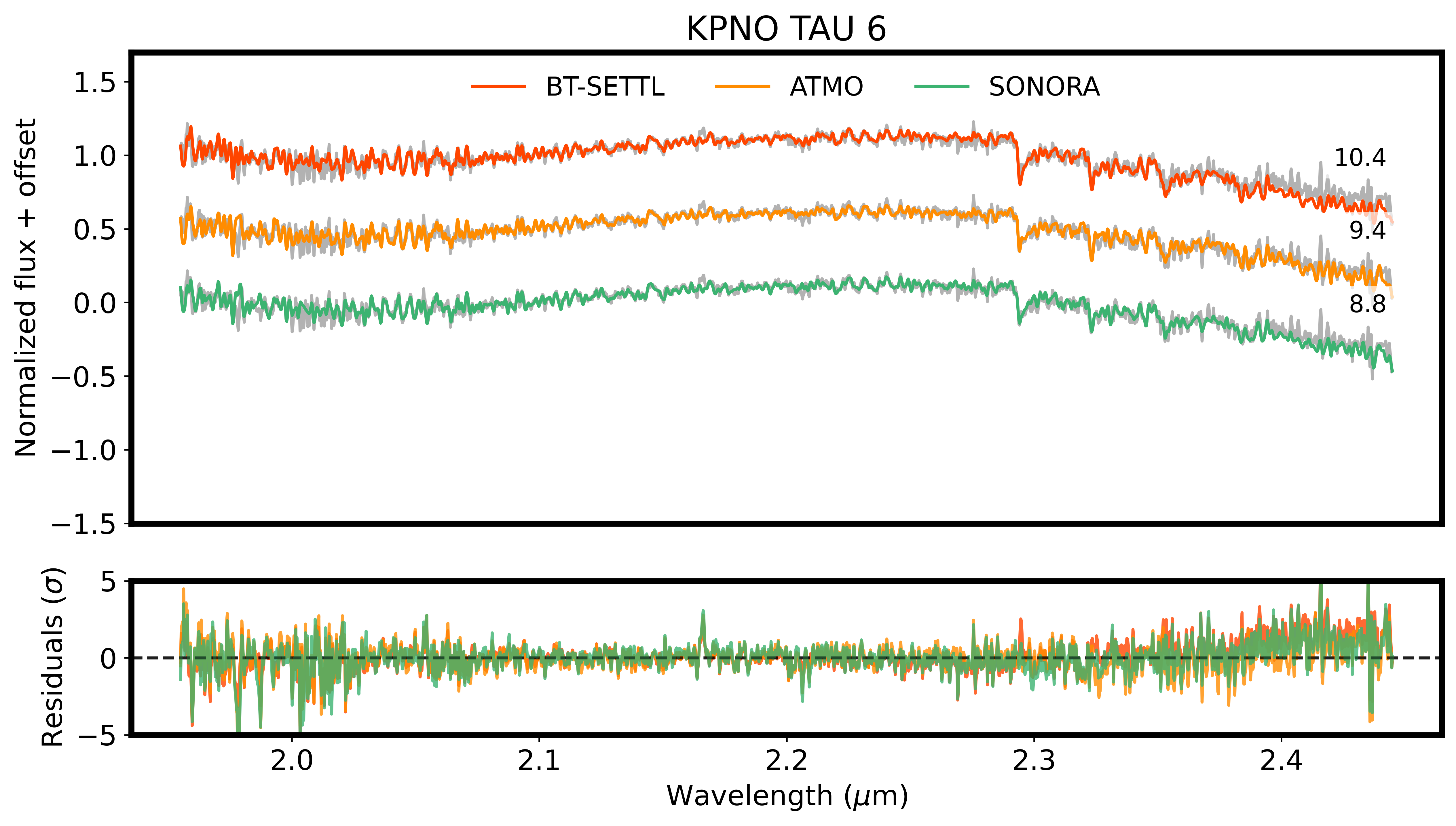}
\end{subfigure}
\hfill
\begin{subfigure}[b]{0.33\textwidth}
    \includegraphics[width=\textwidth]{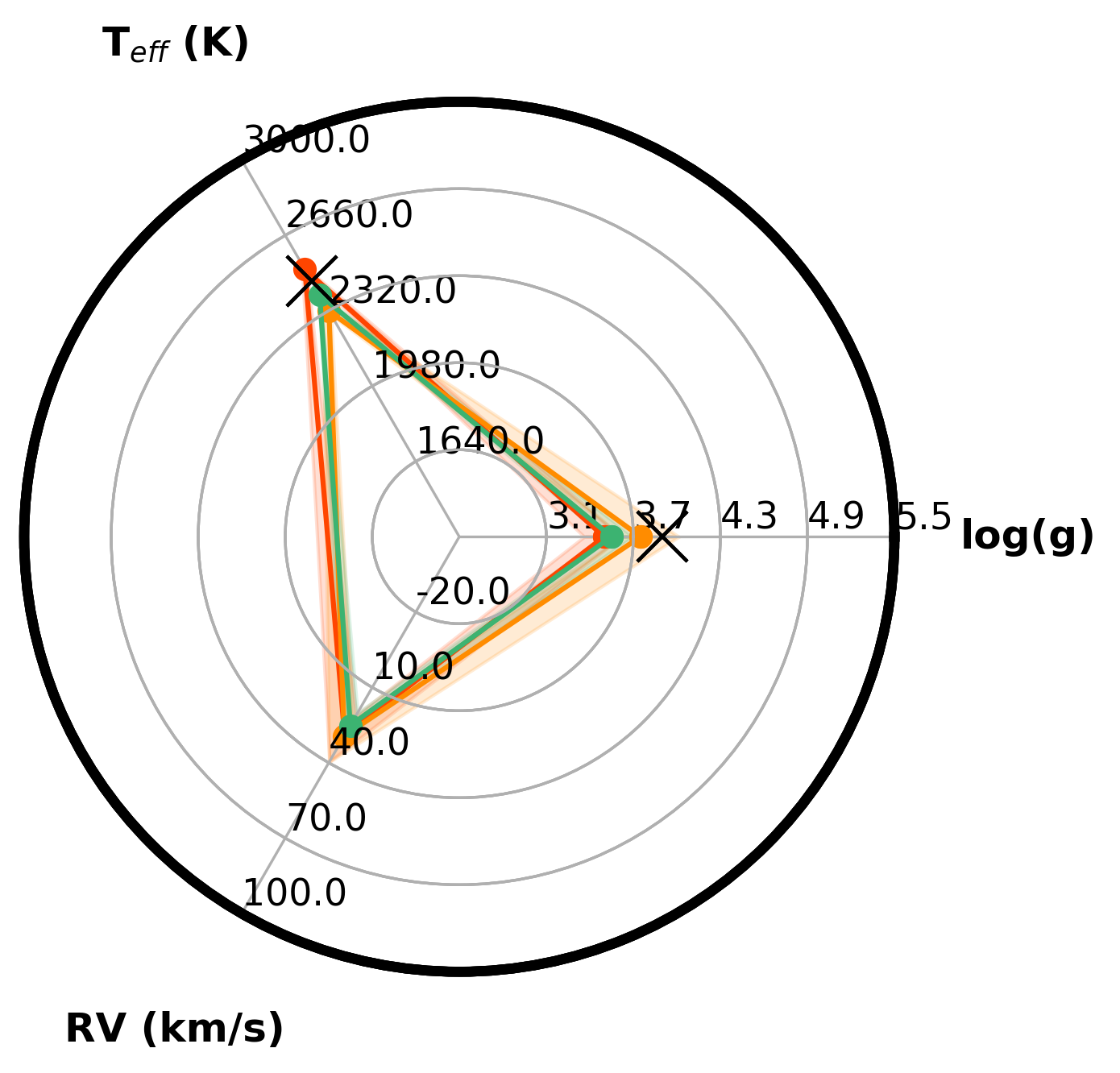}
\end{subfigure}
\caption{Same as Figure \ref{fig:target_CD-35_2722_b} but for KPNO Tau 6}
\label{fig:target_KPNO_Tau_6}
\end{figure*}

KPNO Tau 1, 4, and 6 are young brown brown dwarfs from the Taurus star-forming region and share a late-M spectral type; therefore, we group them. They were initially identified and spectroscopically classified by \citet{Briceno2002} and later characterized in more detail by \citet{Luhman2003}. Their near-infrared spectral types confirm their substellar status, and follow-up studies, such as those by \citet{Bonnefoy2014s} and \citet{Quanz2010}, have supported their classification and atmospheric properties within the broader context of young brown dwarfs. Regarding the comparison of evolutionary models with the derived properties of forward modeling, we observe that KPNO Tau 4 might have a lower \Teff than initially predicted. These objects typically show weak or absent Br$_{\gamma}$ emission, indicating low or episodic accretion activity, as expected for late-M objects at this evolutionary stage. Here we observe a potential accretion event through Br$_{\gamma}$ emission for KPNO Tau 6, as evident from the residuals in Figure \ref{fig:target_KPNO_Tau_6}. Variability, a common feature among such young substellar sources, has also been reported by \citet{Vos2020} for brown dwarfs at the Taurus star-forming region, likely arising from heterogeneous cloud coverage or magnetic phenomena.

\begin{figure*}[ht]
\centering
\begin{subfigure}[b]{0.61\textwidth}
    \includegraphics[width=\textwidth]{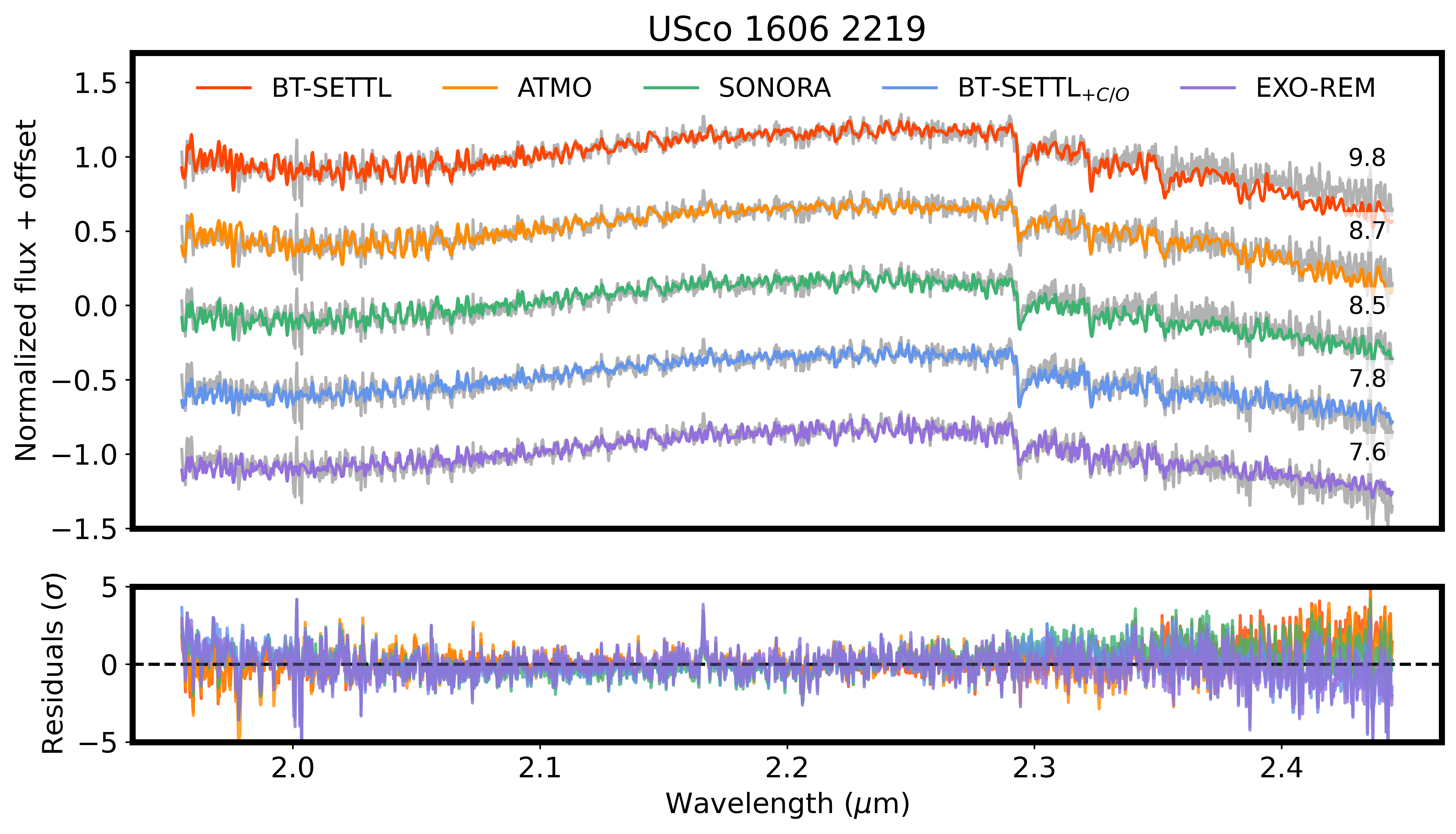}
\end{subfigure}
\hfill
\begin{subfigure}[b]{0.33\textwidth}
    \includegraphics[width=\textwidth]{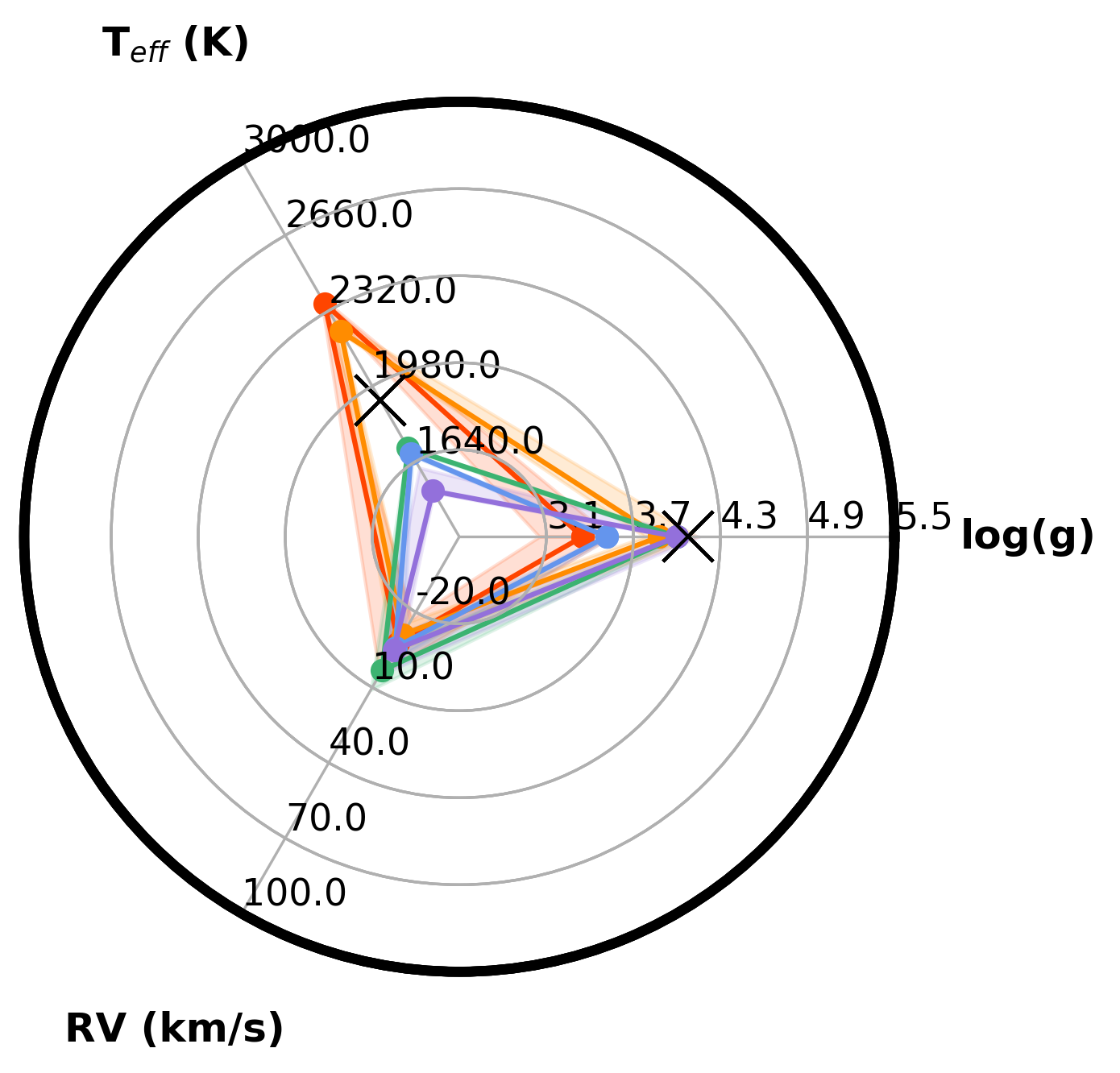}
\end{subfigure}
\caption{Same as Figure \ref{fig:target_CD-35_2722_b} but for USco 1606-2219}
\label{fig:target_USco1606-2219}
\end{figure*}

We also group the Upper Scorpius (USco) brown dwarfs from our sample, all of which are consistent with late-M to early-L type brown dwarfs. These targets are aged between 5 and 11 million years, and are located at approximately 145 parsecs \citep{Lodieu2011, Petrus2020s}. From the mentioned surveys, their spectral classifications are broadly consistent. 
From the SINFONI data, it is worth highlighting that USco1606-2219 shows signs of Br$_{\gamma}$ emission (see the residuals in Figure \ref{fig:target_USco1606-2219}).

Finally, USco CTIO 108 A is an interesting target, since this brown dwarf hosts a lower-mass wide orbit companion USco CTIO 108 b at a projected separation of about 670 au \citep{Bejar2008}. The system's wide separation and low binding energy suggest it is weakly bound, and dynamic interactions with other stars in the region could lead to its eventual disruption, offering valuable insights into the formation and evolution of substellar companions.



\begin{figure*}[ht]
\centering
\begin{subfigure}[b]{0.61\textwidth}
    \includegraphics[width=\textwidth]{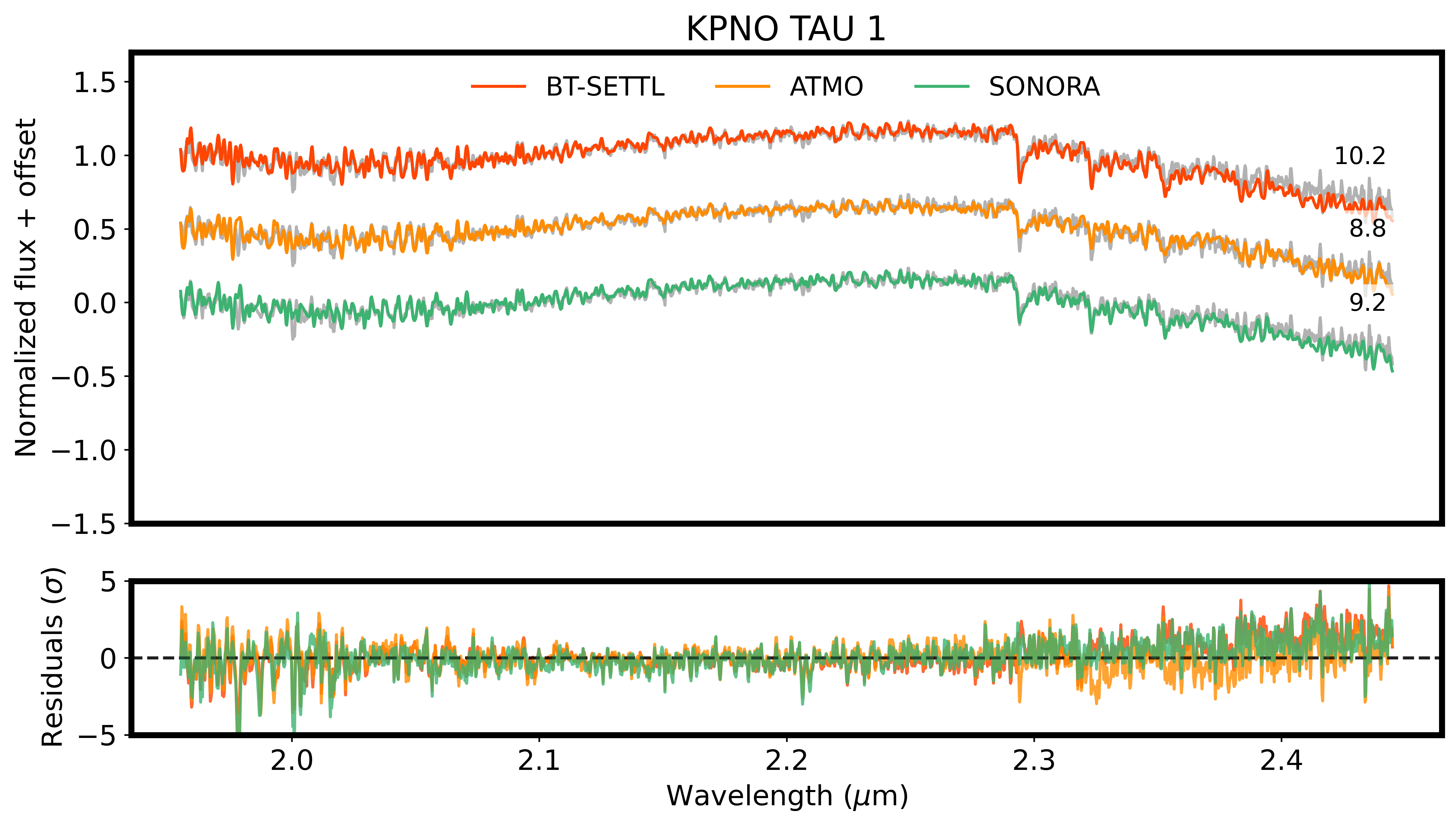}
\end{subfigure}
\hfill
\begin{subfigure}[b]{0.33\textwidth}
    \includegraphics[width=\textwidth]{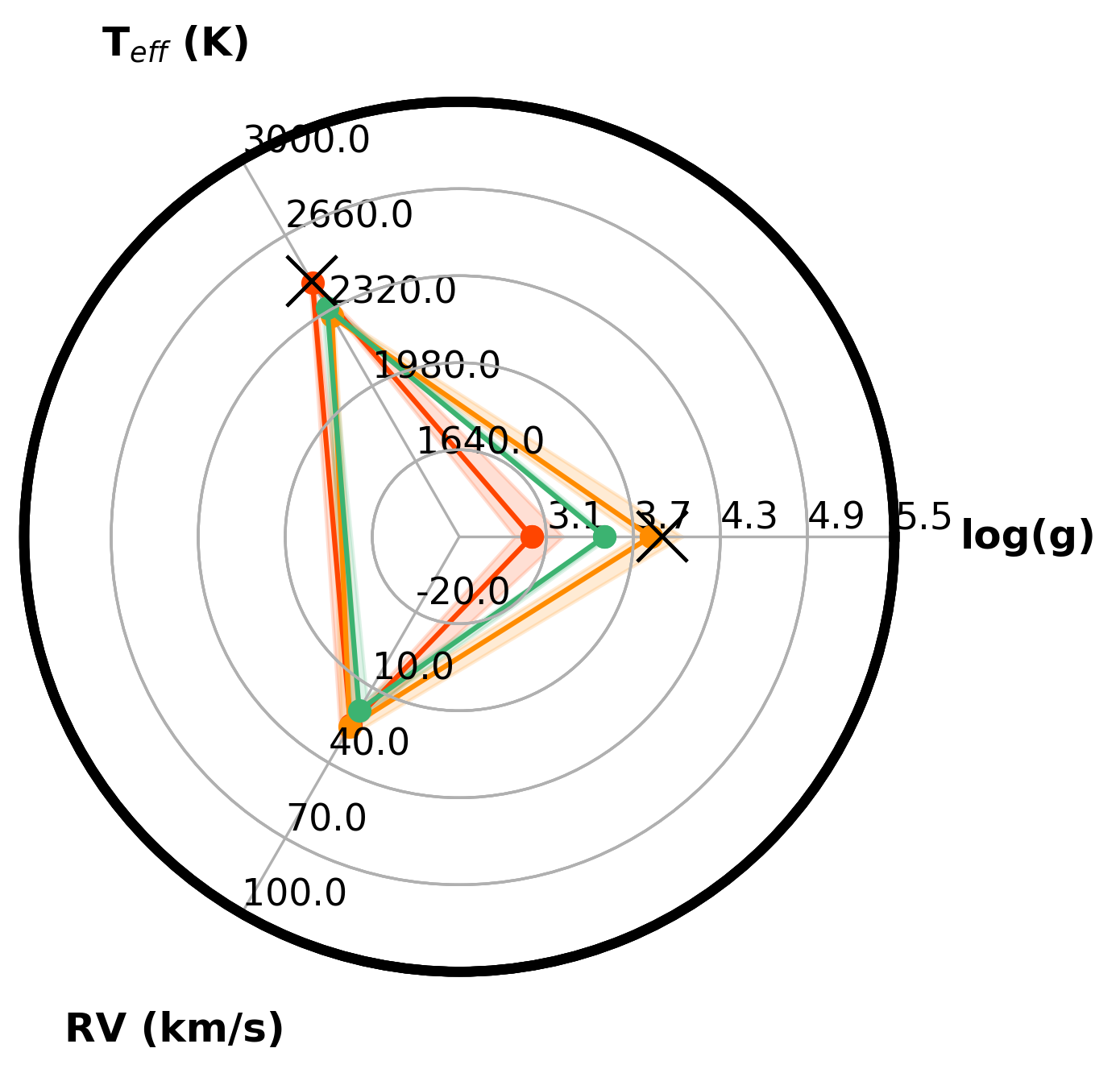}
\end{subfigure}
\caption{Same as Figure \ref{fig:target_CD-35_2722_b} but for KPNO Tau 1}
\label{fig:target_KPNO_Tau_1}
\end{figure*}

\begin{table*}[ht]
\centering
\caption{Same as Table \ref{tab:target_CD-35_2722_b} but for KPNO Tau 1}
\renewcommand{\arraystretch}{1.5} 
\resizebox{\textwidth}{!}{%
\begin{tabular}{lcccccccccccc}
\hline
model& \Teff (K) & \logg (dex) & \met & \co & $\gamma$ & \fsed & RV (km/s) & $\beta$ (km/s) & ln(z) & $\chi ^2 _{red}$ \\ 
\hline
\btex&$2447^{+21}_{-21}$&$3.0^{+0.21}_{-0.12}$&&&&&$25.06^{+6.61}_{-6.39}$&$89.47^{+15.15}_{-14.46}$&-10169.4&10\\ 
\atmo&$2295^{+32}_{-15}$&$3.82^{+0.22}_{-0.09}$&$0.6^{+0.0}_{-0.09}$&$0.69^{+0.01}_{-0.37}$&$1.02^{+0.0}_{-0.01}$&&$25.56^{+5.65}_{-5.84}$&$100^{+12}_{-10}$&-8833.7&9\\ 
\sono&$2332^{+44}_{-14}$&$3.5^{+0.04}_{-0.0}$&&&&$3.95^{+3.28}_{-0.22}$&$19.06^{+5.01}_{-5.22}$&$81.12^{+16.83}_{-13.45}$&-9272.6&9\\ 
\hline
\end{tabular}}
\label{tab:target_KPNO_Tau_1}
\end{table*}

\begin{figure*}[ht]
\centering
\begin{subfigure}[b]{0.61\textwidth}
    \includegraphics[width=\textwidth]{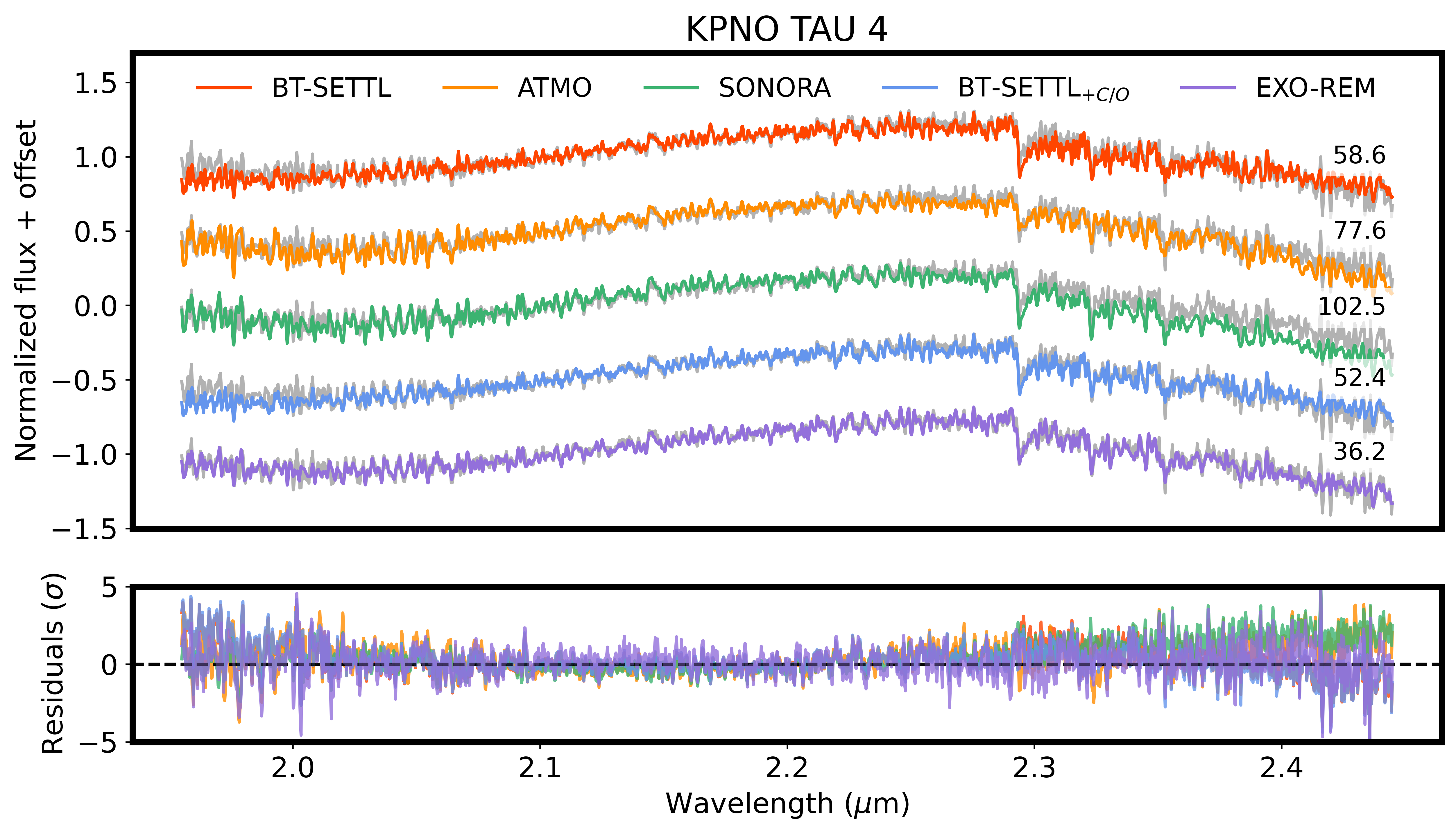}
\end{subfigure}
\hfill
\begin{subfigure}[b]{0.33\textwidth}
    \includegraphics[width=\textwidth]{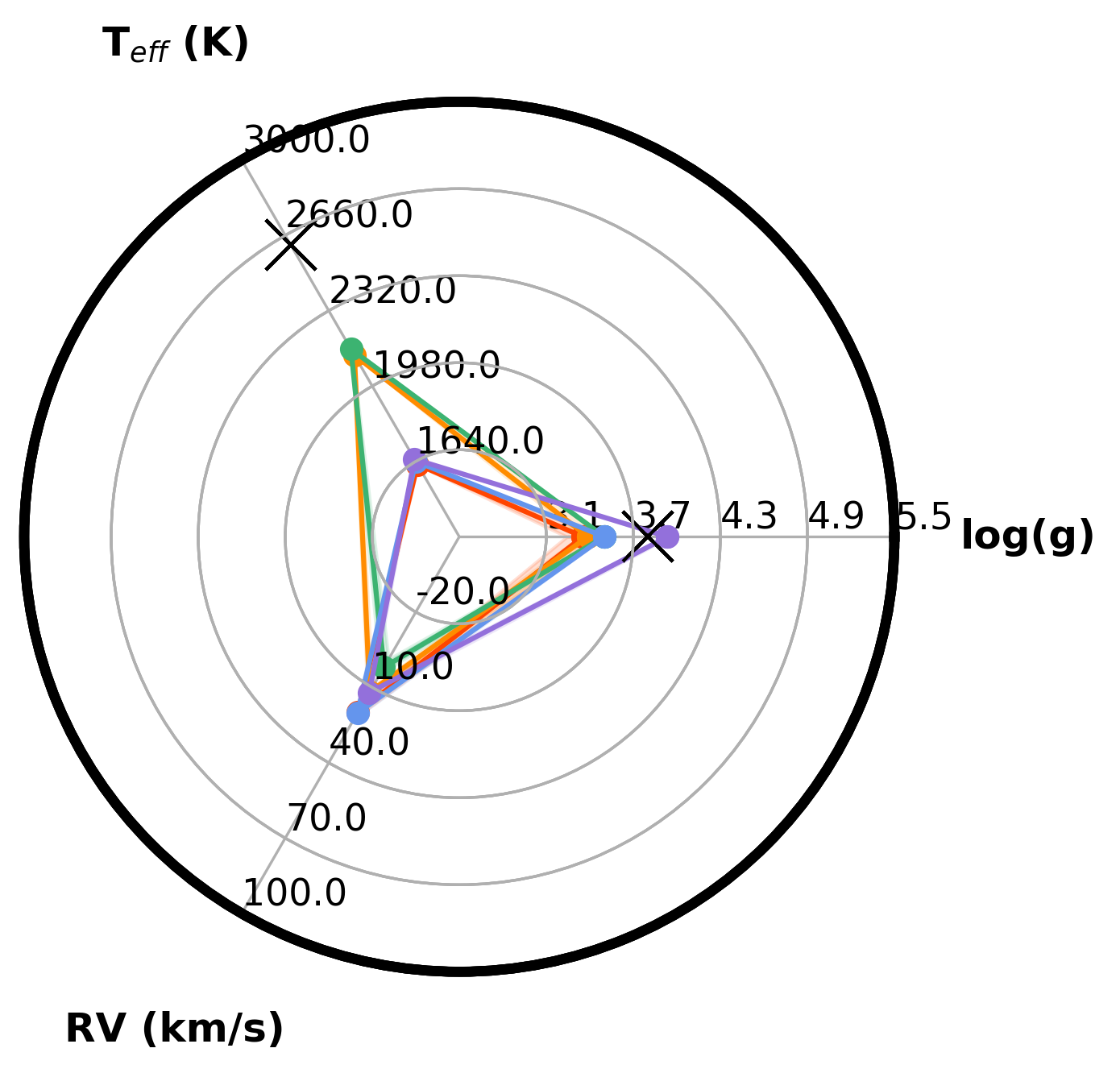}
\end{subfigure}
\caption{Same as Figure \ref{fig:target_CD-35_2722_b} but for KPNO Tau 4}
\label{fig:target_KPNO_Tau_4}
\end{figure*}

\begin{table*}[ht]
\centering
\caption{Same as Table \ref{tab:target_CD-35_2722_b} but for KPNO Tau 4}
\renewcommand{\arraystretch}{1.5} 
\resizebox{\textwidth}{!}{%
\begin{tabular}{lcccccccccccc}
\hline
model& \Teff (K) & \logg (dex) & \met & \co & $\gamma$ & \fsed & RV (km/s) & $\beta$ (km/s) & ln(z) & $\chi ^2 _{red}$ \\ 
\hline
\btex&$1626^{+2}_{-3}$&$3.35^{+0.13}_{-0.11}$&&&&&$19.84^{+2.29}_{-2.4}$&$23.31^{+13.0}_{-12.57}$&-58487.6&59\\ 
\atmo&$2120^{+13}_{-5}$&$3.39^{+0.11}_{-0.05}$&$0.6^{+0.0}_{-0.0}$&$0.7^{+0.0}_{-0.02}$&$1.01^{+0.0}_{-0.0}$&&$11.82^{+1.49}_{-1.85}$&$85^{+3}_{-2}$&-77347.3&78\\ 
\sono&$2146^{+8}_{-7}$&$3.5^{+0.0}_{-0.0}$&&&&$1.0^{+0.01}_{-0.0}$&$1.96^{+2.56}_{-3.18}$&$76.64^{+5.9}_{-5.64}$&-102234.2&102\\ 
\btse&$1639^{+5}_{-5}$&$3.5^{+0.0}_{-0.0}$&&$0.39^{+0.01}_{-0.01}$&&&$20.11^{+2.38}_{-2.37}$&$29.47^{+8.39}_{-10.22}$&-52274.6&52\\ 
\exor&$1650^{+2}_{-3}$&$3.94^{+0.02}_{-0.01}$&$0.6^{+0.03}_{-0.02}$&$0.4^{+0.0}_{-0.0}$&&&$12.01^{+2.04}_{-2.06}$&$73.34^{+4.57}_{-4.51}$&-36090.6&36\\ 
\hline
\end{tabular}}
\label{tab:target_KPNO_Tau_4}
\end{table*}

\begin{table*}[ht]
\centering
\caption{Same as Table \ref{tab:target_CD-35_2722_b} but for KPNO Tau 6}
\renewcommand{\arraystretch}{1.5} 
\resizebox{\textwidth}{!}{%
\begin{tabular}{lcccccccccccc}
\hline
model& \Teff (K) & \logg (dex) & \met & \co & $\gamma$ & \fsed & RV (km/s) & $\beta$ (km/s) & ln(z) & $\chi ^2 _{red}$ \\ 
\hline
\btex&$2507^{+23}_{-19}$&$3.5^{+0.12}_{-0.14}$&&&&&$29.05^{+10.41}_{-7.92}$&$99.9^{+22.78}_{-18.34}$&-10360.6&10\\ 
\atmo&$2319^{+48}_{-16}$&$3.75^{+0.26}_{-0.12}$&$0.58^{+0.02}_{-0.17}$&$0.67^{+0.03}_{-0.36}$&$1.05^{+0.0}_{-0.01}$&&$29.3^{+10.46}_{-6.07}$&$96^{+21}_{-13}$&-9366.5&9\\ 
\sono&$2391^{+9}_{-5}$&$3.55^{+0.15}_{-0.05}$&&&&$6.85^{+1.13}_{-2.86}$&$25.13^{+5.04}_{-4.97}$&$84.65^{+16.82}_{-9.14}$&-8837.6&9\\ 
\hline
\end{tabular}}
\label{tab:target_KPNO_Tau_6}
\end{table*}

\begin{table*}[ht]
\centering
\caption{Same as Table \ref{tab:target_CD-35_2722_b} but for USco 1606-2219}
\renewcommand{\arraystretch}{1.5} 
\resizebox{\textwidth}{!}{%
\begin{tabular}{lcccccccccccc}
\hline
model& \Teff (K) & \logg (dex) & \met & \co & $\gamma$ & \fsed & RV (km/s) & $\beta$ (km/s) & ln(z) & $\chi ^2 _{red}$ \\ 
\hline
\btex&$2351^{+30}_{-22}$&$3.35^{+0.2}_{-0.3}$&&&&&$-7.91^{+12.95}_{-9.61}$&$98.11^{+27.94}_{-21.78}$&-9809.5&10\\ 
\atmo&$2227^{+55}_{-13}$&$3.88^{+0.26}_{-0.1}$&$0.59^{+0.01}_{-0.24}$&$0.35^{+0.33}_{-0.05}$&$1.03^{+0.0}_{-0.01}$&&$-11.06^{+6.67}_{-5.31}$&$107^{+18}_{-12}$&-8706.4&9\\ 
\sono&$1700^{+4}_{-8}$&$4.0^{+0.05}_{-0.02}$&&&&$1.0^{+0.04}_{-0.0}$&$3.16^{+7.79}_{-10.81}$&$69.23^{+21.9}_{-20.74}$&-8539.9&9\\ 
\btse&$1675^{+8}_{-11}$&$3.51^{+0.04}_{-0.01}$&&$0.61^{+0.01}_{-0.02}$&&&$-5.9^{+6.69}_{-5.46}$&$51.82^{+18.11}_{-21.92}$&-7930.5&8\\ 
\exor&$1507^{+104}_{-7}$&$4.0^{+0.15}_{-0.04}$&$0.29^{+0.23}_{-0.05}$&$0.3^{+0.25}_{-0.05}$&&&$-4.85^{+10.0}_{-7.49}$&$64.0^{+16.11}_{-15.98}$&-7583.3&8\\ 
\hline
\end{tabular}}
\label{tab:target_USco1606-2219}
\end{table*}

\begin{figure*}[ht]
\centering
\begin{subfigure}[b]{0.61\textwidth}
    \includegraphics[width=\textwidth]{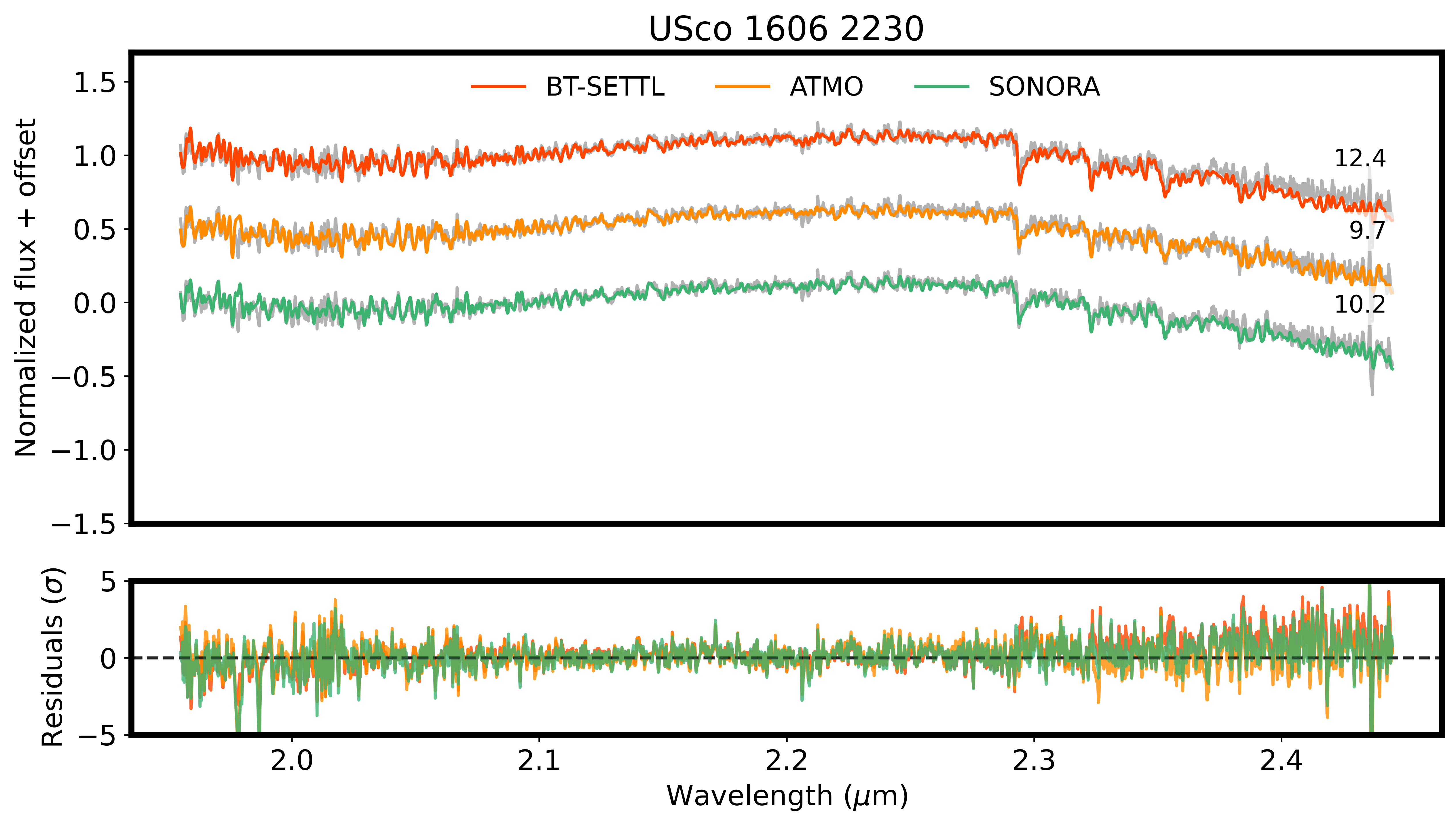}
\end{subfigure}
\hfill
\begin{subfigure}[b]{0.33\textwidth}
    \includegraphics[width=\textwidth]{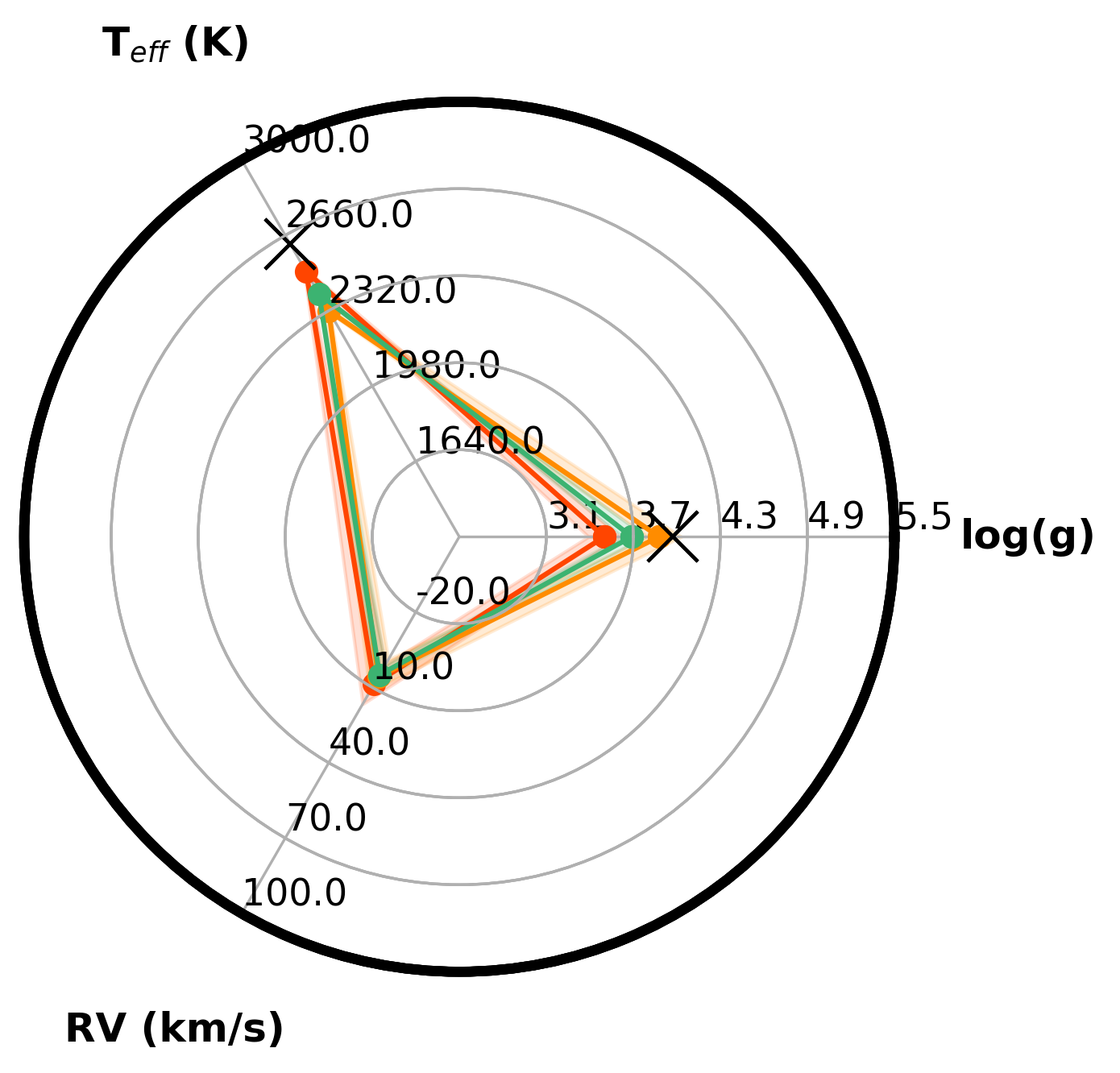}
\end{subfigure}
\caption{Same as Figure \ref{fig:target_CD-35_2722_b} but for USco 1606-2230}
\label{fig:target_USco1606-2230}
\end{figure*}

\begin{table*}[ht]
\centering
\caption{Same as Table \ref{tab:target_CD-35_2722_b} but for USco 1606-2230}
\renewcommand{\arraystretch}{1.5} 
\resizebox{\textwidth}{!}{%
\begin{tabular}{lcccccccccccc}
\hline
model& \Teff (K) & \logg (dex) & \met & \co & $\gamma$ & \fsed & RV (km/s) & $\beta$ (km/s) & ln(z) & $\chi ^2 _{red}$ \\ 
\hline
\btex&$2497^{+18}_{-17}$&$3.5^{+0.11}_{-0.11}$&&&&&$8.64^{+7.98}_{-7.67}$&$96.27^{+14.84}_{-13.27}$&-12421.3&12\\ 
\atmo&$2320^{+61}_{-12}$&$3.88^{+0.15}_{-0.08}$&$0.58^{+0.02}_{-0.46}$&$0.36^{+0.34}_{-0.06}$&$1.04^{+0.0}_{-0.02}$&&$5.75^{+6.04}_{-6.83}$&$96^{+11}_{-8}$&-9690.0&10\\
\sono&$2396^{+4}_{-3}$&$3.69^{+0.09}_{-0.06}$&&&&$6.36^{+1.6}_{-2.34}$&$4.97^{+3.01}_{-3.83}$&$88.93^{+8.64}_{-3.54}$&-10303.1&10\\ 
\hline
\end{tabular}}
\label{tab:target_USco1606-2230}
\end{table*}

\begin{figure*}[ht]
\centering
\begin{subfigure}[b]{0.61\textwidth}
    \includegraphics[width=\textwidth]{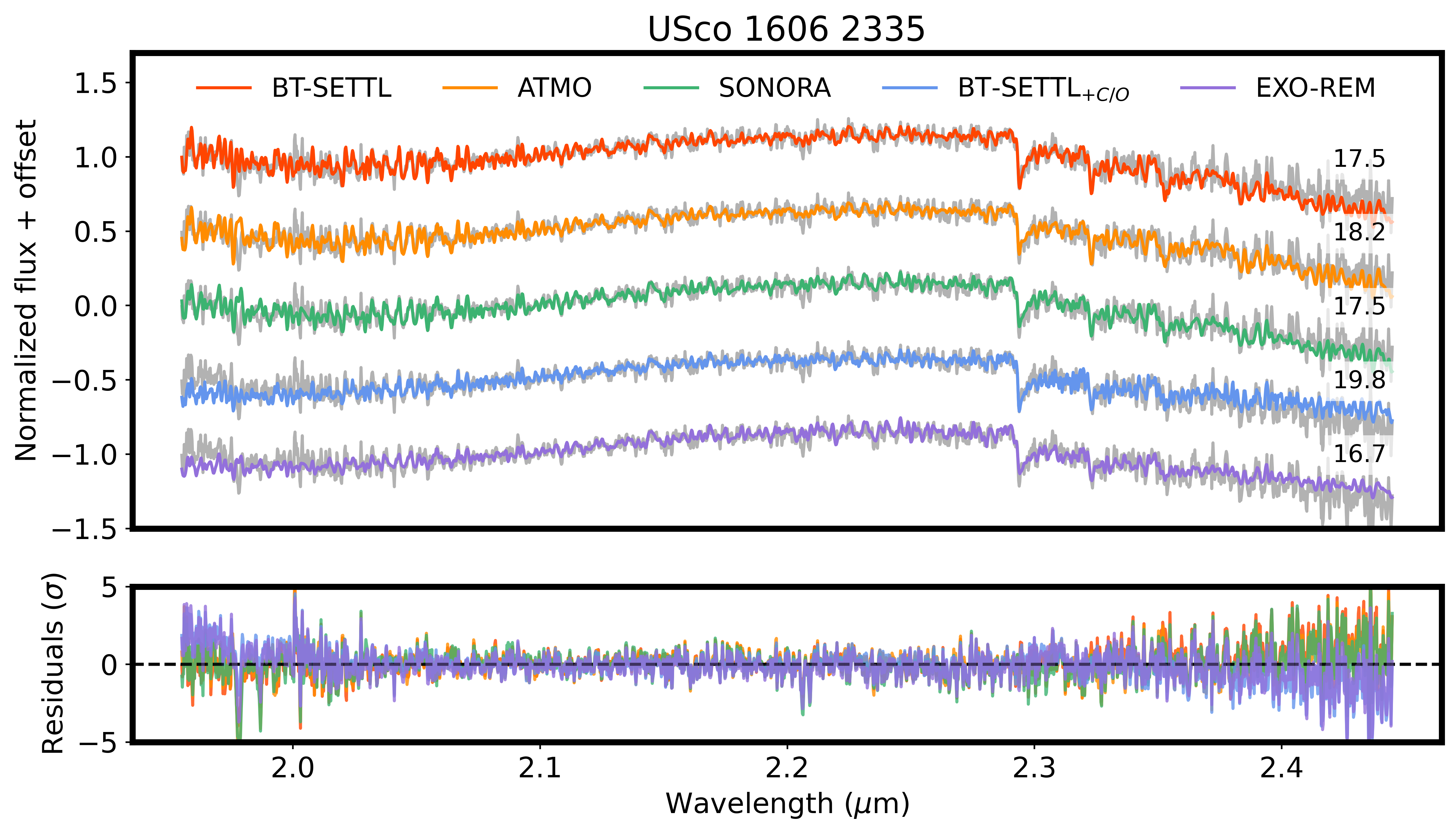}
\end{subfigure}
\hfill
\begin{subfigure}[b]{0.33\textwidth}
    \includegraphics[width=\textwidth]{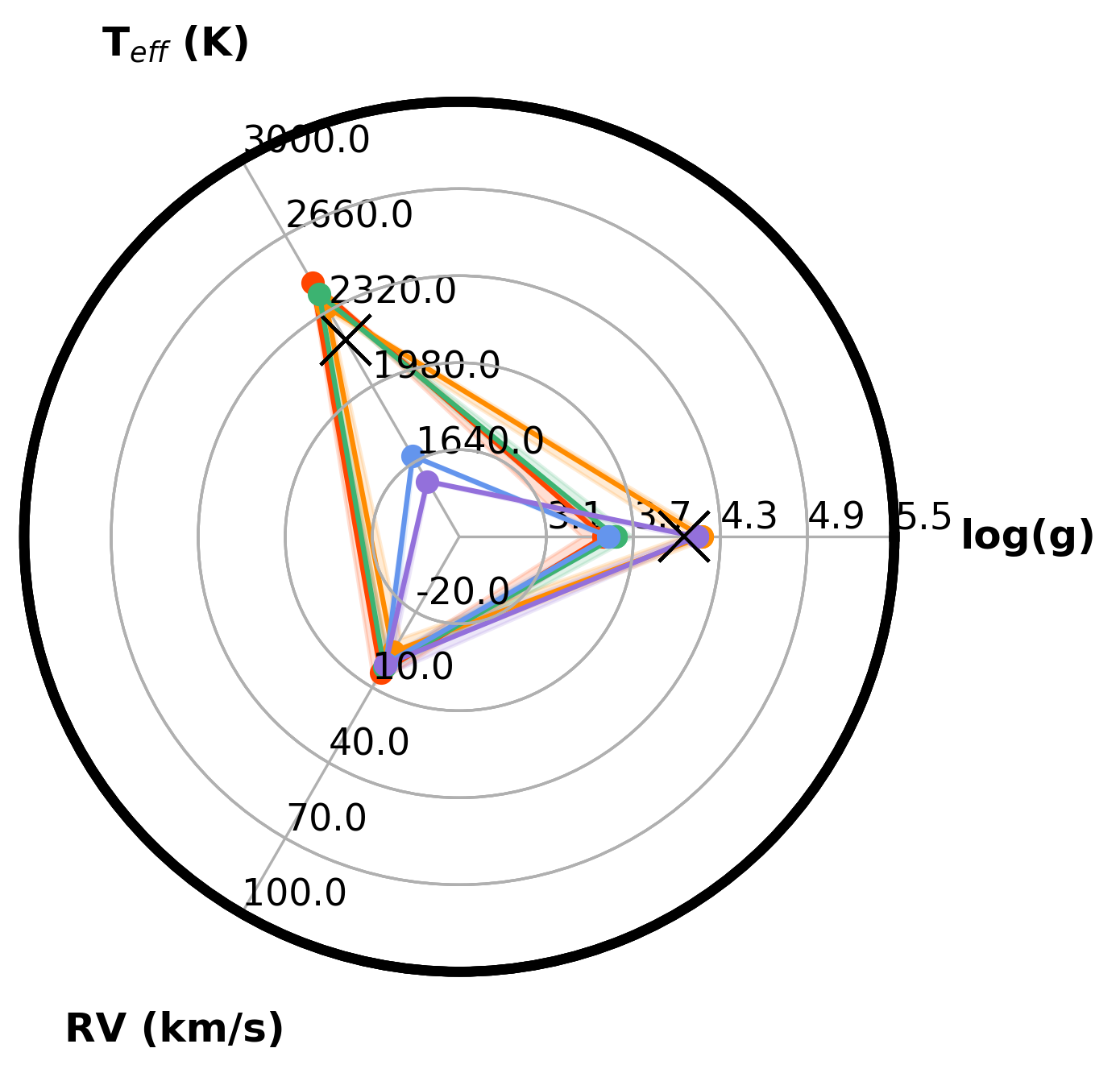}
\end{subfigure}
\caption{Same as Figure \ref{fig:target_CD-35_2722_b} but for USco 1606-2335}
\label{fig:target_USco1606-2335}
\end{figure*}

\begin{table*}[ht]
\centering
\caption{Same as Table \ref{tab:target_CD-35_2722_b} but for USco 1606-2335}
\renewcommand{\arraystretch}{1.5} 
\resizebox{\textwidth}{!}{%
\begin{tabular}{lcccccccccccc}
\hline
model& \Teff (K) & \logg (dex) & \met & \co & $\gamma$ & \fsed & RV (km/s) & $\beta$ (km/s) & ln(z) & $\chi ^2 _{red}$ \\ 
\hline
\btex&$2445^{+17}_{-13}$&$3.5^{+0.04}_{-0.15}$&&&&&$4.12^{+4.79}_{-5.51}$&$76.58^{+12.54}_{-12.38}$&-17467.6&17\\ 
\atmo&$2338^{+12}_{-30}$&$4.17^{+0.07}_{-0.15}$&$0.59^{+0.01}_{-0.09}$&$0.68^{+0.02}_{-0.18}$&$1.04^{+0.01}_{-0.0}$&&$-4.18^{+3.76}_{-5.42}$&$94^{+10}_{-8}$&-18206.7&18\\ 
\sono&$2394^{+6}_{-18}$&$3.58^{+0.12}_{-0.08}$&&&&$1.17^{+0.02}_{-0.03}$&$1.87^{+2.75}_{-2.48}$&$78.61^{+4.28}_{-6.17}$&-17618.8&18\\ 
\btse&$1665^{+3}_{-6}$&$3.53^{+0.05}_{-0.03}$&&$0.68^{+0.01}_{-0.02}$&&&$0.54^{+3.15}_{-4.29}$&$31.41^{+11.95}_{-16.44}$&-19830.3&20\\ 
\exor&$1550^{+2}_{-6}$&$4.14^{+0.03}_{-0.08}$&$0.49^{+0.03}_{-0.07}$&$0.55^{+0.0}_{-0.0}$&&&$0.92^{+5.56}_{-5.91}$&$90.11^{+7.51}_{-7.65}$&-16656.2&17\\ 
\hline
\end{tabular}}
\label{tab:target_USco1606-2335}
\end{table*}

\begin{figure*}[ht]
\centering
\begin{subfigure}[b]{0.61\textwidth}
    \includegraphics[width=\textwidth]{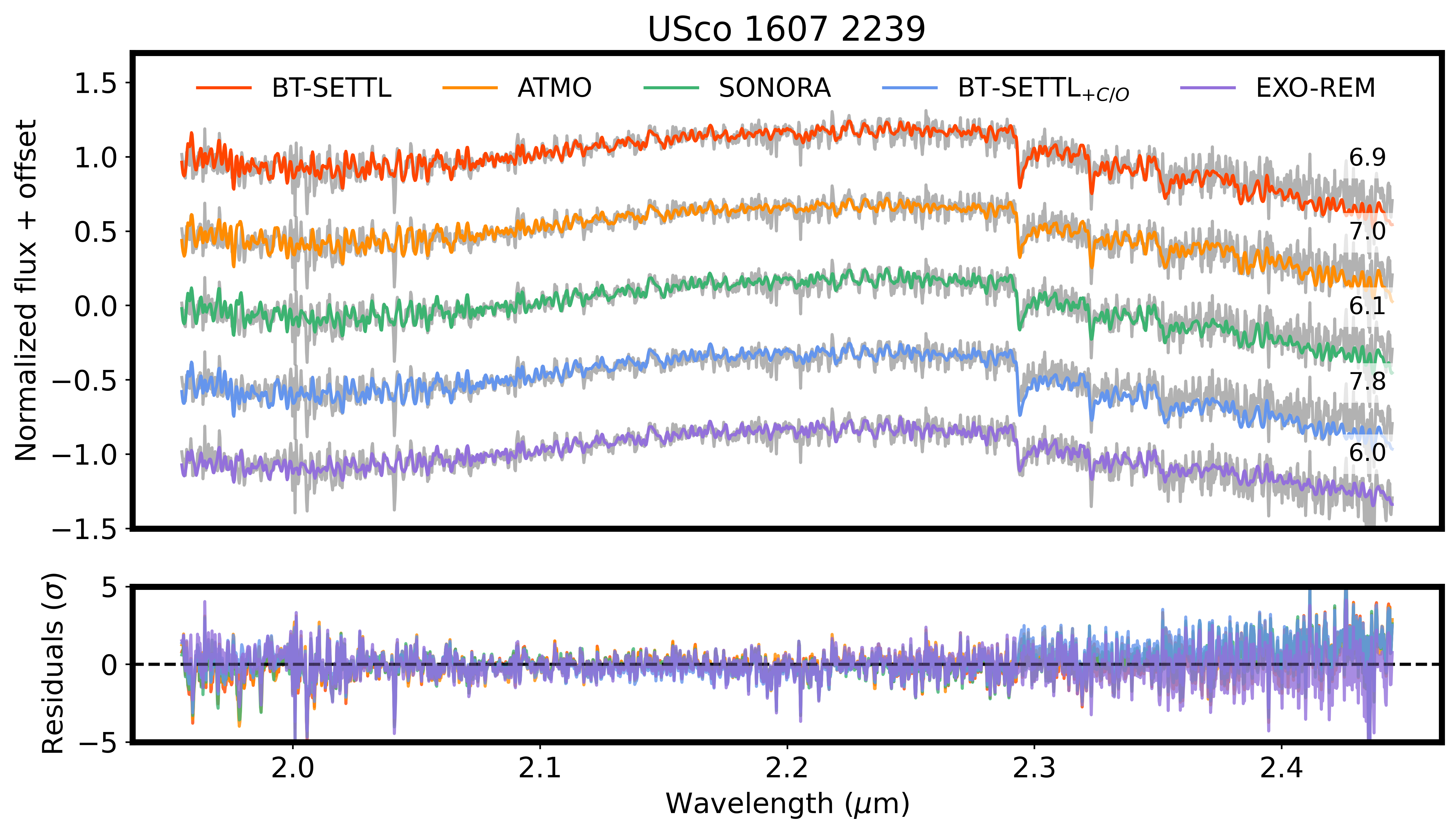}
\end{subfigure}
\hfill
\begin{subfigure}[b]{0.33\textwidth}
    \includegraphics[width=\textwidth]{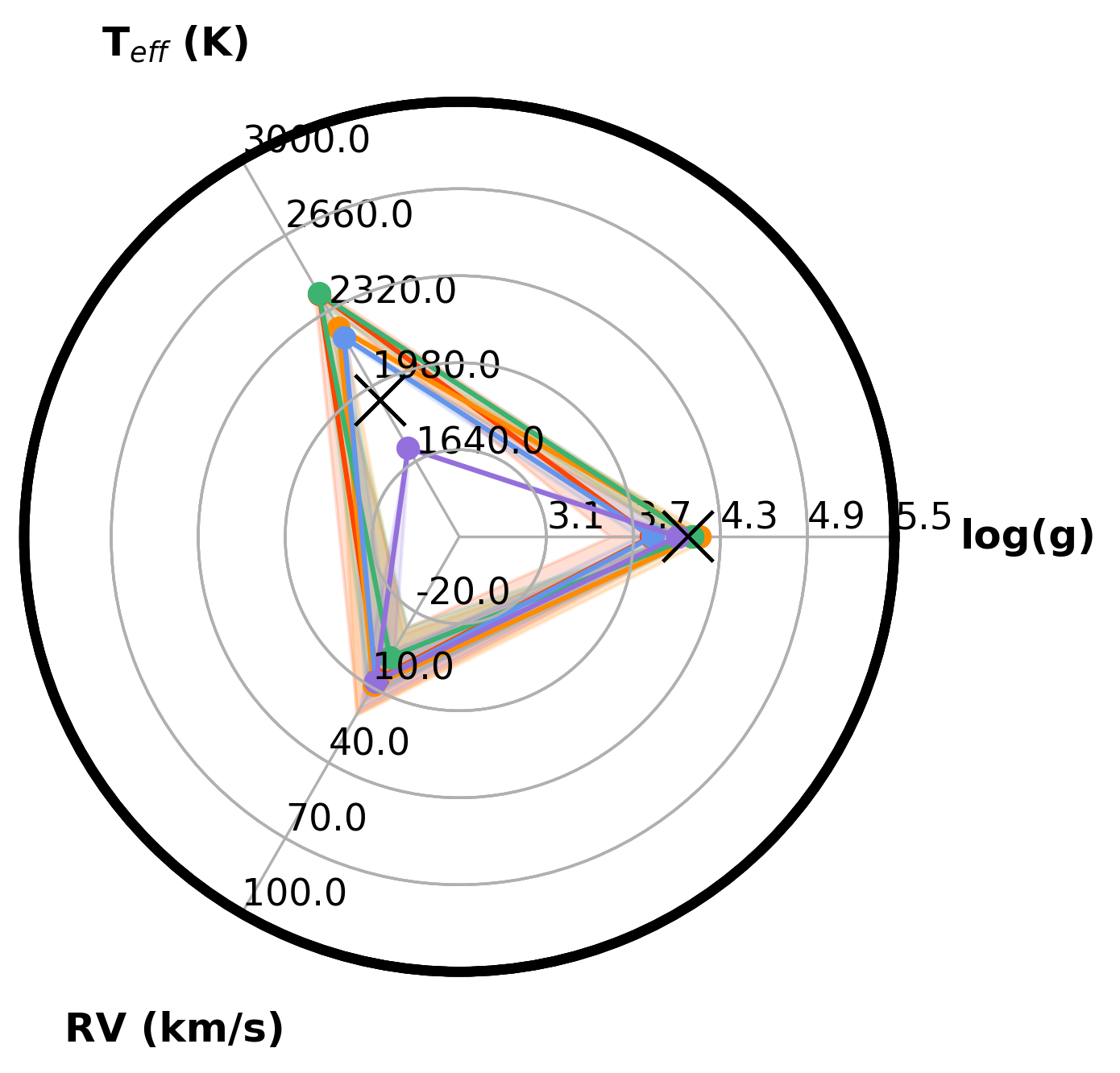}
\end{subfigure}
\caption{Same as Figure \ref{fig:target_CD-35_2722_b} but for USco 1607-2239}
\label{fig:target_USco1607-2239}
\end{figure*}

\begin{table*}[ht]
\centering
\caption{Same as Table \ref{tab:target_CD-35_2722_b} but for USco 1607-2239}
\renewcommand{\arraystretch}{1.5} 
\resizebox{\textwidth}{!}{%
\begin{tabular}{lcccccccccccc}
\hline
model& \Teff (K) & \logg (dex) & \met & \co & $\gamma$ & \fsed & RV (km/s) & $\beta$ (km/s) & ln(z) & $\chi ^2 _{red}$ \\ 
\hline
\btex&$2396^{+39}_{-37}$&$3.82^{+0.32}_{-0.28}$&&&&&$6.6^{+13.84}_{-20.31}$&$102.56^{+30.58}_{-24.07}$&-6912.0&7\\ 
\atmo&$2242^{+25}_{-54}$&$4.15^{+0.09}_{-0.14}$&$0.49^{+0.11}_{-0.25}$&$0.68^{+0.02}_{-0.37}$&$1.04^{+0.01}_{-0.01}$&&$9.08^{+11.98}_{-20.4}$&$114^{+26}_{-19}$&-7005.0&7\\ 
\sono&$2397^{+3}_{-77}$&$4.11^{+0.07}_{-0.11}$&&&&$3.31^{+4.38}_{-0.18}$&$-2.12^{+14.49}_{-12.13}$&$94.26^{+25.93}_{-24.19}$&-6092.2&6\\ 
\btse&$2200^{+0}_{-22}$&$3.83^{+0.18}_{-0.1}$&&$0.54^{+0.01}_{-0.03}$&&&$7.73^{+7.1}_{-10.47}$&$99.63^{+13.47}_{-17.18}$&-7843.6&8\\ 
\exor&$1700^{+4}_{-5}$&$4.01^{+0.08}_{-0.02}$&$-0.08^{+0.06}_{-0.05}$&$0.45^{+0.0}_{-0.01}$&&&$7.59^{+11.07}_{-13.01}$&$93.48^{+20.51}_{-16.57}$&-6002.3&6\\ 
\hline
\end{tabular}}
\label{tab:target_USco1607-2239}
\end{table*}

\begin{figure*}[ht]
\centering
\begin{subfigure}[b]{0.61\textwidth}
    \includegraphics[width=\textwidth]{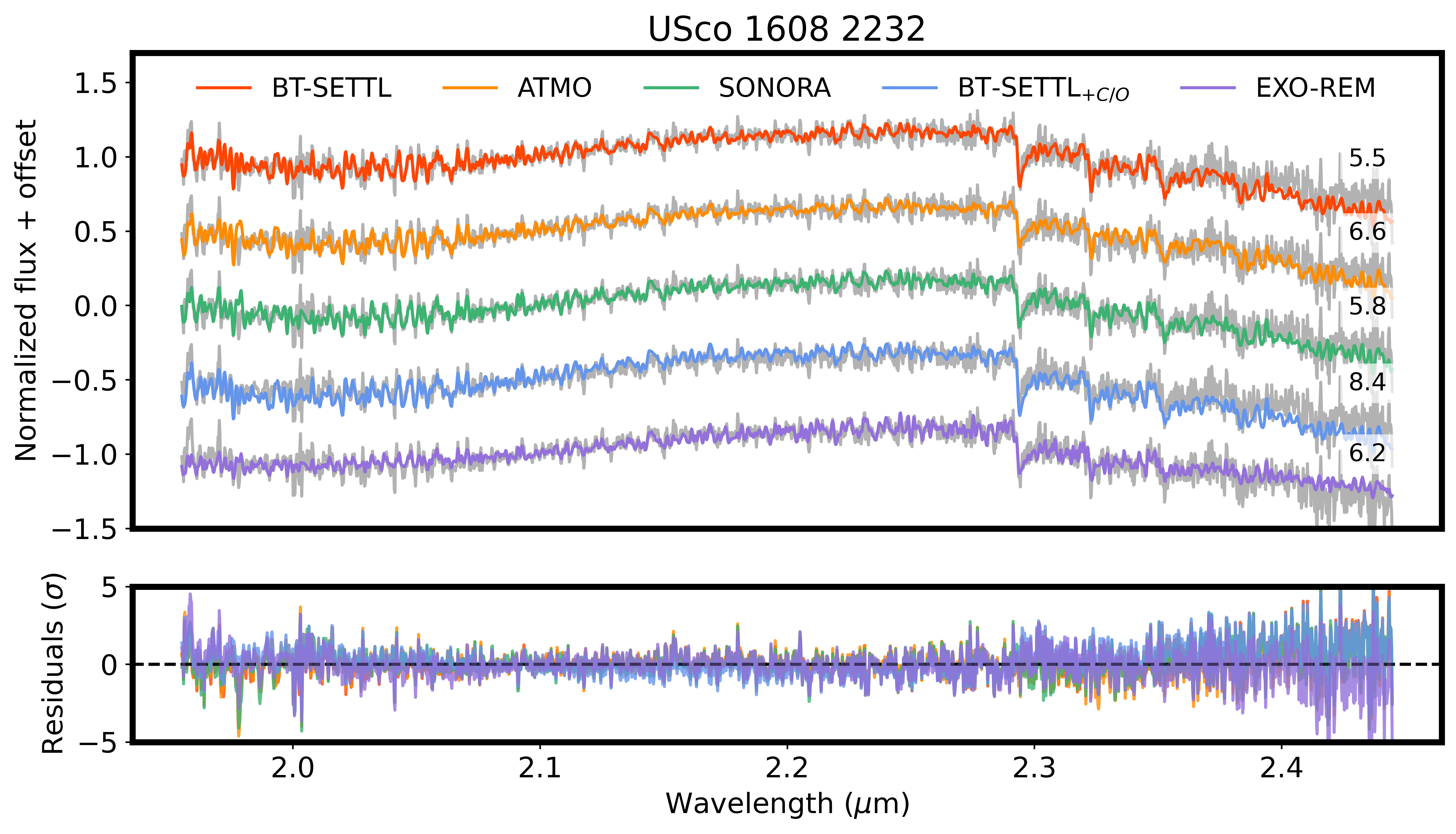}
\end{subfigure}
\hfill
\begin{subfigure}[b]{0.33\textwidth}
    \includegraphics[width=\textwidth]{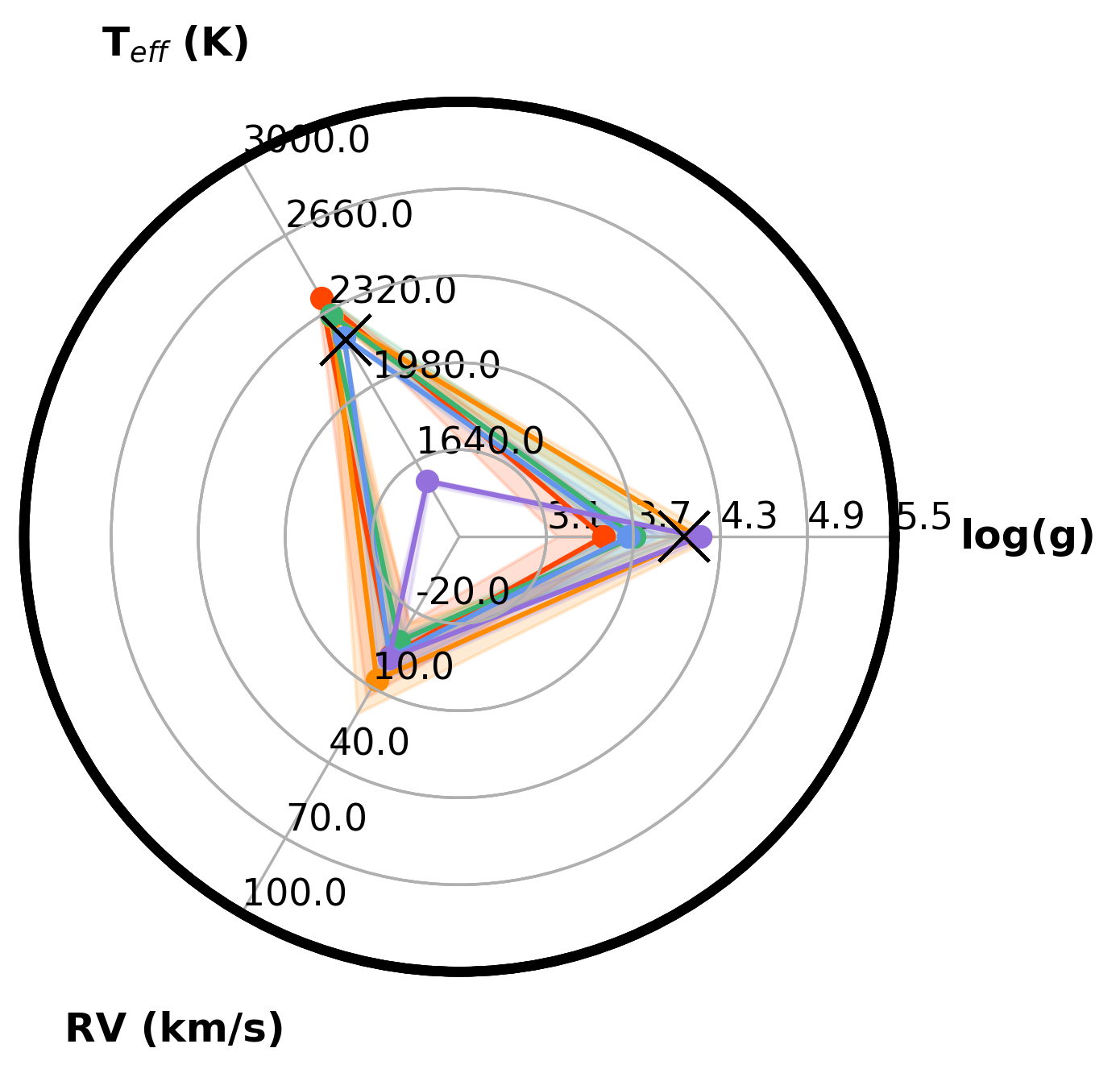}
\end{subfigure}
\caption{Same as Figure \ref{fig:target_CD-35_2722_b} but for USco 1608-2232}
\label{fig:target_USco1608-2232}
\end{figure*}

\begin{table*}[ht]
\centering
\caption{Same as Table \ref{tab:target_CD-35_2722_b} but for USco 1608-2232}
\renewcommand{\arraystretch}{1.5} 
\resizebox{\textwidth}{!}{%
\begin{tabular}{lcccccccccccc}
\hline
model& \Teff (K) & \logg (dex) & \met & \co & $\gamma$ & \fsed & RV (km/s) & $\beta$ (km/s) & ln(z) & $\chi ^2 _{red}$ \\ 
\hline
\btex&$2377^{+35}_{-25}$&$3.49^{+0.22}_{-0.32}$&&&&&$-2.62^{+16.67}_{-13.28}$&$96.62^{+31.23}_{-21.86}$&-5466.9&5\\ 
\atmo&$2294^{+20}_{-50}$&$4.14^{+0.12}_{-0.17}$&$0.59^{+0.01}_{-0.29}$&$0.68^{+0.02}_{-0.37}$&$1.03^{+0.01}_{-0.01}$&&$6.93^{+13.42}_{-22.33}$&$114^{+38}_{-20}$&-6562.5&7\\ 
\sono&$2302^{+93}_{-11}$&$3.71^{+0.33}_{-0.13}$&&&&$1.06^{+2.31}_{-0.05}$&$-8.27^{+7.16}_{-3.11}$&$78.81^{+13.69}_{-10.41}$&-5882.4&6\\ 
\btse&$2200^{+0}_{-2}$&$3.67^{+0.21}_{-0.14}$&&$0.52^{+0.0}_{-0.01}$&&&$-2.07^{+7.73}_{-3.19}$&$95.53^{+12.08}_{-5.67}$&-8469.6&8\\ 
\exor&$1551^{+10}_{-18}$&$4.16^{+0.07}_{-0.18}$&$0.68^{+0.11}_{-0.21}$&$0.55^{+0.01}_{-0.01}$&&&$-1.53^{+9.96}_{-9.73}$&$72.45^{+21.38}_{-18.44}$&-6188.0&6\\ 
\hline
\end{tabular}}
\label{tab:target_USco1608-2232}
\end{table*}

\begin{figure*}[ht]
\centering
\begin{subfigure}[b]{0.61\textwidth}
    \includegraphics[width=\textwidth]{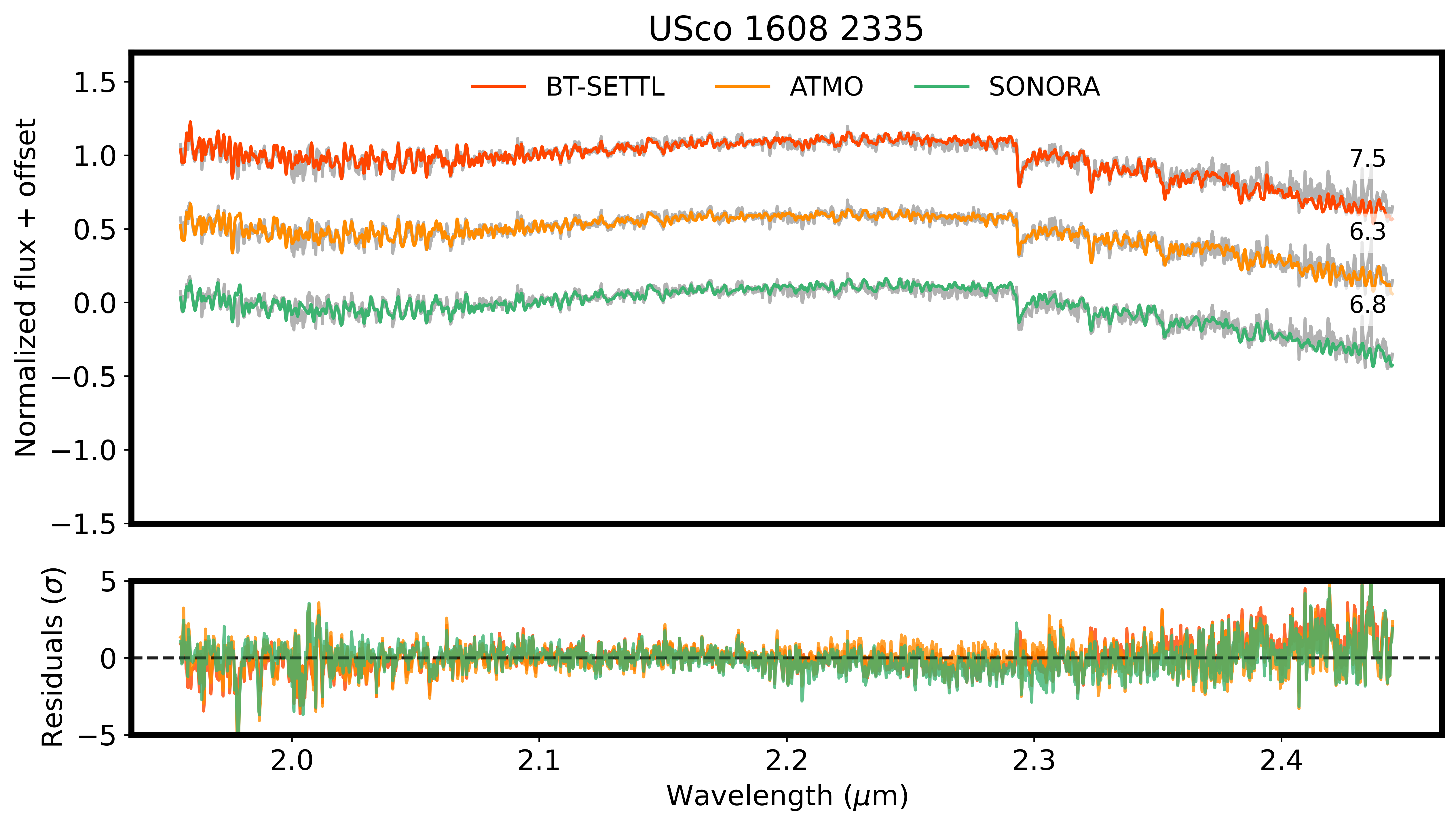}
\end{subfigure}
\hfill
\begin{subfigure}[b]{0.33\textwidth}
    \includegraphics[width=\textwidth]{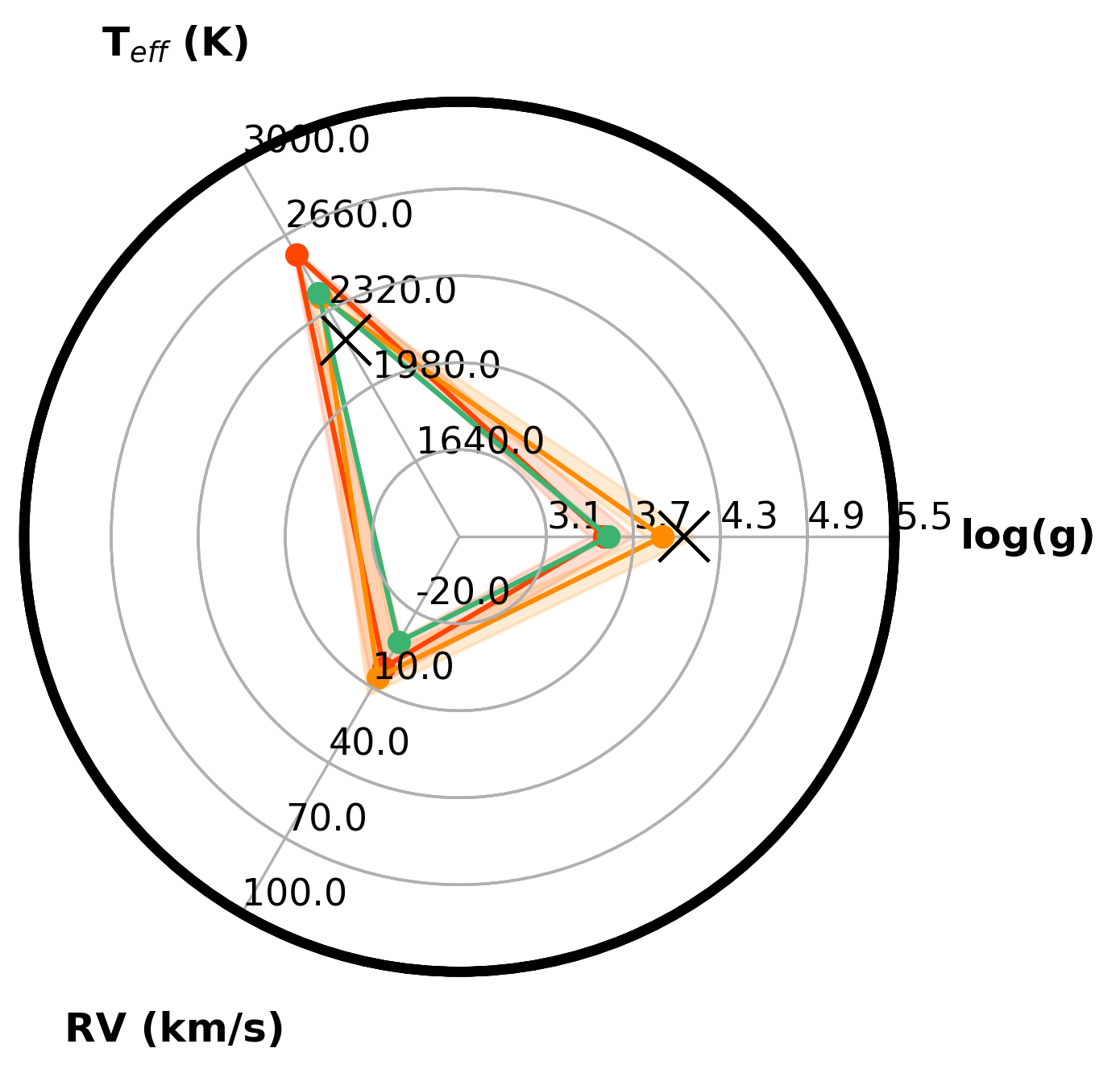}
\end{subfigure}
\caption{Same as Figure \ref{fig:target_CD-35_2722_b} but for USco 1608-2335}
\label{fig:target_USco1608-2335}
\end{figure*}

\begin{table*}[ht]
\centering
\caption{Same as Table \ref{tab:target_CD-35_2722_b} but for USco 1608-2335}
\renewcommand{\arraystretch}{1.5} 
\resizebox{\textwidth}{!}{%
\begin{tabular}{lcccccccccccc}
\hline
model& \Teff (K) & \logg (dex) & \met & \co & $\gamma$ & \fsed & RV (km/s) & $\beta$ (km/s) & ln(z) & $\chi ^2 _{red}$ \\ 
\hline
\btex&$2575^{+27}_{-24}$&$3.5^{+0.2}_{-0.11}$&&&&&$2.07^{+8.32}_{-11.71}$&$76.74^{+18.47}_{-17.05}$&-7475.8&7\\ 
\atmo&$2381^{+74}_{-14}$&$3.9^{+0.23}_{-0.11}$&$0.5^{+0.1}_{-0.37}$&$0.6^{+0.1}_{-0.3}$&$1.05^{+0.0}_{-0.02}$&&$5.93^{+7.23}_{-11.35}$&$92^{+17}_{-14}$&-6317.6&6\\ 
\sono&$2399^{+0}_{-0}$&$3.53^{+0.03}_{-0.03}$&&&&$6.11^{+0.03}_{-0.03}$&$-8.13^{+0.34}_{-0.06}$&$96.05^{+0.6}_{-0.18}$&-6963.2&7\\ 
\hline
\end{tabular}}
\label{tab:target_USco1608-2335}
\end{table*}

\begin{figure*}[ht]
\centering
\begin{subfigure}[b]{0.61\textwidth}
    \includegraphics[width=\textwidth]{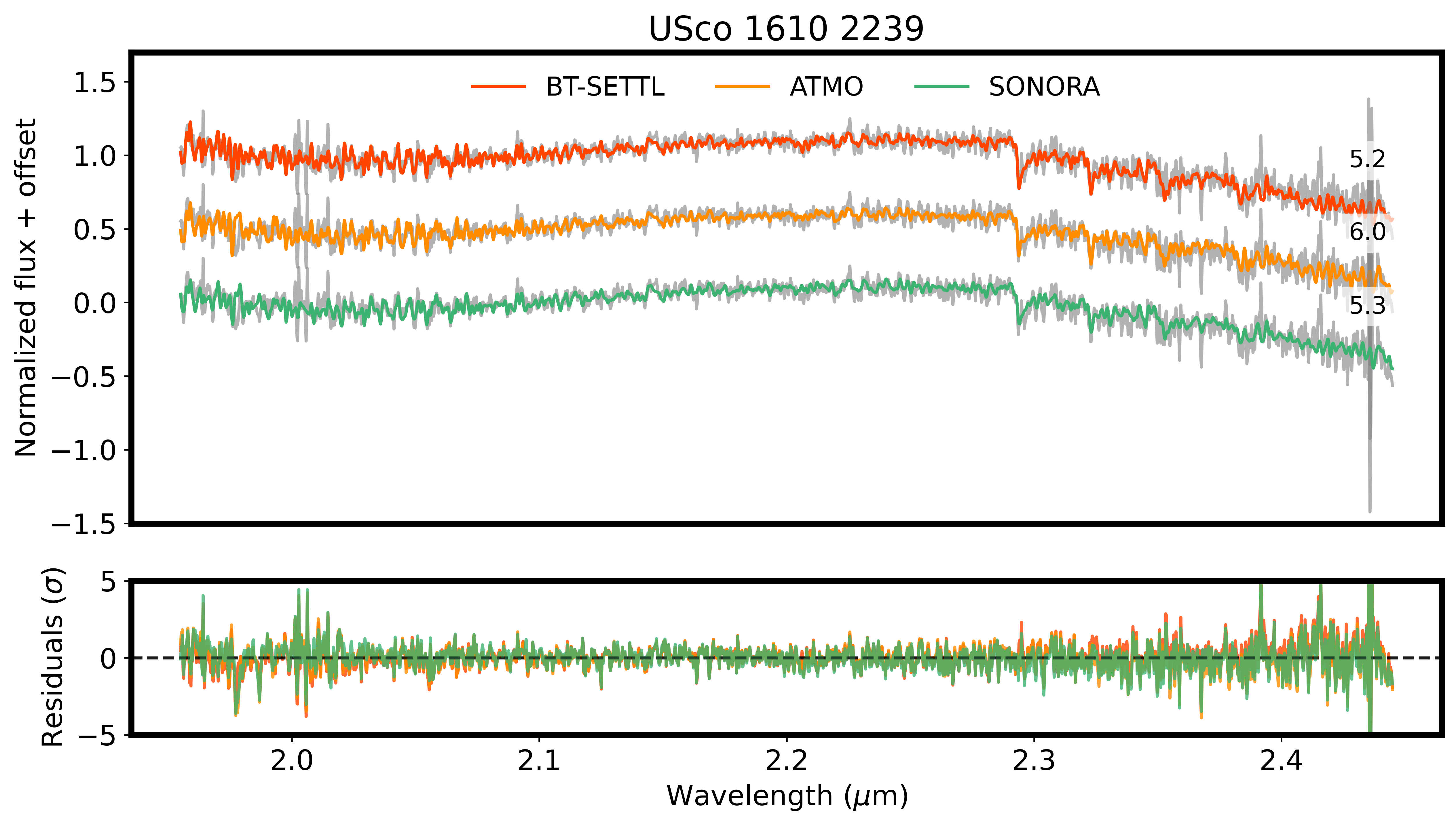}
\end{subfigure}
\hfill
\begin{subfigure}[b]{0.33\textwidth}
    \includegraphics[width=\textwidth]{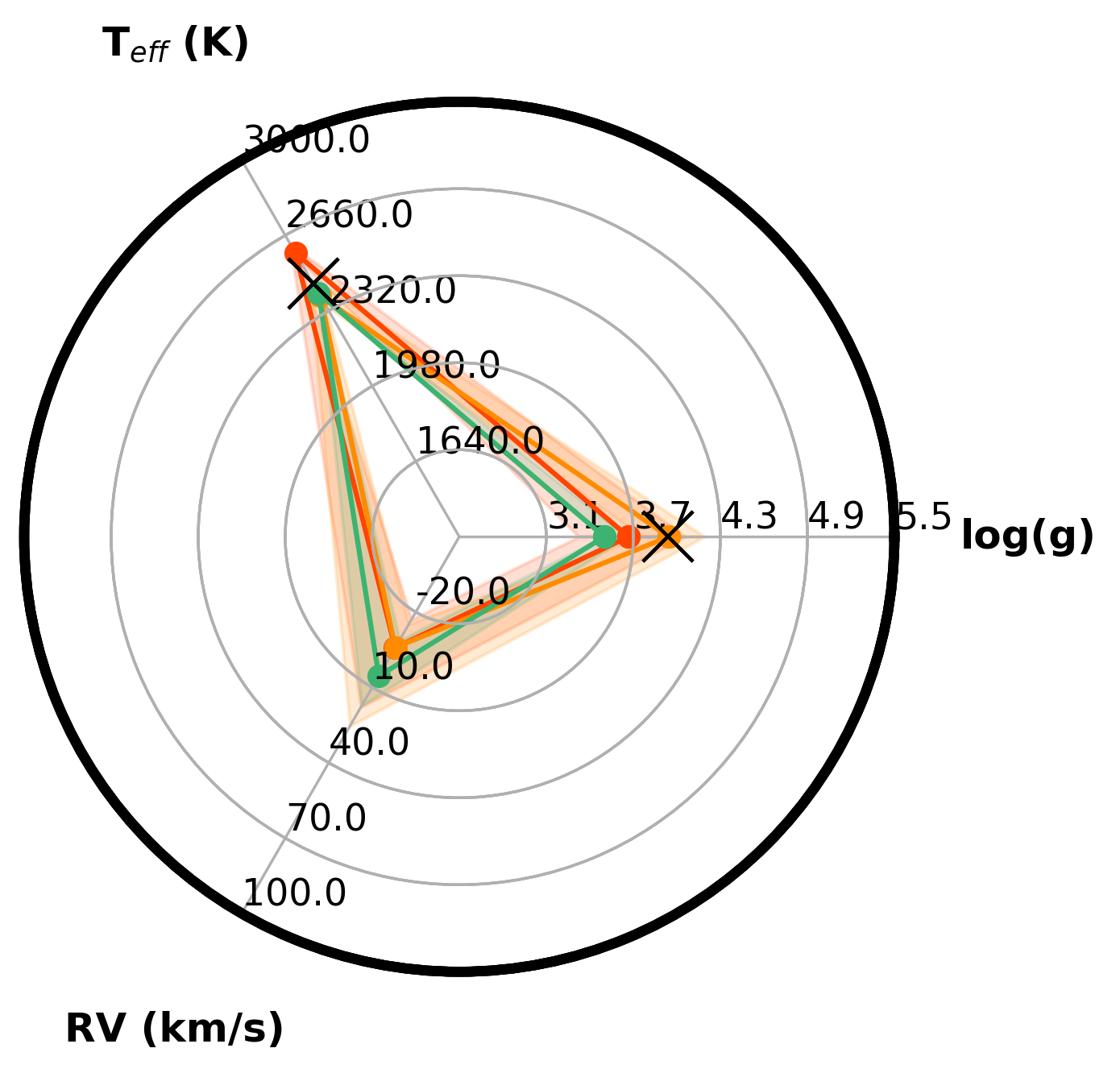}
\end{subfigure}
\caption{Same as Figure \ref{fig:target_CD-35_2722_b} but for USco 1610-2239}
\label{fig:target_USco1610-2239}
\end{figure*}

\begin{table*}[ht]
\centering
\caption{Same as Table \ref{tab:target_CD-35_2722_b} but for USco 1610-2239}
\renewcommand{\arraystretch}{1.5} 
\resizebox{\textwidth}{!}{%
\begin{tabular}{lcccccccccccc}
\hline
model& \Teff (K) & \logg (dex) & \met & \co & $\gamma$ & \fsed & RV (km/s) & $\beta$ (km/s) & ln(z) & $\chi ^2 _{red}$ \\ 
\hline
\btex&$2581^{+41}_{-39}$&$3.67^{+0.39}_{-0.34}$&&&&&$-5.8^{+23.63}_{-11.3}$&$75.59^{+51.28}_{-37.22}$&-5250.4&5\\ 
\atmo&$2386^{+91}_{-40}$&$3.94^{+0.25}_{-0.19}$&$0.58^{+0.02}_{-0.4}$&$0.67^{+0.03}_{-0.36}$&$1.05^{+0.0}_{-0.02}$&&$-5.94^{+31.32}_{-9.27}$&$85^{+56}_{-19}$&-5954.9&6\\ 
\sono&$2399^{+1}_{-11}$&$3.5^{+0.16}_{-0.0}$&&&&$5.67^{+2.32}_{-1.67}$&$5.48^{+11.88}_{-14.27}$&$85.27^{+28.84}_{-25.22}$&-5349.4&5\\ 
\hline
\end{tabular}}
\label{tab:target_USco1610-2239}
\end{table*}

\begin{figure*}[ht]
\centering
\begin{subfigure}[b]{0.61\textwidth}
    \includegraphics[width=\textwidth]{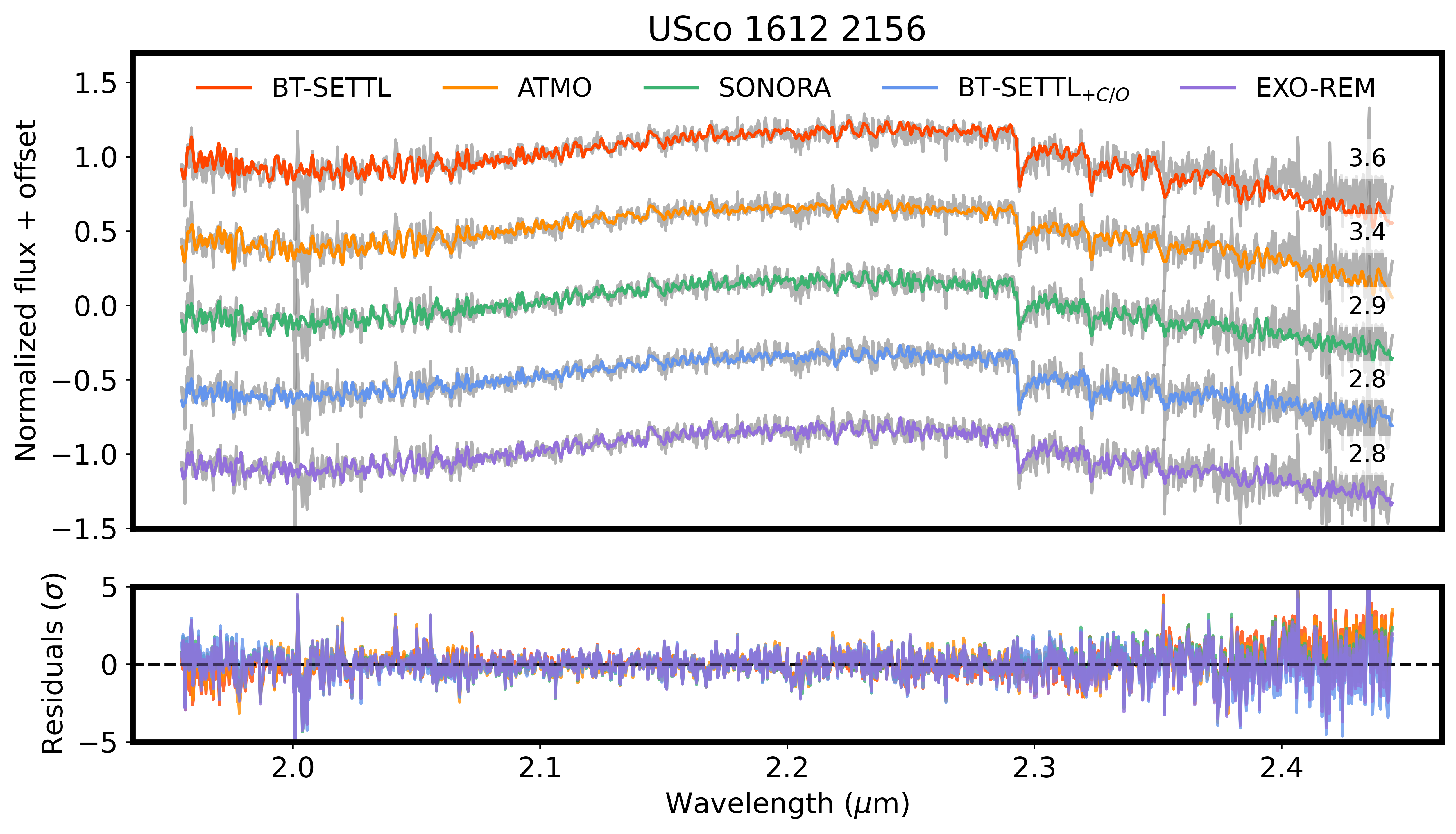}
\end{subfigure}
\hfill
\begin{subfigure}[b]{0.33\textwidth}
    \includegraphics[width=\textwidth]{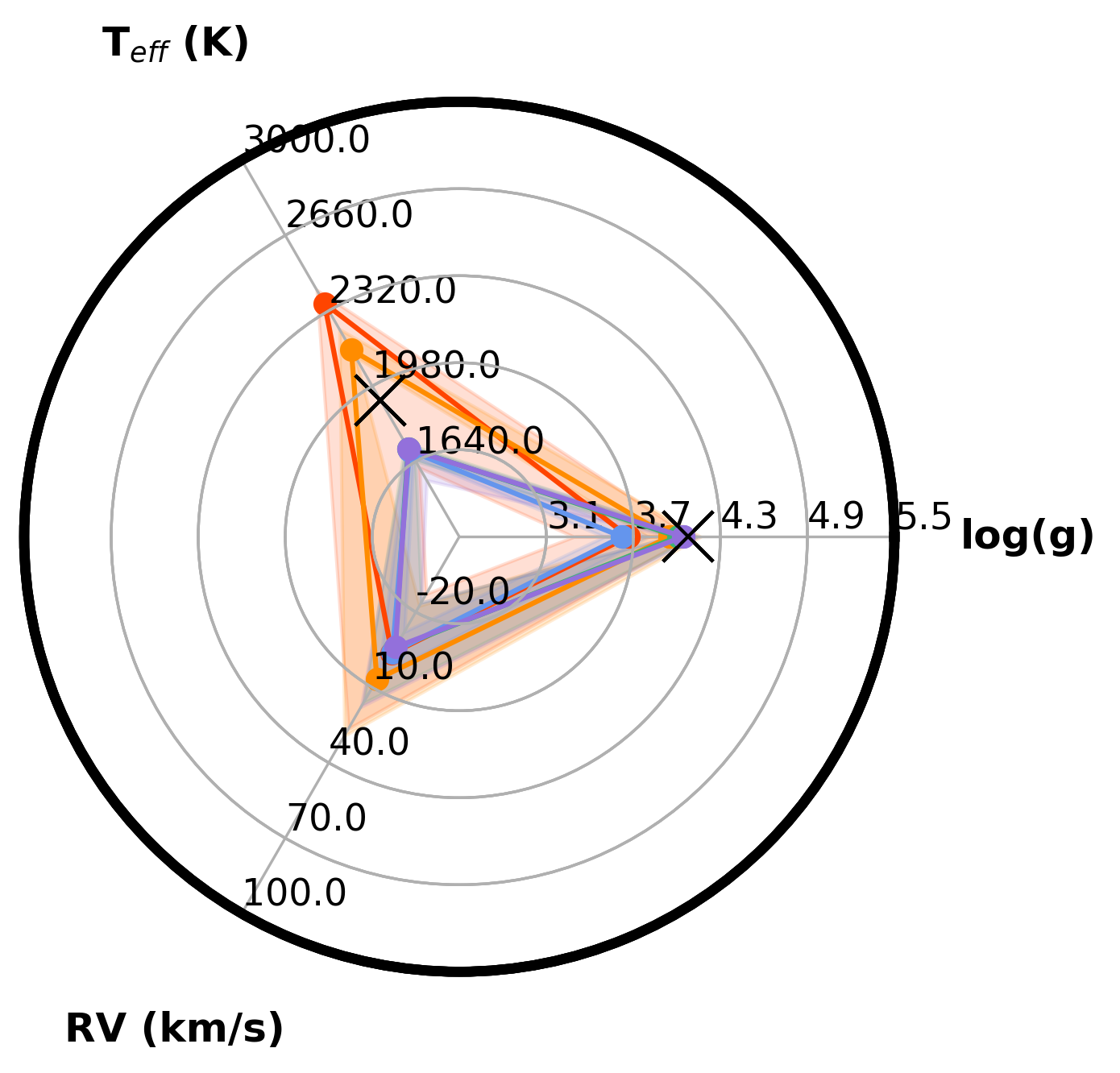}
\end{subfigure}
\caption{Same as Figure \ref{fig:target_CD-35_2722_b} but for USco 1612-2156}
\label{fig:target_USco1612-2156}
\end{figure*}

\begin{table*}[ht]
\centering
\caption{Same as Table \ref{tab:target_CD-35_2722_b} but for USco 1612-2156}
\renewcommand{\arraystretch}{1.5} 
\resizebox{\textwidth}{!}{%
\begin{tabular}{lcccccccccccc}
\hline
model& \Teff (K) & \logg (dex) & \met & \co & $\gamma$ & \fsed & RV (km/s) & $\beta$ (km/s) & ln(z) & $\chi ^2 _{red}$ \\ 
\hline
\btex&$2353^{+61}_{-742}$&$3.67^{+0.43}_{-0.38}$&&&&&$-3.53^{+29.99}_{-23.76}$&$118.08^{+70.33}_{-52.94}$&-3598.3&4\\ 
\atmo&$2143^{+93}_{-33}$&$3.95^{+0.21}_{-0.15}$&$0.19^{+0.38}_{-0.15}$&$0.7^{+0.0}_{-0.39}$&$1.03^{+0.02}_{-0.02}$&&$6.8^{+22.44}_{-29.32}$&$129^{+72}_{-31}$&-3453.1&3\\ 
\sono&$1699^{+23}_{-57}$&$4.01^{+0.12}_{-0.03}$&&&&$1.01^{+0.27}_{-0.01}$&$-5.48^{+21.71}_{-18.35}$&$76.46^{+55.24}_{-53.03}$&-2881.3&3\\ 
\btse&$1689^{+17}_{-30}$&$3.62^{+0.18}_{-0.12}$&&$0.61^{+0.04}_{-0.06}$&&&$-3.94^{+9.84}_{-8.21}$&$73.1^{+22.85}_{-16.94}$&-2882.9&3\\ 
\exor&$1697^{+14}_{-142}$&$4.04^{+0.11}_{-0.08}$&$-0.04^{+0.63}_{-0.09}$&$0.45^{+0.16}_{-0.02}$&&&$-5.99^{+23.73}_{-17.83}$&$89.14^{+53.79}_{-32.66}$&-2856.1&3\\ 
\hline
\end{tabular}}
\label{tab:target_USco1612-2156}
\end{table*}

\begin{figure*}[ht]
\centering
\begin{subfigure}[b]{0.61\textwidth}
    \includegraphics[width=\textwidth]{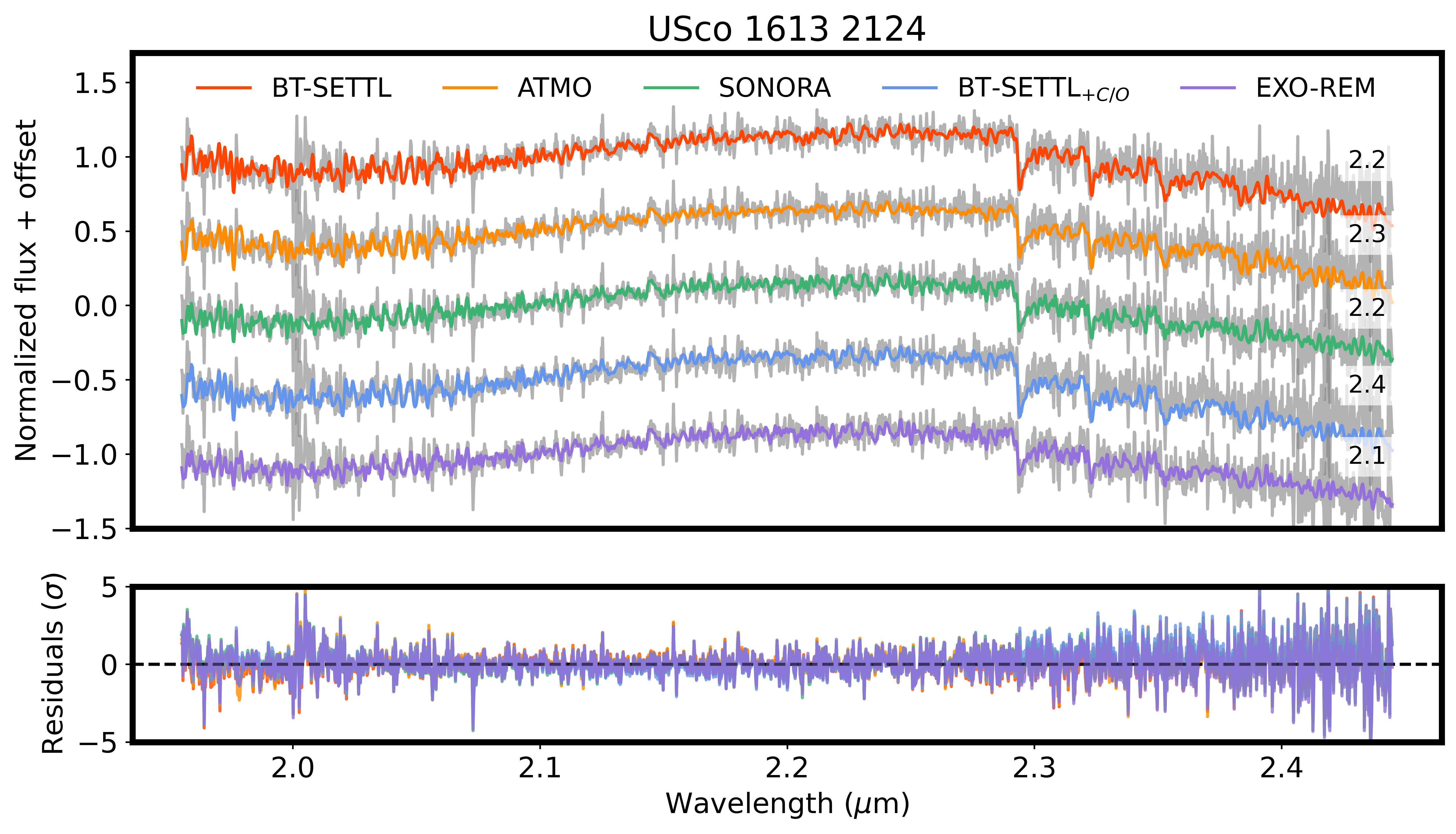}
\end{subfigure}
\hfill
\begin{subfigure}[b]{0.33\textwidth}
    \includegraphics[width=\textwidth]{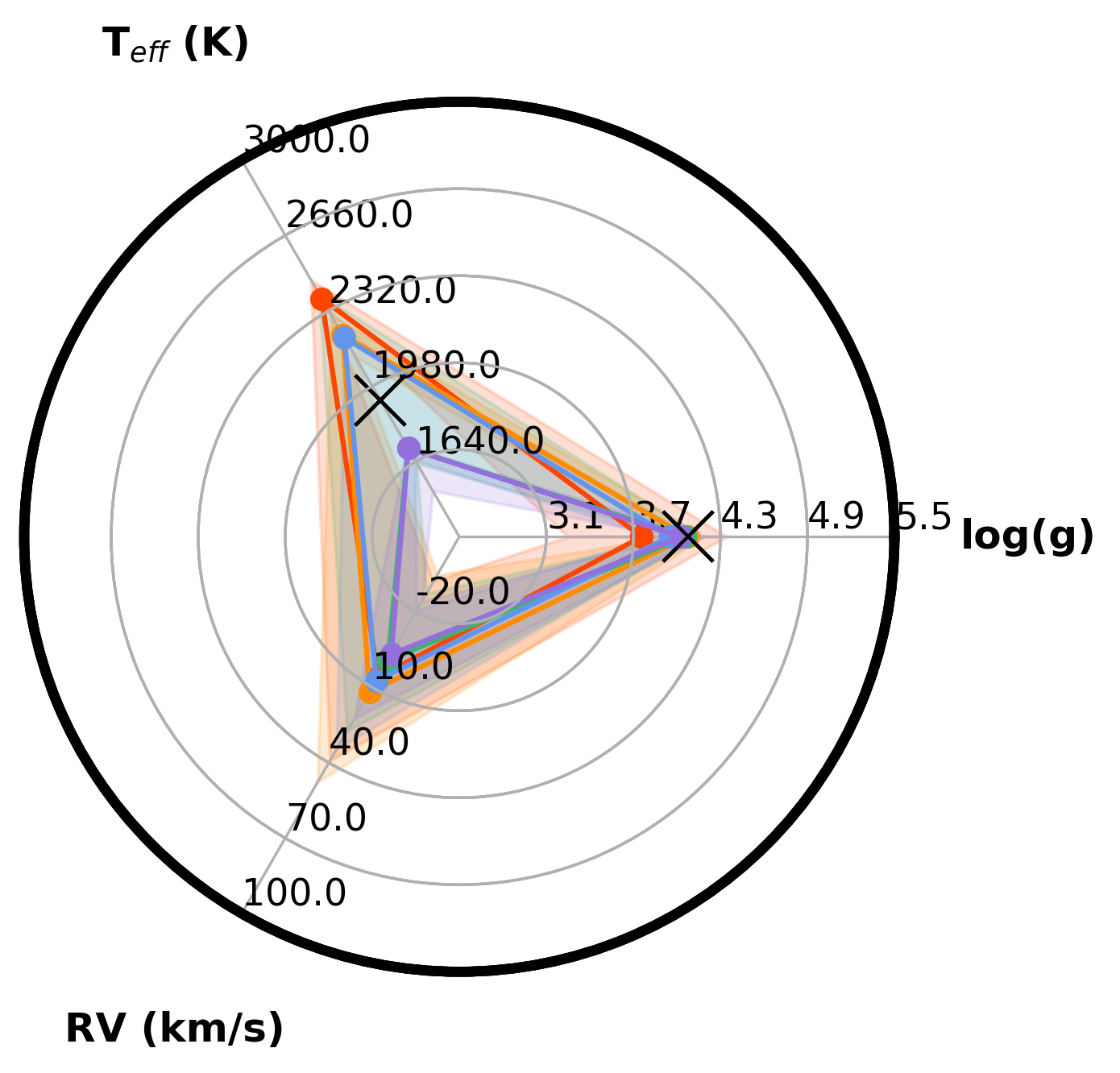}
\end{subfigure}
\caption{Same as Figure \ref{fig:target_CD-35_2722_b} but for USco 1613-2124}
\label{fig:target_USco1613-2124}
\end{figure*}

\begin{table*}[ht]
\centering
\caption{Same as Table \ref{tab:target_CD-35_2722_b} but for USco 1613-2124}
\renewcommand{\arraystretch}{1.5} 
\resizebox{\textwidth}{!}{%
\begin{tabular}{lcccccccccccc}
\hline
model& \Teff (K) & \logg (dex) & \met & \co & $\gamma$ & \fsed & RV (km/s) & $\beta$ (km/s) & ln(z) & $\chi ^2 _{red}$ \\ 
\hline
\btex&$2375^{+84}_{-58}$&$3.76^{+0.58}_{-0.51}$&&&&&$6.12^{+33.04}_{-38.44}$&$98.26^{+84.27}_{-53.34}$&-2223.0&2\\ 
\atmo&$2211^{+88}_{-101}$&$4.07^{+0.12}_{-0.11}$&$0.44^{+0.16}_{-0.46}$&$0.65^{+0.05}_{-0.35}$&$1.04^{+0.01}_{-0.03}$&&$11.82^{+35.95}_{-46.65}$&$116^{+110}_{-40}$&-2331.5&2\\ 
\sono&$1699^{+697}_{-69}$&$4.06^{+0.11}_{-0.07}$&&&&$1.02^{+6.86}_{-0.02}$&$-0.85^{+28.29}_{-26.57}$&$64.32^{+87.73}_{-56.49}$&-2185.8&2\\ 
\btse&$2198^{+2}_{-553}$&$3.96^{+0.16}_{-0.11}$&&$0.52^{+0.07}_{-0.04}$&&&$7.3^{+27.24}_{-27.94}$&$91.31^{+71.97}_{-51.73}$&-2424.6&2\\ 
\exor&$1699^{+10}_{-192}$&$4.03^{+0.15}_{-0.06}$&$-0.07^{+0.86}_{-0.1}$&$0.45^{+0.17}_{-0.02}$&&&$-3.17^{+25.46}_{-21.05}$&$77.16^{+61.05}_{-43.48}$&-2152.1&2\\ 
\hline
\end{tabular}}
\label{tab:target_USco1613-2124}
\end{table*}

\begin{figure*}[ht]
\centering
\begin{subfigure}[b]{0.61\textwidth}
    \includegraphics[width=\textwidth]{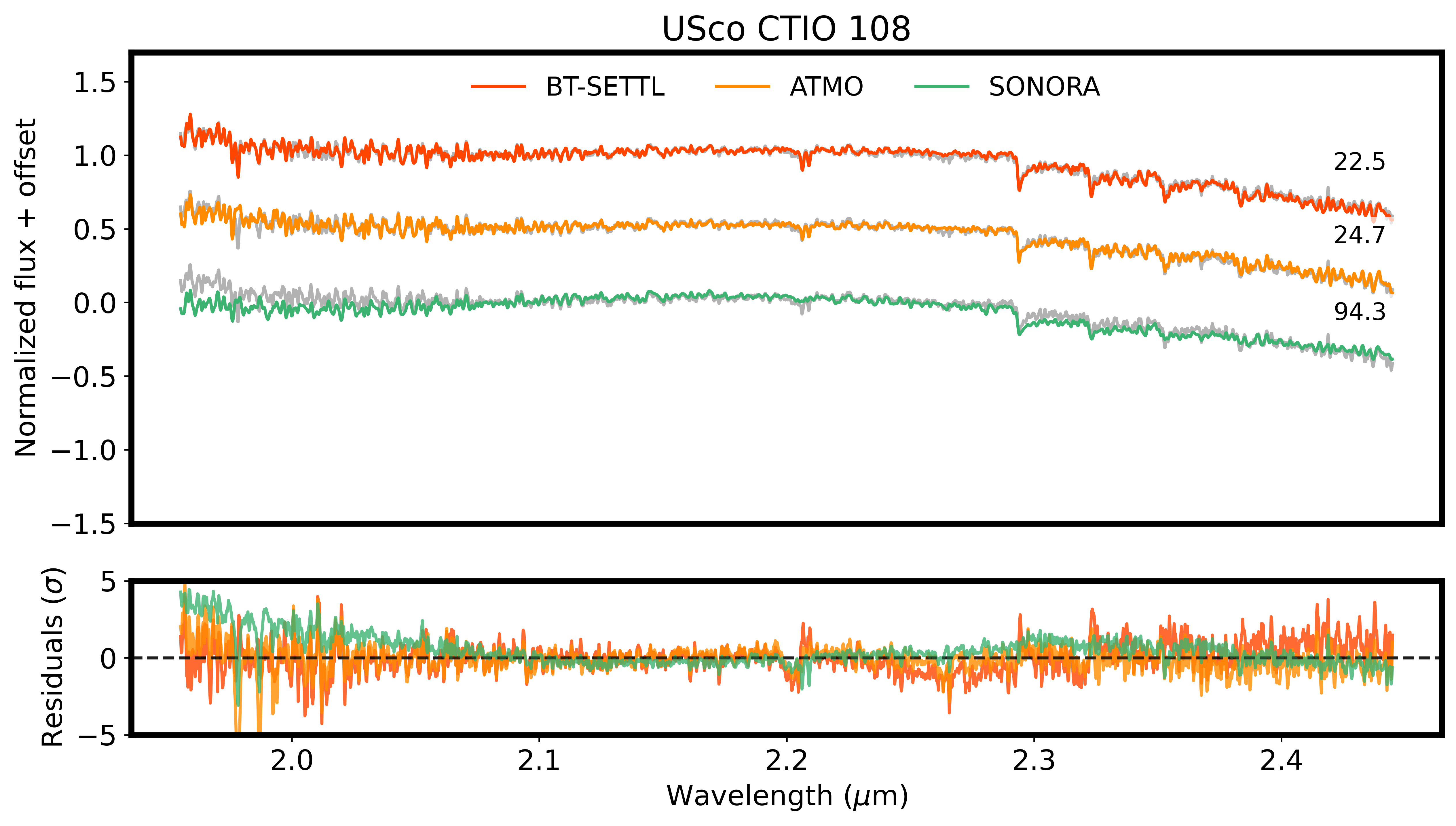}
\end{subfigure}
\hfill
\begin{subfigure}[b]{0.33\textwidth}
    \includegraphics[width=\textwidth]{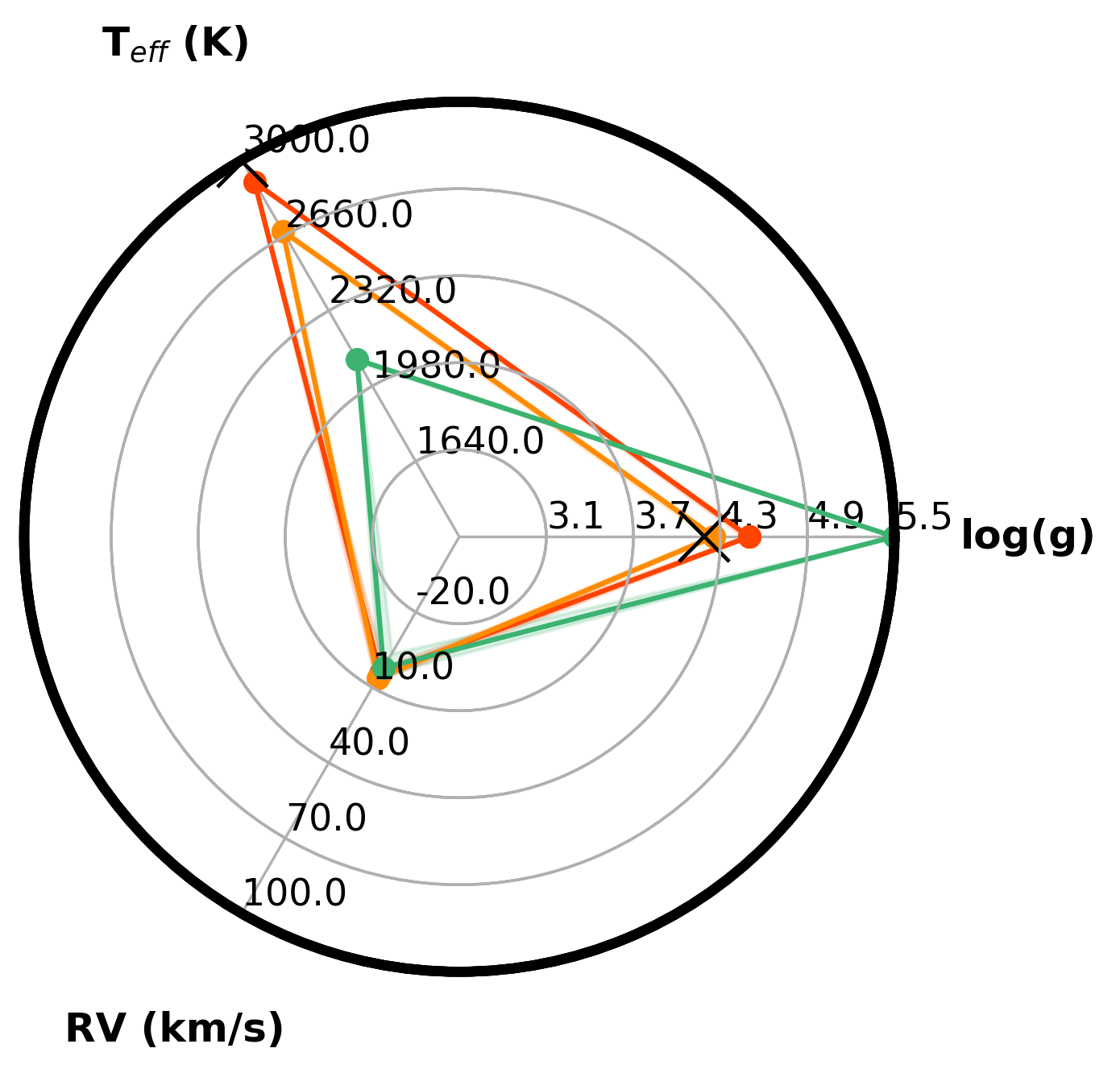}
\end{subfigure}
\caption{Same as Figure \ref{fig:target_CD-35_2722_b} but for USco CTIO 108 A}
\label{fig:target_UScoCTIO108}
\end{figure*}

\begin{table*}[ht]
\centering
\caption{Same as Table \ref{tab:target_CD-35_2722_b} but for USco CTIO 108 A}
\renewcommand{\arraystretch}{1.5} 
\resizebox{\textwidth}{!}{%
\begin{tabular}{lcccccccccccc}
\hline
model& \Teff (K) & \logg (dex) & \met & \co & $\gamma$ & \fsed & RV (km/s) & $\beta$ (km/s) & ln(z) & $\chi ^2 _{red}$ \\ 
\hline
\btex&$2900^{+0}_{-3}$&$4.5^{+0.01}_{-0.04}$&&&&&$4.55^{+2.91}_{-2.83}$&$76.88^{+7.6}_{-7.09}$&-22497.9&23\\ 
\atmo&$2680^{+10}_{-17}$&$4.26^{+0.02}_{-0.04}$&$-0.08^{+0.08}_{-0.05}$&$0.69^{+0.01}_{-0.05}$&$1.05^{+0.0}_{-0.0}$&&$6.25^{+1.48}_{-1.48}$&$75^{+3}_{-3}$&-24661.7&25\\ 
\sono&$2100^{+1}_{-12}$&$5.5^{+0.0}_{-0.0}$&&&&$1.66^{+0.06}_{-0.04}$&$2.2^{+3.47}_{-5.66}$&$82.54^{+9.71}_{-9.17}$&-94087.4&94\\ 
\hline
\end{tabular}}
\label{tab:target_UScoCTIO108}
\end{table*}

\clearpage
\end{appendix}
\end{document}